\documentclass[amsmath, amssymb, preprintnumbers, showpacs, showkeys, aps, prb,superscriptaddress,twocolumn]{revtex4-2}
\usepackage{amssymb}
\usepackage[utf8]{inputenc}
\usepackage{graphicx}
\usepackage{amsmath}
\usepackage{csquotes}
\usepackage{color, xcolor}
\usepackage{placeins}
\usepackage{braket}
\usepackage{comment}
\usepackage[colorlinks=true, linkcolor={red!80!black},citecolor={blue!70!black},urlcolor={blue!80!black},
pdfstartview=FitV,
pdftitle={},
pdfauthor={},
pdfsubject={},
pdfkeywords={},
pdfpagemode=None,
bookmarksopen=true
]{hyperref}

\newcommand{\beq}{\begin{equation}}
	\newcommand{\eneq}{\end{equation}}

\newcommand{\M}{M}
\newcommand{\ii}{i}

\usepackage{amsmath, amssymb, ascmac, mathtools, bm, braket, bbold}
\begin{document}

\title{Exceptional second-order topological insulators}
	
\author{Yutaro Tanaka}
\email{yutaro.tanaka.ay@riken.jp}
\affiliation{RIKEN Center for Emergent Matter Science, Wako, Saitama, 351-0198, Japan}

\author{Daichi Nakamura}
\email{daichi.nakamura@issp.u-tokyo.ac.jp}
\affiliation{Institute for Solid State Physics, University of Tokyo, Kashiwa, Chiba 277-8581, Japan}
	
\author{Ryo Okugawa}
\email{okugawa@rs.tus.ac.jp}
\affiliation{Department of Applied Physics, Tokyo University of Science, Tokyo 125-8585, Japan}

\author{Kohei Kawabata}
\email{kawabata@issp.u-tokyo.ac.jp}
\affiliation{Institute for Solid State Physics, University of Tokyo, Kashiwa, Chiba 277-8581, Japan}
	
%\author{Name}
%\affiliation{affiliation}

\date{\today}
	
\begin{abstract} 
Point-gap topological phases of non-Hermitian systems exhibit exotic boundary states that have no counterparts in Hermitian systems. 
Here, we develop classification of second-order point-gap topological phases protected by reflection symmetry. 
Based on this classification, we propose exceptional second-order topological insulators, exhibiting second-order boundary states stabilized by point-gap topology.
As an illustrative example, we uncover a two-dimensional exceptional second-order topological insulator with point-gapless corner states.
Furthermore, we identify a three-dimensional exceptional second-order topological insulator that features hinge states with isolated exceptional points, representing second-order topological phases intrinsic to non-Hermitian systems. 
Our work enlarges the family of point-gap topological phases in non-Hermitian systems.
\end{abstract}
	
\maketitle
\section{Introduction}
Topological insulators have become a central concept 
in condensed matter physics \cite{RevModPhys.82.3045, RevModPhys.83.1057}. 
A characteristic feature of topological insulators is the bulk-boundary correspondence, where $d$-dimensional topological insulators possess $(d-1)$-dimensional boundary states characterized by bulk topological invariants.
This framework extends to higher-order topological insulators: $d$-dimensional $n$th-order topological insulators support $(d-n)$-dimensional boundary states \cite{PhysRevLett.108.126807,PhysRevLett.110.046404,benalcazar2017quantized, PhysRevB.96.245115, PhysRevLett.119.246401, PhysRevLett.119.246402, schindler2018higher, fang2017rotation}.

Non-Hermiticity arises in effective descriptions of open classical and quantum systems \cite{Gamow1928, PhysRev.56.750,  FESHBACH1958357, FESHBACH1962287, Kane:2014uu, Huber:2016tg, RevModPhys.88.035002, feng2017non, el2018non, kozii2017non,  PhysRevLett.121.026403, PhysRevB.99.201107}, 
enriching topological phases and bulk-boundary correspondence \cite{doi:10.1080/00018732.2021.1876991, RevModPhys.93.015005}. 
Complex energy spectra in non-Hermitian systems give rise to the concept of point gap \cite{PhysRevLett.120.146402, PhysRevX.8.031079, PhysRevX.9.041015}. 
A point gap is defined to be open when a region in the complex energy plane is devoid of the eigenstates with respect to a reference point.
While certain point-gap topological phases are continuously deformable to Hermitian phases~\cite{PhysRevLett.124.086801, PhysRevLett.132.136401}, others are unique to non-Hermitian systems and lead to exotic topological phenomena without Hermitian counterparts, such as non-Hermitian skin effects \cite{PhysRevLett.116.133903, PhysRevB.97.121401, PhysRevX.8.031079, PhysRevLett.121.086803, PhysRevLett.121.136802, PhysRevLett.121.026808, PhysRevX.9.041015, PhysRevB.99.201103, PhysRevB.99.235112, PhysRevLett.123.016805, PhysRevLett.123.066404, PhysRevLett.123.170401, PhysRevLett.123.246801, PhysRevResearch.1.023013, PhysRevLett.124.056802, PhysRevLett.124.086801, PhysRevB.101.195147, PhysRevLett.125.126402, PhysRevResearch.2.022062,  
PhysRevB.103.165123, PhysRevB.104.165117, PhysRevB.105.075128, yokomizo2021non, PhysRevB.105.L180406, PhysRevB.105.245143, PhysRevLett.129.070401, PhysRevLett.129.086601, zhang2022universal, doi:10.1063/5.0097530, PhysRevResearch.4.033122, PhysRevX.13.021007, PhysRevB.109.035131, PhysRevB.109.144203, PhysRevLett.133.136502, peters2024hinge, ishikawa2024z4skin} 
and their higher-order~\cite{PhysRevB.102.205118, PhysRevB.102.241202, PhysRevB.103.045420, PhysRevB.103.205205, PhysRevB.104.035424, PhysRevLett.131.116601, PhysRevB.108.174307} and defect-induced~\cite{PhysRevLett.127.066401, PhysRevB.104.L161106, PhysRevB.104.L241402} extensions.
These unique point-gap topological phases have been experimentally observed in various metamaterials \cite{xiao2020non, weidemann2020topological, helbig2020generalized, brandenbourger2019non, chen2021realization, zhang2021acoustic, PhysRevResearch.2.023265, palacios2021guided}, including two-dimensional (2D) systems \cite{palacios2021guided, PhysRevResearch.2.023265, shang2022experimental, wu2023spin,  PhysRevLett.132.063804, PhysRevLett.132.113802, zhong2024higher}.

Another notable feature of point-gap topological phases is the emergence of boundary states with anomalous complex energy dispersion~\cite{PhysRevResearch.2.023364, PhysRevLett.126.216405, denner2021exceptional,  Denner_2023_infernal, PRXQuantum.4.030315, PhysRevLett.131.256602,  PhysRevB.109.035418,  PhysRevLett.132.136401, Hamanaka2024PRL}, distinct from skin modes.
These point-gap topological phases are called exceptional topological insulators \cite{denner2021exceptional}.
Three-dimensional (3D) exceptional topological insulators exhibit surface states with the characteristic dispersion $k_x + ik_y$ or with an odd number of exceptional points (EPs).
However, it has been largely unclear how non-Hermiticity diversifies higher-order topological phases and enriches boundary phenomena therein.
This stems from the lack of systematic classification for crystalline non-Hermitian topology, in contrast to the counterparts with internal symmetry~\cite{PhysRevX.8.031079, PhysRevX.9.041015}.

In this paper, we develop classification of second-order point-gap topological phases protected by reflection symmetry. 
Based on this classification, we propose exceptional second-order topological insulators, non-Hermitian topological phases featuring point-gapless corner and hinge states.
Our classification generally identifies possible second-order point-gap topology and uncovers previously unexplored second-order non-Hermitian topological phenomena.
As a prototypical example, we introduce a 2D exceptional second-order topological insulator in class D that hosts corner states protected by pseudo-reflection symmetry.
Furthermore, we present a 3D exceptional second-order topological insulator in class AIII $+$ $\mathcal{S}_{-}$ supporting point-gapless hinge states with a single EP, which has no counterparts in Hermitian second-order topological insulators.
Our work expands the family of topological phases in non-Hermitian systems.

This paper is organized as follows. 
In Sec.~\ref{sec:classification}, we generally classify reflection-symmetric second-order point-gap topology in 2D and 3D non-Hermitian systems. In Sec.~\ref{sec:2DETI}, we introduce a 2D exceptional second-order topological insulator in class D $+$ $\tilde{\mathcal{M}}_+$ as a prototypical example of the second-order point-gap topological phases. In Sec.~\ref{sec:3DETI}, we construct a 3D exceptional second-order topological insulator in class AIII $+$ $\mathcal{S}_{-}$ $+$ ${\mathcal{M}}_{++}$ from 2D exceptional topological insulators. Conclusion and discussion are given in Sec.~\ref{sec:conclusion}.

%intrinsic without breaking of reflection symmetry
\begin{table}[t]
	\centering
	\caption{Classification of reflection-symmetric second-order point-gap topology in two and three dimensions.
    The nontrivial topological phases are shown with compatible reflection symmetry ${\cal \M}$ and pseudo-reflection symmetry $\tilde{\cal \M}$ (or equivalently, reflection symmetry$^{\dag}$).
    In the complex Altland-Zirnbauer class, the subscripts of $\mathcal{S}$ specify the commutation ($+$) or anticommutation ($-$) relation with chiral symmetry, and the subscripts of ${\cal \M}$ and $\tilde{\cal \M}$ specify the commutation ($+$) or anticommutation ($-$) relation with chiral and/or sublattice symmetry.
    In the real Altland-Zirnbauer($^{\dag}$) class, the subscripts of ${\cal \M}$ and $\tilde{\cal \M}$ specify the commutation ($+$) or anticommutation ($-$) relation with time-reversal and/or particle-hole symmetry($^{\dag}$).
    The classification is intrinsic (i.e., termination-independent) without local breaking of reflection symmetry~\cite{physrevb.97.205135}.
    }
	\label{tab: AZ}
     \begin{tabular}{ccc} \hline \hline
    Class & $d=2$ & $d=3$ \\ \hline 
    A & $\mathbb{Z}$ (${\cal \M}$) & $\mathbb{Z}$ ($\tilde{{\cal \M}}$)  \\
    AIII & $\mathbb{Z}$ (${\cal M}_-$, $\tilde{\cal M}_+$) & $\mathbb{Z}$ (${\cal M}_+$, $\tilde{\cal M}_-$) \\
    AIII + $\mathcal{S}_+$ & $\mathbb{Z}$ (${\cal M}_{\pm +}$, $\tilde{\cal M}_{\pm +}$) & $\mathbb{Z}$ (${\cal M}_{\pm -}$, $\tilde{\cal M}_{\pm -}$) \\
    A + $\mathcal{S}$ & $\mathbb{Z} \oplus \mathbb{Z}$ (${\cal M}_+$) & $\mathbb{Z} \oplus \mathbb{Z}$ ($\tilde{{\cal M}}_-$)  \\
    AIII + $\mathcal{S}_-$ & $\mathbb{Z} \oplus \mathbb{Z}$ (${\cal M}_{-+}$, $\tilde{\cal M}_{+-}$) & $\mathbb{Z} \oplus \mathbb{Z}$ (${\cal M}_{++}$, $\tilde{\cal M}_{--}$) \\ \hline
    AI & $\mathbb{Z}$ ($\mathcal{\M}_+$) & $\mathbb{Z}$ ($\tilde{\mathcal{\M}}_-$) \\
    BDI & $\mathbb{Z}_2$ (${\cal \M}_{++}$, $\tilde{\cal \M}_{-+}$), & $\mathbb{Z}$ (${\cal \M}_{++}$, $\tilde{\cal \M}_{-+}$) \\
    &  $\mathbb{Z}$ (${\cal \M}_{+-}$, $\tilde{\cal \M}_{++}$) & 
    \\
    D & $\mathbb{Z}_2$ (${\cal \M}_{+}$, $\tilde{\cal \M}_+$), $2\mathbb{Z}$ (${\cal \M}_{-}$) & $\mathbb{Z}_2$ (${\cal \M}_+$), $\mathbb{Z}$ ({$\tilde{\cal \M}_+$}) 
    \\
    DIII & $\mathbb{Z}_2$ (${\cal \M}_{-+}$, $\tilde{\cal \M}_{++}$) 
    & $\mathbb{Z}_2$ (${\cal \M}_{\pm+}$, $\tilde{\cal \M}_{+\pm}$), \\
    & $2\mathbb{Z}$ (${\cal \M}_{+-}$, $\tilde{\cal \M}_{--}$), 
    &  $2\mathbb{Z}$ (${\cal \M}_{--}$, $\tilde{\cal \M}_{-+}$) \\
    AII & $\mathbb{Z}_2$ (${\cal M}_{-}$), $2\mathbb{Z}$ (${\cal M}_{+}$) & $\mathbb{Z}_2$ ($\tilde{{\cal M}}_{+}$), $2\mathbb{Z}$ ($\tilde{{\cal M}}_{-}$) \\
    CII & $2\mathbb{Z}$ (${\cal M}_{+-}$, $\tilde{{\cal M}}_{++}$) &
    $\mathbb{Z}_2$ (${\cal \M}_{--}$, $\tilde{\cal \M}_{+-}$)  \\
    & & $2\mathbb{Z}$ (${\cal \M}_{++}$, $\tilde{\cal \M}_{-+}$), \\
    C & $2\mathbb{Z}$ (${\cal \M}_{-}$) & $2\mathbb{Z}$ ($\tilde{\cal \M}_{+}$) \\
    CI & $\mathbb{Z}$ (${\cal \M}_{+-}$, $\tilde{\cal \M}_{--}$) &  $2\mathbb{Z}$ (${\cal \M}_{--}$, $\tilde{\cal \M}_{-+}$) \\ \hline
    AI$^{\dag}$ &  $2\mathbb{Z}$ (${\cal M}_-$) & $2\mathbb{Z}$ ($\tilde{{\cal M}}_-$) \\
    BDI$^{\dag}$ & $\mathbb{Z}$ (${\cal M}_{-+}$, $\tilde{{\cal M}}_{++}$) & $2\mathbb{Z}$ (${\cal M}_{--}$, $\tilde{{\cal M}}_{-+}$) \\
    D$^{\dag}$ & $\mathbb{Z}$ (${\cal \M}_+$) & $\mathbb{Z}$ ($\tilde{\cal \M}_{+}$) \\
    DIII$^{\dag}$ & $\mathbb{Z}_2$ (${\cal M}_{++}$, $\tilde{\cal \M}_{-+}$), & $\mathbb{Z}$ (${\cal M}_{++}$, $\tilde{\cal \M}_{-+}$) \\
    & $\mathbb{Z}$ (${\cal M}_{-+}$, $\tilde{\cal \M}_{--}$) &  \\
    AII$^{\dag}$ & $\mathbb{Z}_2$ (${\cal M}_{+}$, $\tilde{\cal M}_{-}$), $2\mathbb{Z}$ (${\cal M}_{-}$) & $\mathbb{Z}_2$ (${\cal M}_{+}$), $\mathbb{Z}$ ($\tilde{{\cal M}}_{-}$) \\
    CII$^{\dag}$ & $\mathbb{Z}_2$ (${\cal \M}_{+-}$, $\tilde{\cal \M}_{--}$), & $\mathbb{Z}_2$ (${\cal \M}_{+\pm}$, $\tilde{\cal \M}_{\pm-}$), \\
    & $2\mathbb{Z}$ (${\cal \M}_{-+}$, $\tilde{\cal \M}_{++}$) &  $2\mathbb{Z}$ (${\cal \M}_{--}$, $\tilde{\cal \M}_{-+}$) \\
    C$^{\dag}$ &$\mathbb{Z}_2$ (${\cal M}_{-}$), $2\mathbb{Z}$ (${\cal M}_{+}$) & $\mathbb{Z}_2$ ($\tilde{\cal M}_{-}$), $2\mathbb{Z}$ ($\tilde{\cal M}_{+}$) \\
    CI$^{\dag}$ & $2\mathbb{Z}$ (${\cal \M}_{-+}$, $\tilde{\cal \M}_{--}$) & 
    $\mathbb{Z}_2$ (${\cal \M}_{--}$, $\tilde{\cal \M}_{+-}$), \\ 
    & & $2\mathbb{Z}$ (${\cal \M}_{++}$, $\tilde{\cal \M}_{-+}$)
    \\
    \hline \hline
  \end{tabular}
\end{table}

\section{Classification of second-order point-gap topological phases} \label{sec:classification}
We consider two types of reflection symmetry with respect to the mirror plane perpendicular to the $i$ ($i=x,y,z$) direction: 
\begin{align}
    &\mathcal{M}H(k_i, \boldsymbol{k}_{\parallel})\mathcal{M}^{-1}=H(-k_i, \boldsymbol{k}_{\parallel}), \quad \mathcal{M}^2=1, \label{eq:reflection} \\
    &\tilde{\mathcal{M}}H^\dagger (k_i, \boldsymbol{k}_{\parallel})\tilde{\mathcal{M}}^{-1}= H (-k_i, \boldsymbol{k}_{\parallel}),  \quad \tilde{\mathcal{M}}^2=1,\label{eq:pseudoreflection}
\end{align}
where $\mathcal{M}$ and $\tilde{\mathcal{M}}$ are unitary matrices, and $\boldsymbol{k}_{\parallel}$ denotes the wavevector except for the $k_i$ component. 
We refer to Eq.~(\ref{eq:reflection}) and Eq.~(\ref{eq:pseudoreflection}) as reflection symmetry and pseudo-reflection symmetry, respectively.
The latter is also called reflection symmetry$^{\dag}$ in the terminology of Ref.~\cite{PhysRevX.9.041015}.
While these two types of reflection symmetry coincide with each other in Hermitian systems, 
this is not the case in non-Hermitian systems, leading to the rich topological classification.

Non-Hermitian Hamiltonians $H$ are defined to have a point gap when their complex spectra do not cross a reference energy $E \in \mathbb{C}$ (i.e., $\det \left[ H-E \right] \neq 0$)~\cite{PhysRevX.8.031079, PhysRevX.9.041015}.
To elucidate point-gap topology of $H$, we introduce Hermitized Hamiltonians ${\sf H}$ by
\begin{equation}
    {\sf H} \coloneqq \begin{pmatrix}
        0 & H \\
        H^{\dag} & 0
    \end{pmatrix}.
\end{equation} 
Point-gap topology of $H$ coincides with Hermitian topology of ${\sf H}$.
We identify the relevant symmetry class of ${\sf H}$ for each internal symmetry class of $H$ with additional reflection or pseudo-reflection symmetry (see Appendix~\ref{subsec: classification}). 
We thus classify reflection-symmetric second-order point-gap topology in 2D and 3D non-Hermitian systems (Table \ref{tab: AZ}) from the classification of Hermitian reflection-symmetric second-order topology~\cite{PhysRevLett.119.246401, physrevb.97.205135}.
Table~\ref{tab: AZ} specifies possible reflection-symmetric second-order point-gap topological phases and predicts previously unexplored second-order non-Hermitian topological phenomena.
While Table~\ref{tab: AZ} includes second-order skin effects~\cite{PhysRevB.102.205118, PhysRevB.102.241202}, we also identify exceptional second-order topological insulators featuring point-gapless corner and hinge states, as summarized in Table~\ref{tab:esoti} (see also Appendix~\ref{subsec: classification}).
As illustrative examples, we below study a 2D exceptional second-order topological insulator in class D $+$ $\tilde{\mathcal{M}}_+$ and a 3D exceptional second-order topological insulator in class AIII $+$ $\mathcal{S}_{-}$ $+$ ${\mathcal{M}}_{++}$.

\section{2D exceptional second-order topological insulator}\label{sec:2DETI}
We discuss exceptional second-order topological insulators in class D, where non-Hermitian Hamiltonians respect particle-hole symmetry
\begin{equation}
    \mathcal{C} H^{T}(\boldsymbol{k}) {\mathcal{C}}^{-1} = -H(-\boldsymbol{k}), \ \ \mathcal{C} \mathcal{C}^{*}=1,
\end{equation}
with a unitary matrix $\mathcal{C}$.
The second-order point-gap topological phases are 
$\mathbb{Z}_2$ classified %in the presence of 
with
pseudo-reflection ($\tilde{\mathcal{M}}_+$) symmetry in Eq.~(\ref{eq:pseudoreflection}) commuting with $\mathcal{C}$.

\begin{table}
	\centering
	\caption{Symmetry classes hosting exceptional second-order topological insulators in two and three dimensions protected by reflection or pseudo-reflection symmetry.
    In the complex Altland-Zirnbauer class, the subscripts of $\mathcal{S}$ specify the commutation ($+$) or anticommutation ($-$) relation with chiral symmetry, and the subscripts of ${\cal \M}$ and $\tilde{\cal \M}$ specify the commutation ($+$) or anticommutation ($-$) relation with chiral and/or sublattice symmetry.
    In the real Altland-Zirnbauer($^{\dag}$) class, the subscripts of ${\cal \M}$ and $\tilde{\cal \M}$ specify the commutation ($+$) or anticommutation ($-$) relation with time-reversal and/or particle-hole symmetry($^{\dag}$).
    }
	\label{tab:esoti}
     \begin{tabular}{ccc} \hline \hline
    Class & $d=2$ & $d=3$ \\ \hline 
    A & 0 & $\tilde{{\cal \M}}$  \\
    AIII &  $\tilde{\cal M}_+$ & ${\cal M}_+$, $\tilde{\cal M}_-$ \\
    AIII + $\mathcal{S}_+$ & ${\cal M}_{+ +}$, $\tilde{\cal M}_{+ +}$ & ${\cal M}_{\pm -}$, $\tilde{\cal M}_{\pm -}$ \\
    A + $\mathcal{S}$ & ${\cal M}_+$ & $\tilde{{\cal M}}_-$  \\
    AIII + $\mathcal{S}_-$ & ${\cal M}_{-+}$, $\tilde{\cal M}_{+-}$ & ${\cal M}_{++}$, $\tilde{\cal M}_{--}$ \\ \hline
    AI & 0 & $\tilde{\mathcal{\M}}_-$ \\
    BDI & ${\cal \M}_{++}$,  $\tilde{\cal \M}_{++}$, $\tilde{\cal \M}_{-+}$ & ${\cal \M}_{++}$, $\tilde{\cal \M}_{-+}$ \\
    D & ${\cal \M}_{+}$, $\tilde{\cal \M}_+$ & ${\cal \M}_+$, {$\tilde{\cal \M}_+$} 
    \\
    DIII & ${\cal \M}_{-+}$, $\tilde{\cal \M}_{++}$,  & ${\cal \M}_{\pm+}$, ${\cal \M}_{--}$,  \\
    & $\tilde{\cal \M}_{--}$ & $\tilde{\cal \M}_{+\pm}$,  $\tilde{\cal \M}_{-+}$ \\
    AII & 0 & $\tilde{{\cal M}}_{+}$, $\tilde{{\cal M}}_{-}$ \\
    CII &  $\tilde{{\cal M}}_{++}$ & ${\cal \M}_{++}$, $\tilde{\cal \M}_{-+}$,  \\
    & & ${\cal \M}_{--}$, $\tilde{\cal \M}_{+-}$ \\
    C & 0 & $\tilde{\cal \M}_{+}$ \\
    CI &  $\tilde{\cal \M}_{--}$ &  ${\cal \M}_{--}$, $\tilde{\cal \M}_{-+}$ \\ \hline
    AI$^{\dag}$ & 0 & $\tilde{{\cal M}}_-$ \\
    BDI$^{\dag}$ &  $\tilde{{\cal M}}_{++}$ &  ${\cal M}_{--}$, $\tilde{{\cal M}}_{-+}$ \\
    D$^{\dag}$ & 0 & $\tilde{\cal \M}_{+}$ \\
    DIII$^{\dag}$ & $\tilde{\cal \M}_{-+}$, $\tilde{\cal \M}_{--}$ & ${\cal M}_{++}$, $\tilde{\cal \M}_{-+}$ \\
    AII$^{\dag}$ &  0 & $\tilde{{\cal M}}_{-}$ \\
    CII$^{\dag}$ & $\tilde{\cal \M}_{++}$, & ${\cal \M}_{+\pm}$, $\tilde{\cal \M}_{\pm-}$, \\
    & $\tilde{\cal \M}_{--}$ &  ${\cal \M}_{--}$, $\tilde{\cal \M}_{-+}$ \\
    C$^{\dag}$ & 0 & 
    $\tilde{\cal M}_{+}$,  $\tilde{\cal M}_{-}$ \\
    CI$^{\dag}$ &  $\tilde{\cal \M}_{--}$ &  ${\cal \M}_{++}$, ${\cal \M}_{--}$,  \\ 
    & & $\tilde{\cal \M}_{+-}$,  $\tilde{\cal \M}_{-+}$
    \\
    \hline \hline
  \end{tabular}
\end{table}

To capture boundary states, %of the exceptional second-order topological insulator, 
we analyze their low-energy continuum Hamiltonian in the presence of pseudo-reflection symmetry with respect to the $y$-$z$ plane.
We begin with a chiral edge state $k_x$ of a 2D first-order point-gap topological phase in class D, 
which is $\mathbb{Z}_2$ classified~\cite{PhysRevX.9.041015}.
To realize a second-order point-gap topological phase, we double this $\mathbb{Z}_2$ boundary state to trivialize the first-order topology.
In the simultaneous presence of particle-hole symmetry with $\mathcal{C}=\sigma_0$ and pseudo-reflection symmetry with $\tilde{\mathcal{M}}_+ = \sigma_x$, a generic non-Hermitian doubled boundary state is given as 
\begin{equation}\label{eq:boundary}
    H(k_x) = k_x \sigma_z + i\delta \sigma_y \quad (\delta \in \mathbb{R}),
\end{equation}
where $\sigma_0$ is the $2\times 2$ identity matrix, and $\sigma_i$ ($i=x,y,z$) are the Pauli matrices.
The energy dispersion is $E(k_x)=\pm \sqrt{k_x^2 - \delta^2}$, showing the stability of the gapless points $k_x=\pm \delta$.
Conversely, in the absence of pseudo-reflection symmetry, a particle-hole-symmetric perturbation $\delta' \sigma_y$ ($\delta' \in \mathbb{R}$) induces a point gap.
Consequently, the boundary states remain gapless only in the presence of pseudo-reflection symmetry, implying the emergence of point-gapless states around 
pseudo-reflection-invariant corners.

\begin{figure}
\includegraphics[width=1.\columnwidth]{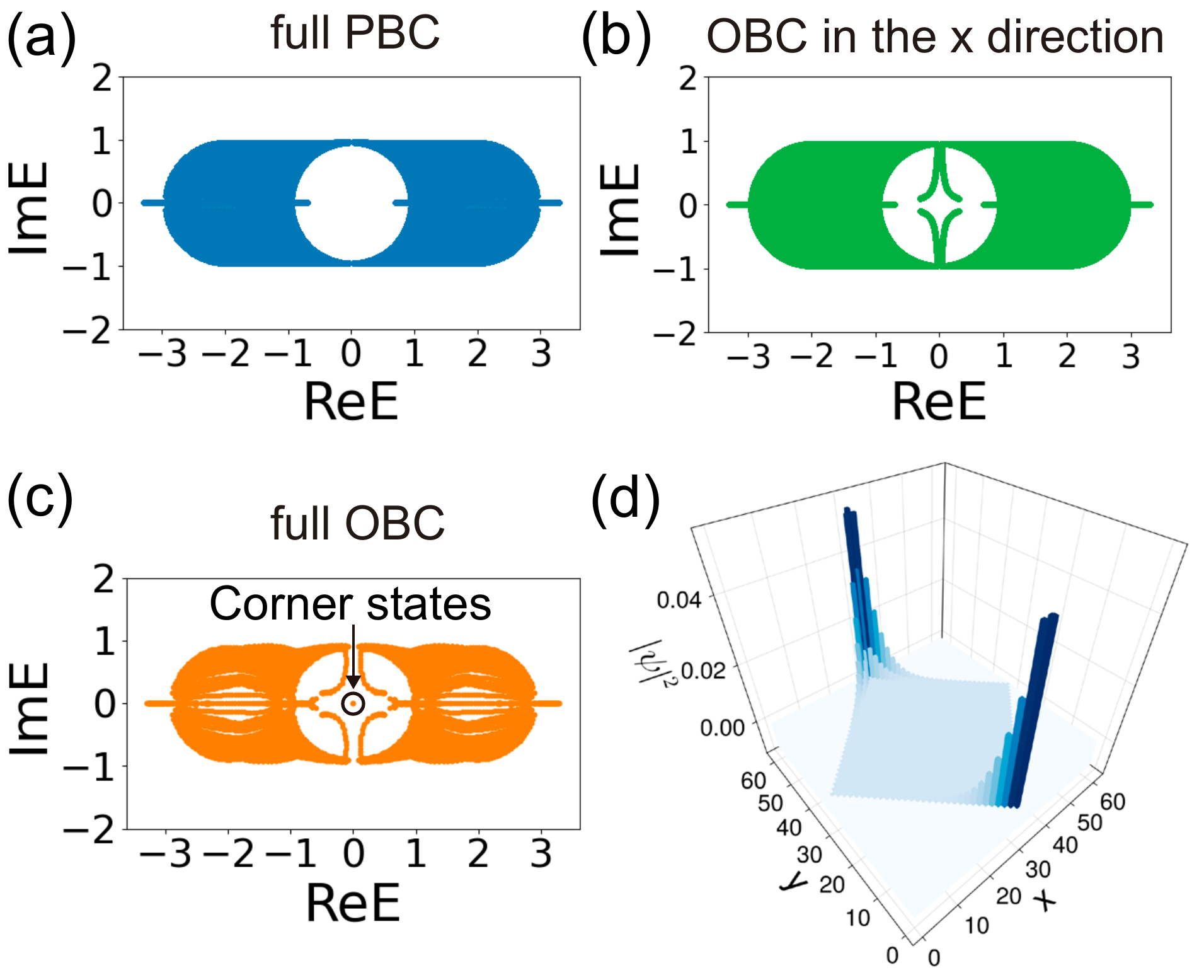}
	\caption{(a-c)~Complex energy spectra of the Hamiltonian (\ref{eq:classD_2D_pseud}) under the (a)~full periodic boundary conditions (PBC), (b)~open boundary conditions (OBC) in the $x$ direction and PBC in the $y$ direction, and (c)~full OBC for a square crystal with the edges perpendicular to the $(1,1)$ and $(1, -{1})$ directions.
    (d)~Real-space distribution of one of the right eigenstates encircled by the circle in (c).
    The parameters are $m_1=0.1$ and $m_2=0.3$.
    The system size is 51 in the $x$ direction with the momentum resolution $\Delta k_y=2\pi/10000$ in (b).
    The system size is 61 in the $x$ and $y$ directions in (c) and (d). 
 }
	\label{fig:2DclassD_pseudo}
\end{figure}

We introduce a tight-binding model on a square lattice, 
\begin{align}\label{eq:classD_2D_pseud}
    H_{\tilde{\mathcal{M}}}(\boldsymbol{k}) = &\sin k_x \tau_x \sigma_z +(1-\cos k_x- \cos k_y)\tau_y  \nonumber \\
    &+ i \sin k_y \sigma_z+im_1\tau_x \sigma_y + m_2 \tau_y \sigma_x,
\end{align}
where $\tau_{i}$ ($i=x,y,z$) are the Pauli matrices.
This model respects particle-hole symmetry with $\mathcal{C}=\sigma_0$ and pseudo-reflection symmetry with $\tilde{\mathcal{M}}_+=\sigma_x$.
The energy spectra of this Hamiltonian 
under the full periodic boundary conditions host a point gap at $E=0$ [Fig.~\ref{fig:2DclassD_pseudo}(a)].  
Under the open boundary conditions in the $x$ direction, the point gap at $E=0$ does not close [Fig.~\ref{fig:2DclassD_pseudo}(b)] since the edges along the $y$ direction are not invariant under the reflection $x\rightarrow -x$.
On the other hand, the corner between the edges perpendicular to the $(1,1)$ and $(1,-1)$ directions remains invariant under the reflection.
Figures~\ref{fig:2DclassD_pseudo}(c) and \ref{fig:2DclassD_pseudo}(d) illustrate the spectra under this boundary conditions,  
demonstrating the point-gap closing at $E=0$ with the emergence of corner states.
These corner states are protected by pseudo-reflection symmetry and localized at the reflection-invariant positions. 
We show that the above second-order point-gap topological phases can be continuously deformed to 
a line-gap topological phase while keeping the point gap and symmetry  (see Appendix~\ref{subsec: cont_deform}).

%\section{Layer construction}
\section{3D exceptional second-order topological insulator}\label{sec:3DETI}
We construct 3D point-gap topological phases from 2D ones, which we refer to as layer construction. 
The layer construction is used for 3D Hermitian topological phases protected by spatial symmetry \cite{PhysRevB.94.125405, PhysRevB.94.155148, PhysRevX.7.011020, PhysRevB.96.205106, song2018quantitative, PhysRevX.9.011012}.
We generalize this method to non-Hermitian systems and construct a 3D exceptional second-order topological insulator protected by reflection symmetry. 

We begin with the following tight-binding model of a 2D exceptional topological insulator as layers to be stacked:
\begin{equation}
    H_{\rm 2D}^{(\pm)}(\boldsymbol{k})=\begin{pmatrix}
    0 & h^{(\pm)}_{1}(\boldsymbol{k}) \\
    h^{(\pm)}_{2}(\boldsymbol{k}) & 0\\
    \end{pmatrix},
        \label{eq: 2D ETI}
\end{equation}
with the off-diagonal parts $h^{(\pm)}_{1}(\boldsymbol{k})$ and $h^{(\pm)}_{2}(\boldsymbol{k})$ 
\begin{align}
    h^{(\pm)}_{1}(\boldsymbol{k})=\pm[& (1 - \cos k_x -\cos k_y)\sigma_x + \sin k_x \sigma_y \nonumber \\
    &+i \sin k_y \sigma_0], \nonumber \\
    h^{(\pm)}_{2}(\boldsymbol{k})=\pm[& (-3 - \cos k_x -\cos k_y)\sigma_x + \sin k_x \sigma_y \nonumber \\
    &+i \sin k_y \sigma_0+i\Delta \sigma_z].
\end{align}
Below, we choose $\Delta=0.01$. 
Both $H_{\rm 2D}^{(+)}(\boldsymbol{k})$ and $H_{\rm 2D}^{(-)}(\boldsymbol{k})$ respect chiral and sublattice symmetries,
\begin{align}
    \Gamma H^{\dagger}(\boldsymbol{k}) {\Gamma}^{-1} = -H(\boldsymbol{k}), \label{eq: CS} \\
    \mathcal{S} H (\boldsymbol{k}) {\mathcal{S}}^{-1} = - H(\boldsymbol{k}), \label{eq: SLS}
\end{align}
with $\Gamma = \tau_x\sigma_z$ and $\mathcal{S} = \tau_z \sigma_0$. 
Hence, they belong to class AIII $+$ $\mathcal{S}_-$ and exhibit 2D point-gap topology and concomitant boundary states with a single EP 
(see Appendix~\ref{appendix: 2DETI})~\cite{PhysRevLett.132.136401, Denner_2023_infernal}. 
This topological phase is characterized by a pair of the Chern numbers of the two Hermitian matrices $ih^{(\pm)}_{1}\sigma_z$ and $ih^{(\pm)}_{2}\sigma_z$:~$({\rm Ch}_1, {\rm Ch}_2)\coloneqq({\rm Ch}[ih^{(\pm)}_{1}\sigma_z], {\rm Ch}[ih^{(\pm)}_{2}\sigma_z])$ \cite{PhysRevX.9.041015}. These Chern numbers for $H_{\rm 2D}^{(+)}(\boldsymbol{k})$ and $H_{\rm 2D}^{(-)}(\boldsymbol{k})$ are given by $({\rm Ch}_1, {\rm Ch}_2)=(+1,0)$ and $({\rm Ch}_1, {\rm Ch}_2)=(-1,0)$, respectively.

\begin{figure}
\includegraphics[width=1.\columnwidth]{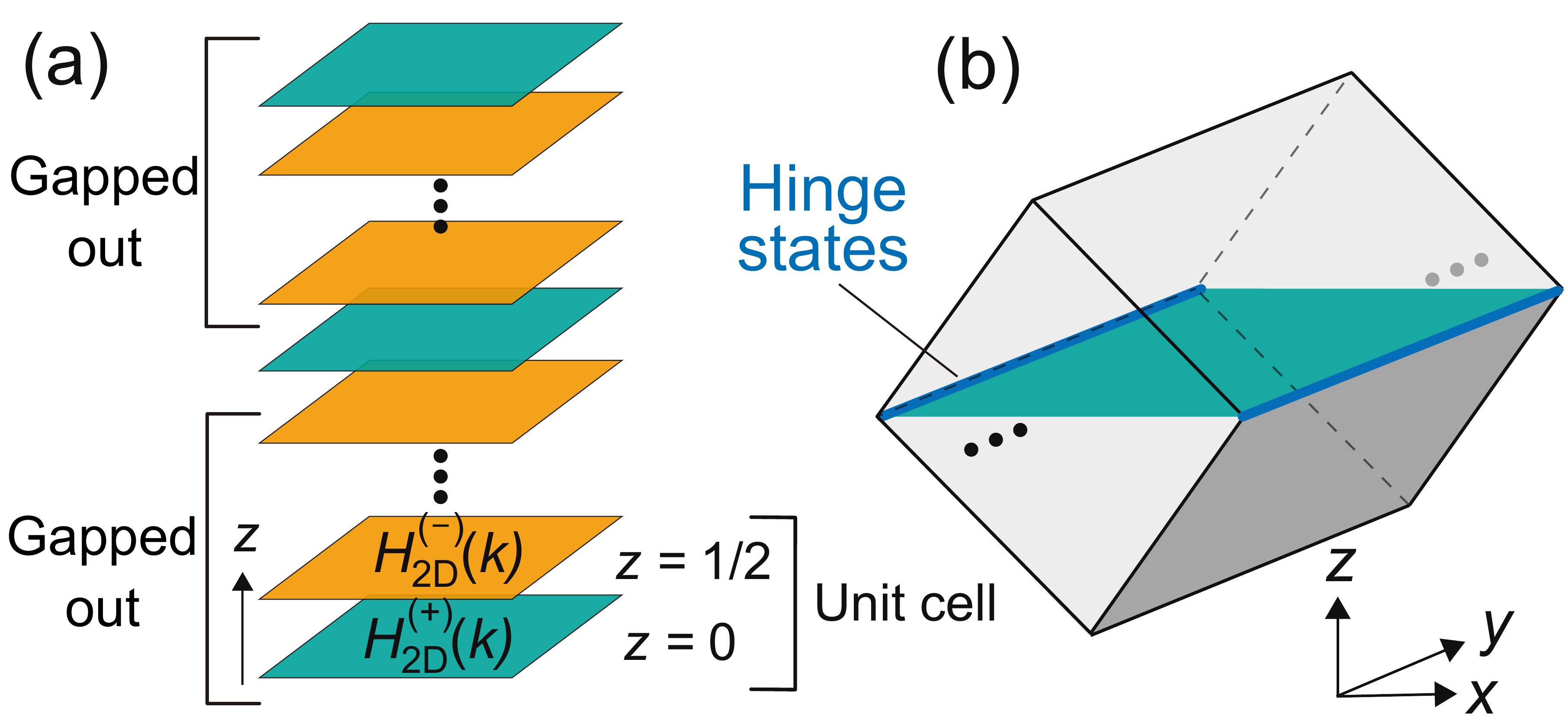}
	\caption{(a)~Layer construction for realizing a three-dimensional (3D) exceptional second-order topological insulator with reflection symmetry. The colors of the layers indicate the sign  $(\pm)$ of $H^{(\pm)}_{\rm 2D}(\boldsymbol{k})$. The dotted box indicates the unit cell. (b) 3D reflection-symmetric exceptional second-order topological insulator with hinge states by the layer construction.}
	\label{fig:LC}
\end{figure}

We alternately stack $H^{(+)}_{\rm 2D}(\boldsymbol{k})$ and $H^{(-)}_{\rm 2D}(\boldsymbol{k})$ layers along the $z$ direction, as shown in Fig.~\ref{fig:LC}(a). 
Each unit cell contains one $H^{(+)}_{\rm 2D}(\boldsymbol{k})$ layer and one $H^{(-)}_{\rm 2D}(\boldsymbol{k})$, positioned 
at $z=n$ and $z=n+1/2$, respectively, where $n$ is a nonnegative integer, and the size of the unit cell in the $z$ direction is $1$.
By introducing the interlayer coupling $H_z (\boldsymbol{k})$, the Bloch Hamiltonian of the stacked layers is 
\begin{equation}\label{eq:intrinsc3d}
    H_{\mathcal{M}}(\boldsymbol{k})=\begin{pmatrix}
        H_{\rm 2D}^{(+)}(\boldsymbol{k}) & H_z (\boldsymbol{k}) \\
        H_z^{\dagger} (\boldsymbol{k}) & H_{\rm 2D}^{(-)}(\boldsymbol{k})
    \end{pmatrix},
\end{equation}
where $H_z (\boldsymbol{k})$ is a $4\times 4$ matrix expressed as 
\begin{equation}
    H_z (\boldsymbol{k})=   (1+e^{-ik_z}) (t_1\tau_x \sigma_x+i t_2 \tau_y \sigma_z).
\end{equation}
The Pauli matrices $\tau_i$ and $\sigma_i$ correspond to the degrees of freedom in each layer. 
Henceforth, we assume that the number of layers is odd so that reflection symmetry with respect to the $x$-$y$ plane will be satisfied.
While we have focused on reflection symmetry, we also show that inversion symmetry leads to a 3D exceptional second-order topological insulator in Appendix~\ref{appendix: 3DESOTI_inversion}.

This construction leads to a reflection-symmetric exceptional second-order topological insulator with point-gapless hinge states. 
As shown in Fig.~\ref{fig:LC}(a), 
while pairs of the layers are trivialized due to their Chern numbers of different signs, a single layer at the reflection center maintains topological edge states, resulting in the one-dimensional hinge states [Fig.~\ref{fig:LC}(b)]. 
Inheriting the edge states of the original 2D exceptional topological insulators [i.e., Eq.~(\ref{eq: 2D ETI})], the hinge state also possesses a single EP. 
This discussion extends the layer construction of second-order topological insulators from Chern insulators  \cite{PhysRevB.97.205136, PhysRevB.98.115150, PhysRevB.98.205129, PhysRevB.98.245102, PhysRevResearch.2.043274} to exceptional second-order topological insulators.

The Hamiltonian $H_{\mathcal{M}}(\boldsymbol{k})$ respects reflection symmetry in Eq.~(\ref{eq:reflection}) with respect to the $x$-$y$ plane. 
The symmetry representation is 
$\mathcal{M}={\rm diag}(1,e^{-ik_z})$, which acts on the sublattice degrees of freedom 
$z=0$ and $z=1/2$.
The model $H_{\mathcal{M}}(\boldsymbol{k})$ also respects chiral symmetry in Eq.~(\ref{eq: CS})
and sublattice symmetry in Eq.~(\ref{eq: SLS}). 
Therefore, $H_{\mathcal{M}}(\boldsymbol{k})$ belongs to class AIII $+$ $\mathcal{S_{-}}$ in the 38-fold symmetry classes \cite{PhysRevX.9.041015}. 
Since both $\Gamma$ and  
$\mathcal{S}$ commute with $\mathcal{M}$, the second-order point-gap topology in $H_{\mathcal{M}}(\boldsymbol{k})$ is classified as $\mathbb{Z}\oplus \mathbb{Z}$ (see class AIII $+$ $\mathcal{S_{-}}$ with $\mathcal{M}_{++}$ in Table~\ref{tab: AZ}).

\begin{figure}
\includegraphics[width=1.\columnwidth]{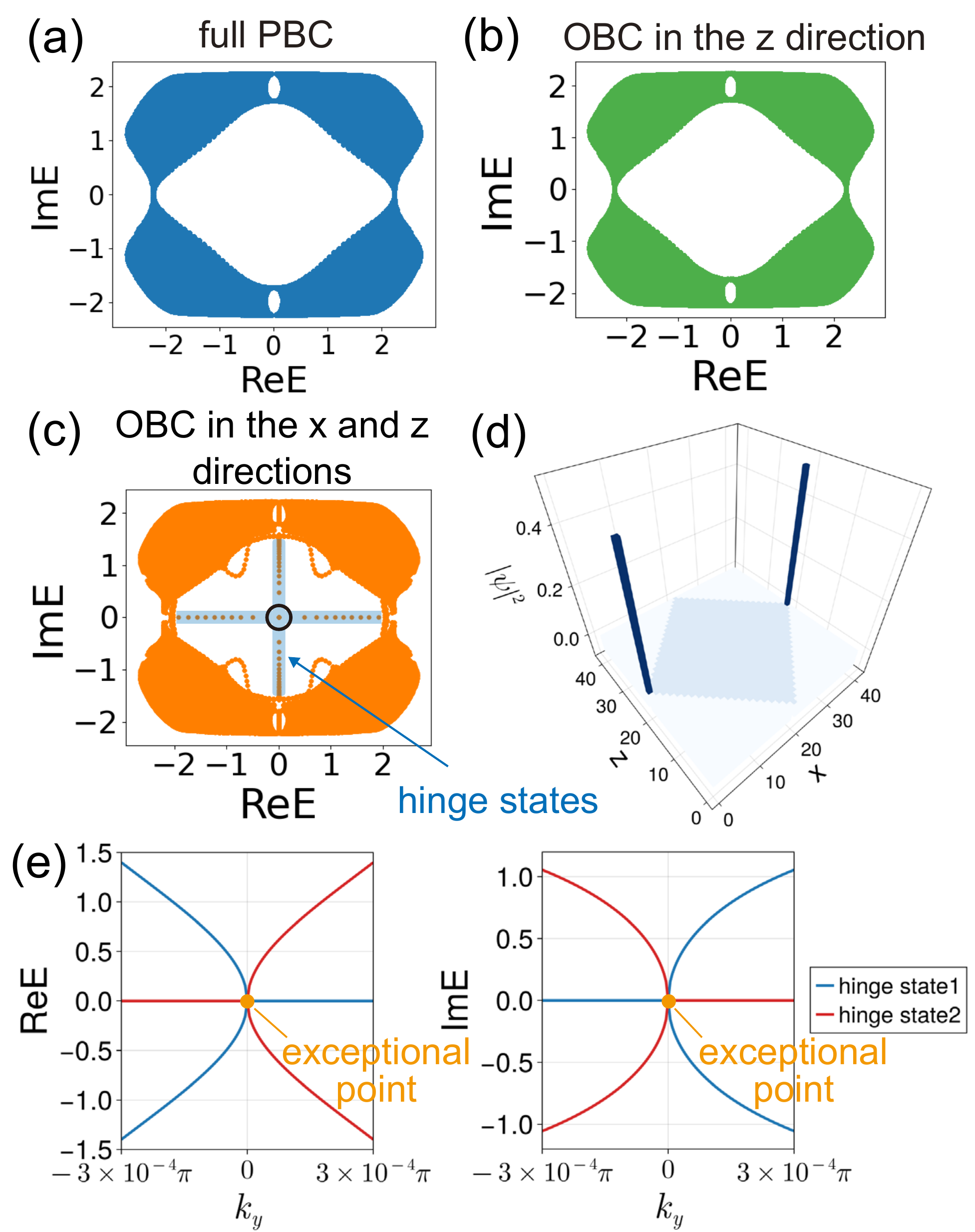}
	\caption{(a-c)~Complex energy spectra of the Hamiltonian (\ref{eq:intrinsc3d}) under the (a)~full periodic boundary conditions (PBC), (b)~open boundary conditions (OBC) in the $z$ direction and PBC in the $x$ and $y$ directions, and (c)~OBC in the $x$ and $z$ directions and PBC in the $y$ direction. 
    (d)~Real-space distribution of one of the right eigenstates encircled by the circle in (c). 
    (e) Energy dispersion of the hinge states. 
    The parameters are $t_1=0.4$ and $t_2=0.3$. 
    The system size in the $x$ direction is $41$ in (c) and (d), and the number of layers is $81$ in (b-d). 
    The momentum resolutions in the $k_i$ ($i=x,y$) direction are $\Delta k_i=2\pi/80$ in the whole Brillouin zone in (b) and (c). 
    For $-\pi/1000 \leq k_y \leq \pi/1000$, the momentum resolution is set to $\Delta k_y=2\pi/40000$ to obtain the energy spectra of the hinge states.
 }
	\label{fig:3d_intrinsic}
\end{figure} 

We show that a 3D exceptional second-order topological insulator indeed emerges in the Hamiltonian in Eq.~(\ref{eq:intrinsc3d}). 
On the reflection-invariant plane $k_z=k_0$ $(=0,\pi)$, the Hamiltonian $H_{\mathcal{M}}(k_x, k_y, k_0)$ is block diagonalizable: $H_+ (k_x, k_y,  k_0)\oplus H_- (k_x, k_y, k_0)$, where $H_+ (k_x, k_y, k_0)$ and $H_- (k_x, k_y, k_0)$ are the matrices labeled by the reflection-symmetry eigenvalues $\pm1$, respectively.
Since $\mathcal{M}$ commutes with both chiral and sublattice symmetries, each sector belongs to class AIII $+$ $\mathcal{S}_-$.
Owing to sublattice symmetry, $H_{\pm}(k_x, k_y, k_0)$ is given by 
\begin{equation}
    H_{\pm}(k_x, k_y, k_0)=\begin{pmatrix}
    0 & h_{1,\pm}(k_x, k_y, k_0) \\
    h_{2, \pm}(k_x, k_y, k_0) & 0 \\
    \end{pmatrix},
\end{equation}
and
effectively reduces to 
quasi-2D systems in class AIII $+$ $\mathcal{S}_-$. 
Their topological invariants are given by a pair of the Chern numbers of the Hermitian matrices $ih_{1, \pm}\sigma_z$ and $ih_{2, \pm}\sigma_z$:  $({\rm Ch}_{\pm, 1}, {\rm Ch}_{\pm, 2}) \coloneqq ({\rm Ch}[ih_{1, \pm}\sigma_z], {\rm Ch}[ih_{2, \pm}\sigma_z])$ \cite{PhysRevX.9.041015}.
From these Chern numbers, we introduce the mirror Chern numbers  
\begin{equation}
    {\rm Ch}_{\mathcal{M}, i}=\frac{1}{2}({\rm Ch}_{+, i} - {\rm Ch}_{-, i}),
\end{equation}
with $i=1,2$. 
This result is consistent with the $\mathbb{Z}\oplus \mathbb{Z}$ classification for class AIII $+$ $\mathcal{S}_-$ in Table~\ref{tab: AZ}. 
The pair of the mirror Chern numbers for our model takes $({\rm Ch}_{\mathcal{M}, 1}, {\rm Ch}_{\mathcal{M}, 2})=(1,0)$ in the $k_z=\pi$ plane. 

We calculate the energy spectra of Eq.~(\ref{eq:intrinsc3d}) to obtain the boundary states corresponding to the nontrivial mirror Chen number for the point gap at $E=0$ [Fig.~\ref{fig:3d_intrinsic}(a)]. 
No boundary states are present under the OBC in the $z$ direction and the PBC in both the $x$ and $y$ directions with the surface perpendicular to the $z$ direction since reflection symmetry is not preserved [Fig.~\ref{fig:3d_intrinsic}(b)]. 
By contrast, the hinge between the surfaces perpendicular to the ($1,0,1$) and ($1,0,-{1}$) directions is invariant under the reflection, leading to the emergence of the boundary states [Fig.~\ref{fig:3d_intrinsic}(c)].
These boundary states are localized at the hinges that are invariant under the reflection $z\rightarrow -z$ [Fig.~\ref{fig:3d_intrinsic}(d)]. 
We identify their effective Hamiltonian as
\begin{equation}
    H(k_y)=\frac{i}{2}(\tau_x+i\tau_y)k_y + \frac{i v}{2}(\tau_x-i\tau_y)\sigma_z \quad  (v\in \mathbb{R}),
\end{equation}
with the energy dispersion $E = \pm \sqrt{vk_y}, \pm i \sqrt{vk_y}$ [Fig.~\ref{fig:3d_intrinsic}(e)], distinct from conventional exceptional topological insulators~\cite{denner2021exceptional}.
Each hinge state supports a single EP at $k_y = 0$, demonstrating that this 3D exceptional second-order topological insulator cannot be continuously deformed into Hermitian topological phases 
and hence is intrinsic to non-Hermitian systems.
Hinge states can appear even in the absence of sublattice symmetry, consistent with the $\mathbb{Z}$ classification for class AIII with $\mathcal{M}_+$ in Table~\ref{tab: AZ}.
However, these hinge states are not necessarily intrinsic to non-Hermitian systems since they cannot support a single EP~\cite{PhysRevLett.132.136401}.
Furthermore, we show that this model exhibits an exotic surface state with two EPs that cannot be removed without breaking reflection symmetry also on surfaces perpendicular to the $x$ direction (see Appendix~\ref{appendix: 3DETCI}).
Such versatility is a unique characteristic of second-order exceptional topological insulators due to crystalline symmetry protection.

\section{Conclusion and discussion}\label{sec:conclusion}
We develop classification of reflection-symmetric second-order point-gap topological phases. 
From this classification, we introduce 2D and 3D exceptional second-order topological insulators as distinct classes of point-gap topological phases protected by reflection symmetry. 
2D exceptional second-order topological insulators are feasible in metamaterials, such as photonic crystals, active matter, and electrical circuits, given the experimental realization of various point-gap topological phases in 2D systems \cite{palacios2021guided, PhysRevResearch.2.023265, shang2022experimental, wu2023spin,  PhysRevLett.132.063804, PhysRevLett.132.113802, zhong2024higher}.
While 3D point-gap topological phases have yet to be experimentally demonstrated, 
our layer construction approach offers a promising route for experimentally realizing 3D exceptional second-order topological insulators, based on the availability of 2D point-gap topological phases.
The unique complex spectrum of 
hinge states in 3D exceptional second-order topological insulators can lead to the distinctive nonunitary dynamics.
Single EPs therein may also enable practical applications in laser-mode selectivity~\cite{Feng2014Science, Hodaei2014Science}, chiral transport~\cite{Doppler2016Nature, Xu2016Nature}, and enhanced sensitivity~\cite{Hodaei2017Nature, Chen2017Nature}. 
Our findings thus expand the family of point-gap topological phases incorporating spatial symmetry.

\begin{acknowledgments}
D.N. and K.K. thank Ken Shiozaki for helpful discussion.
D.N. thanks Masatoshi Sato for helpful discussion.
We appreciate the long-term workshop ``Recent Developments and Challenges in Topological Phases" (YITP-T-24-03) held at Yukawa Institute for Theoretical Physics (YITP), Kyoto University.
Y.T. is supported by RIKEN Special Postdoctoral Researchers Program and Japan Society for the Promotion of Science (JSPS) KAKENHI Grant No.~JP24K22868. 
Y.T. and D.N. are supported by JST CREST Grant No.~JPMJCR19T2.
D.N. is supported by JSPS KAKENHI Grant No.~JP24K22857.
R.O. is supported by JSPS KAKENHI Grants No.~JP23K13033 and No.~JP24K00586. 
K.K. is supported by MEXT KAKENHI Grant-in-Aid for Transformative Research Areas A ``Extreme Universe" No.~JP24H00945.
\end{acknowledgments}

\appendix

\section{Classification of reflection-symmetric second-order point-gap topology}
    \label{subsec: classification}

We develop a classification of reflection-symmetric second-order point-gap topology.
We consider non-Hermitian Hamiltonians $H$ in the 38-fold internal-symmetry classification~\cite{PhysRevX.9.041015} with additional reflection symmetry.
Owing to non-Hermiticity, we consider two different types of reflection symmetry, 
\begin{equation}
    \mathcal{\M} H \mathcal{\M}^{-1} = H, \quad \mathcal{\M}^2 = 1,
    \label{aeq: reflection}
\end{equation}
and 
\begin{equation}
    \tilde{\cal \M} H^{\dag} \tilde{\cal \M}^{-1} = H, \quad \tilde{\cal \M}^2 = 1,
    \label{aeq: pseudo-reflection}
\end{equation}
with unitary operators ${\cal \M}$ and $\tilde{\cal \M}$.
We call the former reflection symmetry for non-Hermitian Hamiltonians and the latter pseudo-reflection symmetry, or equivalently, reflection symmetry$^{\dag}$, following the terminology in Ref.~\cite{PhysRevX.9.041015}.
The combination of pseudo-Hermiticity and reflection symmetry gives rise to pseudo-reflection symmetry.

To capture point-gap topology, it is useful to introduce the Hermitized Hamiltonians ${\sf H}$ by
\begin{equation}
    {\sf H} \coloneqq \begin{pmatrix}
        0 & H \\
        H^{\dag} & 0
    \end{pmatrix}.
        \label{aeq: Hermitization}
\end{equation}
Point-gap topology of the original non-Hermitian Hamiltonians $H$ is reduced to topology of the Hermitized Hamiltonians $H$, and vice versa.
Furthermore, symmetry for $H$ leads to symmetry for ${\sf H}$.
Specifically, reflection symmetry in Eq.~(\ref{aeq: reflection}) and pseudo-reflection symmetry in Eq.~(\ref{aeq: pseudo-reflection}) lead to the following symmetries represented by
\begin{align}
{\sf \M} {\sf H} {\sf \M}^{-1} &= {\sf H},\quad {\sf \M} \coloneqq \begin{pmatrix}
        {\cal \M} & 0 \\
        0 & {\cal \M}
\end{pmatrix}, \quad {\sf \M}^2 = 1, \label{aeq: reflection - H} \\
{\sf \M} {\sf H} {\sf \M}^{-1} &= {\sf H},\quad{\sf \M} \coloneqq \begin{pmatrix}
        0 & \tilde{\cal \M} \\
        \tilde{\cal \M} & 0
    \end{pmatrix}, \quad {\sf \M}^2 = 1,    \label{aeq: pseudo-reflection - H}
\end{align}
respectively.
Additionally, the Hermitized Hamiltonians ${\sf H}$ by construction respect chiral symmetry 
\begin{equation}
    {\sf \Sigma} {\sf H} {\sf \Sigma}^{-1} = - {\sf H},\quad {\sf \Sigma} \coloneqq \begin{pmatrix}
        1 & 0 \\
        0 & -1
    \end{pmatrix}.
        \label{aeq: Hermitized chiral}
\end{equation}
For each symmetry class of non-Hermitian Hamiltonians $H$, we specify the relevant symmetry class of the Hermitized Hamiltonians ${\sf H}$.
Then, from the classification of reflection-symmetric second-order topology in Ref.~\cite{PhysRevLett.119.246401}, corresponding to the intrinsic (or equivalently, termination-independent) classification without locally broken reflection symmetry in Ref.~\cite{physrevb.97.205135}, we further identify second-order point-gap topology for these symmetry classes.
In some symmetry classes, blocks of ${\sf H}$ respect reflection antisymmetry,
\begin{equation}
\bar{\sf \M} {\sf H} \bar{\sf \M}^{-1} = - {\sf H},\quad \bar{\sf \M}^2 = 1,
\end{equation}
although we begin with reflection symmetry or pseudo-reflection symmetry for $H$.
We summarize our classification for the complex Altland-Zirnbauer class in Table~\ref{tab: complex AZ}, real Altland-Zirnbauer class in Table~\ref{tab: real AZ}, and real Altland-Zirnbauer$^{\dag}$ class in Table~\ref{tab: real AZ dag}.
The classification of the other remaining classes is left for future work.
It should be noted that this classification includes both second-order skin effect and exceptional topological insulators.
%For example, while the $\mathbb{Z}$ classification for class A + ${\cal M}$ and $d=2$ describes the mirror skin effect~\cite{PhysRevResearch.2.022062}, the $\mathbb{Z} \oplus \mathbb{Z}$ classification for class AIII + $\mathcal{S}_-$ + $\mathcal{\M}_{++}$ describes the second-order exceptional topological insulators studied in the present work (see Table~\ref{tab: complex AZ}).
Some classes should be accompanied by second-order skin effect, in a similar manner to Refs.~\cite{PhysRevB.102.241202, PhysRevB.102.205118}.
By contrast, other classes correspond to second-order exceptional topological insulators studied in the present work, including 2D class D + $\tilde{\cal M}_+$ ($\mathbb{Z}_2$ classification) and 3D class AIII + ${\cal S}_-$ + ${\cal M}_{++}$ ($\mathbb{Z} \oplus \mathbb{Z}$ classification).
Below, we explicitly describe the identification of the relevant symmetry classes for representative cases.

%\clearpage
%%%%%%%%%%
\subsection{Class AIII + ${\cal S}_\pm$ + ${\cal \M}_{\pm\pm}$/$\tilde{\cal \M}_{\pm\pm}$}

%%%%%
\subsubsection{Reflection}
    \label{subsubsec: reflection - AIII + Spm}

As a prime example of the complex Altland-Zirnbauer class, we study class AIII + ${\cal S}_\pm$ + ${\cal \M}_{\pm\pm}$.
By definition, non-Hermitian Hamiltonians $H$ respect chiral symmetry and sublattice symmetry, 
\begin{align}
    \Gamma H^{\dag} \Gamma^{-1} &= - H,\quad \Gamma^2 = 1, \label{aeq: CS} \\
    \mathcal{S} H \mathcal{S}^{-1} &= - H,\quad {\cal S}^2 = 1, \label{aeq: SLS} %\\
    %\mathcal{\M} H \mathcal{\M}^{-1} &= H, \quad \mathcal{\M}^2 = 1,
\end{align}
in addition to reflection symmetry in Eq.~(\ref{aeq: reflection}).
These symmetries are required to satisfy the algebra
\begin{align}
    &\Gamma \mathcal{S} = \epsilon_{\rm \Gamma S} \mathcal{S} \Gamma,\quad \Gamma \mathcal{\M} = \epsilon_{\rm \Gamma \M} \mathcal{\M} \Gamma,  \nonumber
    \\
    &\mathcal{S} \mathcal{\M} = \epsilon_{\rm S\M} \mathcal{\M} \mathcal{S} \quad \left( \epsilon_{\rm \Gamma S}, \epsilon_{\rm \Gamma \M}, \epsilon_{\rm S\M} \in \{ \pm 1\}\right).
\end{align}
Now, we introduce the Hermitized Hamiltonians ${\sf H}$ in Eq.~(\ref{aeq: Hermitization}), for which the above symmetries are represented by
\begin{align}
    {\sf \Gamma} {\sf H} {\sf \Gamma}^{-1} &= - {\sf H},\quad {\sf \Gamma} \coloneqq \begin{pmatrix}
        0 & \Gamma \\
        \Gamma & 0
    \end{pmatrix}, \quad {\sf \Gamma}^2 = 1, \label{aeq: CS - H} \\
    {\sf S} {\sf H} {\sf S}^{-1} &= - {\sf H},\quad {\sf S} \coloneqq \begin{pmatrix}
        {\cal S} & 0 \\
        0 & {\cal S}
    \end{pmatrix}, \quad {\sf S}^2 = 1, \label{aeq: SLS - H} %\\
    %{\sf \M} {\sf H} {\sf \M}^{-1} &= {\sf H},\quad {\sf \M} \coloneqq \begin{pmatrix}
    %    {\cal \M} & 0 \\
    %    0 & {\cal \M}
    %\end{pmatrix}, \quad {\sf \M}^2 = 1,
\end{align}
and Eq.~(\ref{aeq: reflection - H}), with
\begin{equation}
    {\sf \Gamma}{\sf S} = \epsilon_{\rm \Gamma S} {\sf S}{\sf \Gamma},\quad {\sf \Gamma}{\sf \M} = \epsilon_{\rm \Gamma \M} {\sf \M}{\sf \Gamma},\quad {\sf S}{\sf \M} = \epsilon_{\rm S\M} {\sf \M}{\sf S}.
\end{equation}
By construction, the Hermitized Hamiltonians ${\sf H}$ respect additional chiral symmetry ${\sf \Sigma}$ in Eq.~(\ref{aeq: Hermitized chiral}), satisfying
\begin{equation}
    {\sf \Gamma}{\sf \Sigma} = - {\sf \Sigma}{\sf \Gamma},\quad {\sf S}{\sf \Sigma} ={\sf \Sigma}{\sf S},\quad {\sf \M}{\sf \Sigma} = {\sf \Sigma}{\sf \M}.
\end{equation}
Because of the simultaneous presence of the two chiral symmetries described by ${\sf S}$ and ${\sf \Sigma}$, the combined operation gives unitary symmetry that commutes with ${\sf H}$:
\begin{equation}
    \left( {\sf S\Sigma} \right) {\sf H} \left( {\sf S\Sigma} \right)^{-1} = {\sf H},\quad \left( {\sf S\Sigma} \right)^2 = 1.
        \label{aeq: S-Sigma}
\end{equation}
Consequently, the Hermitized Hamiltonians ${\sf H}$ can be block-diagonalized.
This unitary symmetry is related to the other symmetries given by
\begin{align}
    &{\sf \Gamma} \left( {\sf S\Sigma} \right) = - \epsilon_{\rm \Gamma S} \left( {\sf S\Sigma} \right) {\sf \Gamma}, \quad 
    {\sf S} \left( {\sf S\Sigma} \right) = + \left( {\sf S\Sigma} \right) {\sf S}, \nonumber \\
    &{\sf \Sigma} \left( {\sf S\Sigma} \right) = + \left( {\sf S\Sigma} \right) {\sf \Sigma}, \quad 
    {\sf \M} \left( {\sf S\Sigma} \right) = + \epsilon_{\rm S\M} \left( {\sf S\Sigma} \right) {\sf \M}.
\end{align}
\begin{itemize}
    \item $\epsilon_{\rm \Gamma S} = +1$ (class AIII + ${\cal S}_+$ + ${\cal \M}_{\pm\pm}$).---While each block of ${\sf H}$ respects chiral symmetry ${\sf S}$, it no longer respects chiral symmetry ${\sf \Gamma}$. 
    Therefore, the blocks belong to class AIII.
    Furthermore, for $\epsilon_{\rm S\M} = +1$, reflection symmetry $\mathsf{\M}$ is respected, and its commutation or anticommutation relation with chiral symmetry ${\sf S}$ is specified by $\epsilon_{\rm S\M} = +1$.
    On the other hand, for $\epsilon_{\rm S\M} = -1$, reflection antisymmetry $\mathsf{\Gamma\M}$ ($i\mathsf{\Gamma\M}$) is respected for $\epsilon_{\rm \Gamma\M} = +1$ ($\epsilon_{\rm \Gamma\M} = -1$), and its commutation or anticommutation relation with chiral symmetry ${\sf S}$ is specified by $\epsilon_{\rm \Gamma S}\epsilon_{\rm S\M} = -1$.
    
    \item $\epsilon_{\rm \Gamma S} = -1$ (class AIII + ${\cal S}_-$ + ${\cal \M}_{\pm\pm}$).---Each block of ${\sf H}$ respects both chiral symmetries ${\sf S}$ and ${\sf \Gamma}$.
    Then, the combination of ${\sf S}$ and ${\sf \Gamma}$ further gives unitary symmetry that commutes with ${\sf H}$:
    \begin{equation}
        \left( \ii{\sf S\Gamma} \right) {\sf H} \left( \ii{\sf S\Gamma} \right)^{-1} = {\sf H},\quad \left( \ii{\sf S\Gamma} \right)^2 = 1.
    \end{equation}
    The algebra between this unitary symmetry and the other symmetries is given by
    \begin{align}
    &{\sf \Gamma} \left( \ii{\sf S\Gamma} \right) = - \left( \ii{\sf S\Gamma} \right) {\sf \Gamma}, \quad
    {\sf S} \left( \ii{\sf S\Gamma} \right) = - \left( \ii{\sf S\Gamma} \right) {\sf S}, \nonumber \\
    &{\sf \M} \left( \ii{\sf S\Gamma} \right) = + \epsilon_{\rm \Gamma \M} \epsilon_{\rm S\M} \left( \ii{\sf S\Gamma} \right) {\sf \M}.
    \end{align}
    Each block diagonalized by the two unitary symmetries ${\sf S\Sigma}$ and $i{\sf S\Gamma}$ respects no internal symmetry and thus belongs to class A.
    Moreover, for $\epsilon_{\rm S\M} = +1$, reflection symmetry $\mathsf{\M}$ (antisymmetry $i\mathsf{\Gamma\M}$) survives in each of the blocks for $\epsilon_{\rm \Gamma\M} = +1$ ($\epsilon_{\rm \Gamma\M} = -1$).
    For $\epsilon_{\rm S\M} = -1$, on the other hand, each block of $\ii{\sf S\Gamma}$ respects unitary symmetry $\mathsf{S\Sigma}$ and reflection symmetry $\mathsf{M}$ that anticommute with each other, specified by ``A + $\mathsf{U}$ + $\mathsf{M}_{-}$" in Table~\ref{tab: complex AZ}.
\end{itemize}

%\clearpage
%%%%%
\subsubsection{Pseudo-reflection}

We also study class AIII + ${\cal S}_\pm$ + $\tilde{\cal \M}_{\pm\pm}$.
Non-Hermitian Hamiltonians $H$ respect chiral symmetry in Eq.~(\ref{aeq: CS}), sublattice symmetry in Eq.~(\ref{aeq: SLS}), and pseudo-reflection symmetry (or equivalently, reflection symmetry$^{\dag}$) in Eq.~(\ref{aeq: pseudo-reflection}), %, respectively:
%\begin{align}
%    \Gamma H^{\dag} \Gamma^{-1} &= - H,\quad \Gamma^2 = 1, \\
%    \mathcal{S} H \mathcal{S}^{-1} &= - H,\quad {\cal S}^2 = 1, \\
%    \mathcal{\M} H^{\dag} \mathcal{\M}^{-1} &= H, \quad \mathcal{\M}^2 = 1.
%\end{align}
%These symmetries are required to satisfy the algebra
satisfying
\begin{align}
    &\Gamma \mathcal{S} = \epsilon_{\rm \Gamma S} \mathcal{S} \Gamma,\quad \Gamma \tilde{\cal \M} = \epsilon_{\rm \Gamma \M} \tilde{\cal \M} \Gamma, \nonumber \\ 
    &\mathcal{S} \tilde{\cal \M} = \epsilon_{\rm S\M} \tilde{\cal \M} \mathcal{S} \quad \left( \epsilon_{\rm \Gamma S}, 
    \epsilon_{\rm \Gamma \M}, \epsilon_{\rm S\M} \in \{ \pm 1\} \right).
\end{align}
Correspondingly, the Hermitized Hamiltonians ${\sf H}$ in Eq.~(\ref{aeq: Hermitization}) respect Eqs.~(\ref{aeq: CS - H}), (\ref{aeq: SLS - H}), and (\ref{aeq: pseudo-reflection - H}),
%\begin{align}
%    {\sf \Gamma} {\sf H} {\sf \Gamma}^{-1} &= - {\sf H},\quad {\sf \Gamma} \coloneqq \begin{pmatrix}
%        0 & \Gamma \\
%        \Gamma & 0
%    \end{pmatrix}, \quad {\sf \Gamma}^2 = 1, \\
%    {\sf S} {\sf H} {\sf S}^{-1} &= - {\sf H},\quad {\sf S} \coloneqq \begin{pmatrix}
%        {\cal S} & 0 \\
%        0 & {\cal S}
%    \end{pmatrix}, \quad {\sf S}^2 = 1, \\
%    {\sf \M} {\sf H} {\sf \M}^{-1} &= {\sf H},\quad {\sf \M} \coloneqq \begin{pmatrix}
%        0 & {\cal \M} \\
%        {\cal \M} & 0
%    \end{pmatrix}, \quad {\sf \M}^2 = 1
%\end{align}
with
\begin{equation}
    {\sf \Gamma}{\sf S} = \epsilon_{\rm \Gamma S} {\sf S}{\sf \Gamma},\quad 
    {\sf \Gamma}{\sf \M} = \epsilon_{\rm \Gamma \M} {\sf \M}{\sf \Gamma},\quad 
    {\sf S}{\sf \M} = \epsilon_{\rm S\M} {\sf \M}{\sf S}.
\end{equation}
The additional chiral symmetry ${\sf \Sigma}$ in Eq.~(\ref{aeq: Hermitized chiral}) satisfies
\begin{equation}
    {\sf \Gamma}{\sf \Sigma} = - {\sf \Sigma}{\sf \Gamma},\quad {\sf S}{\sf \Sigma} ={\sf \Sigma}{\sf S},\quad {\sf \M}{\sf \Sigma} = - {\sf \Sigma}{\sf \M}.
\end{equation}
%Because of the simultaneous presence of the two independent chiral symmetries specified by ${\sf S}$ and ${\sf \Sigma}$, their combination gives unitary symmetry that commutes with ${\sf H}$:
%\begin{equation}
%    \left( {\sf S\Sigma} \right) {\sf H} \left( {\sf S\Sigma} \right)^{-1} = {\sf H},\quad \left( {\sf S\Sigma} \right)^2 = 1.
%\end{equation}
%The algebra between this unitary symmetry and the other symmetries is given by
In a similar manner to Sec.~\ref{subsubsec: reflection - AIII + Spm}, ${\sf H}$ is block-diagonalized by the unitary symmetry ${\sf S\Sigma}$ as in Eq.~(\ref{aeq: S-Sigma}), satisfying
\begin{align}
    &{\sf \Gamma} \left( {\sf S\Sigma} \right) = - \epsilon_{\rm \Gamma S} \left( {\sf S\Sigma} \right) {\sf \Gamma}, \quad 
    {\sf S} \left( {\sf S\Sigma} \right) = + \left( {\sf S\Sigma} \right) {\sf S}, \nonumber \\ 
    &{\sf \Sigma} \left( {\sf S\Sigma} \right) = + \left( {\sf S\Sigma} \right) {\sf \Sigma}, \quad 
    {\sf \M} \left( {\sf S\Sigma} \right) = - \epsilon_{\rm S\M} \left( {\sf S\Sigma} \right) {\sf \M}.
\end{align}
\begin{itemize}
    \item $\epsilon_{\rm \Gamma S} = +1$ (class AIII + ${\cal S}_+$ + $\tilde{\cal \M}_{\pm\pm}$).---While each block of ${\sf H}$ respects chiral symmetry ${\sf S}$, it no longer respects chiral symmetry ${\sf \Gamma}$. 
    Therefore, the blocks belong to class AIII.
    For $\epsilon_{\rm S\M} = +1$, reflection antisymmetry ${\sf \Gamma\M}$ ($i{\sf \Gamma\M}$) is respected for $\epsilon_{\rm \Gamma\M} = +1$ ($\epsilon_{\rm \Gamma\M} = -1$), and its commutation or anticommutation relation with chiral symmetry ${\sf S}$ is specified by $\epsilon_{\rm \Gamma S}\epsilon_{\rm S\M} = +1$.
    On the other hand, for $\epsilon_{\rm S\M} = -1$, reflection symmetry is respected, and its commutation or anticommutation relation with chiral symmetry ${\sf S}$ is specified by $\epsilon_{\rm S\M} = -1$.
    
    %\item $\epsilon_{\rm \Gamma S} = -1$ (class AIII + ${\cal S}_-$ + $\tilde{\cal \M}_{\pm\pm}$).---Each block in ${\sf H}$ remains to respect both chiral symmetries ${\sf S}$ and ${\sf \Gamma}$.
    %Moreover, reflection symmetry survives in each of the blocks only for $\epsilon_{\rm S\M} = -1$.
    %Then, the combination of the two independent chiral symmetries ${\sf S}$ and ${\sf \Gamma}$ give unitary symmetry that commutes with ${\sf H}$:
    %\begin{equation}
    %    \left( \ii{\sf S\Gamma} \right) {\sf H} \left( \ii{\sf S\Gamma} \right)^{-1} = {\sf H},\quad \left( \ii{\sf S\Gamma} \right)^2 = 1.
    %\end{equation}
    %The algebra between this unitary symmetry and the other symmetries is given by
    %\begin{equation}
    %{\sf \Gamma} \left( \ii{\sf S\Gamma} \right) = - \left( \ii{\sf S\Gamma} \right) {\sf \Gamma}, \quad
    %{\sf S} \left( \ii{\sf S\Gamma} \right) = - \left( \ii{\sf S\Gamma} \right) {\sf S}, \quad
    %{\sf \M} \left( \ii{\sf S\Gamma} \right) = + \epsilon_{\rm \Gamma \M} \epsilon_{\rm S\M} \left( \ii{\sf S\Gamma} \right) {\sf \M}.
    %\end{equation}
    %Thus, the Hermitian Hamiltonian block-diagonalized by ${\sf S\Sigma}$ and ${\sf S\Gamma}$ respects no internal symmetry (i.e., class A) but reflection symmetry only for $\epsilon_{\rm \Gamma \M} \epsilon_{\rm S\M} = +1$.

    \item $\epsilon_{\rm \Gamma S} = -1$ (class AIII + ${\cal S}_-$ + $\tilde{\cal \M}_{\pm\pm}$).---Each block of ${\sf H}$ respects both chiral symmetries ${\sf S}$ and ${\sf \Gamma}$.
    Then, the combination of ${\sf S}$ and ${\sf \Gamma}$ further gives unitary symmetry that commutes with ${\sf H}$:
    \begin{equation}
        \left( \ii{\sf S\Gamma} \right) {\sf H} \left( \ii{\sf S\Gamma} \right)^{-1} = {\sf H},\quad \left( \ii{\sf S\Gamma} \right)^2 = 1.
    \end{equation}
    The algebra between this unitary symmetry and the other symmetries is given by
    \begin{align}
    &{\sf \Gamma} \left( \ii{\sf S\Gamma} \right) = - \left( \ii{\sf S\Gamma} \right) {\sf \Gamma}, \quad
    {\sf S} \left( \ii{\sf S\Gamma} \right) = - \left( \ii{\sf S\Gamma} \right) {\sf S}, \nonumber \\
    &{\sf \M} \left( \ii{\sf S\Gamma} \right) = + \epsilon_{\rm \Gamma \M} \epsilon_{\rm S\M} \left( \ii{\sf S\Gamma} \right) {\sf \M}.
    \end{align}
    Each block diagonalized by the two unitary symmetries ${\sf S\Sigma}$ and $i{\sf S\Gamma}$ respects no internal symmetry and thus belongs to class A.
    Moreover, for $\epsilon_{\rm S\M} = -1$, reflection symmetry $\mathsf{\M}$ (antisymmetry $i\mathsf{\Gamma\M}$) survives in each of the blocks for $\epsilon_{\rm \Gamma\M} = -1$ ($\epsilon_{\rm \Gamma\M} = +1$).
    For $\epsilon_{\rm S\M} = +1$, on the other hand, each block of $\ii{\sf S\Gamma}$ respects unitary symmetry $\mathsf{S\Sigma}$ and reflection symmetry $\mathsf{M}$ that anticommute with each other, specified by ``A + $\mathsf{U}$ + $\mathsf{M}_{-}$" in Table~\ref{tab: complex AZ}.
\end{itemize}

%%%%%%%%%%
\subsection{Class BDI + ${\cal \M}_{\pm\pm}$/$\tilde{\cal \M}_{\pm\pm}$}

%%%%%
\subsubsection{Reflection}
    \label{subsubsec: reflection - BDI}

As a prime example of the real Altland-Zirnbauer class, we study class BDI + ${\cal \M}_{\pm\pm}$.
By definition, non-Hermitian Hamiltonians $H$ respect time-reversal symmetry and particle-hole symmetry, 
\begin{align}
    \mathcal{T} H^{*} \mathcal{T}^{-1} &= H,\quad \mathcal{T}\mathcal{T}^* = + 1, \label{aeq: TRS} \\
    \mathcal{C} H^{T} \mathcal{C}^{-1} &= - H,\quad \mathcal{C}\mathcal{C}^* = + 1, \label{aeq: PHS} %\\
    %\mathcal{\M} H \mathcal{\M}^{-1} &= H, \quad \mathcal{\M}^2 = 1.
\end{align}
in addition to reflection symmetry in Eq.~(\ref{aeq: reflection}).
These symmetries are required to respect
\begin{align}
    &\mathcal{T} \mathcal{C}^{*} = \mathcal{C} \mathcal{T}^{*},\quad \mathcal{T} \mathcal{\M}^* = \epsilon_{\rm T\M} \mathcal{\M} \mathcal{T}, \nonumber \\
    &\mathcal{C} \mathcal{\M}^* = \epsilon_{\rm C\M} \mathcal{\M} \mathcal{C} \quad \left( \epsilon_{\rm T\M}, \epsilon_{\rm C\M} \in \{ \pm 1\} \right).
\end{align}
As a consequence of these symmetries, the Hermitized Hamiltonians ${\sf H}$ in Eq.~(\ref{aeq: Hermitization}) respect
\begin{align}
    {\sf T} {\sf H}^{*} {\sf T}^{-1} &= {\sf H},\quad {\sf T} \coloneqq \begin{pmatrix}
        {\cal T} & 0 \\
        0 & {\cal T}
    \end{pmatrix}, \quad {\sf T}{\sf T}^* = +1, \label{aeq: TRS - H} \\
    {\sf C} {\sf H}^{T} {\sf C}^{-1} &= - {\sf H},\quad {\sf C} \coloneqq \begin{pmatrix}
        0 & {\cal C} \\
        {\cal C} & 0
    \end{pmatrix}, \quad {\sf C}{\sf C}^* = +1, \label{aeq: PHS - H} %\\
    %{\sf \M} {\sf H} {\sf \M}^{-1} &= {\sf H},\quad {\sf \M} \coloneqq \begin{pmatrix}
    %    {\cal \M} & 0 \\
    %    0 & {\cal \M}
    %\end{pmatrix}, \quad {\sf \M}^2 = 1
\end{align}
and Eq.~(\ref{aeq: reflection - H}) with
\begin{equation}
    {\sf T}{\sf C}^* = {\sf C}{\sf T}^*,\quad {\sf T}{\sf \M}^* = \epsilon_{\rm T\M} {\sf \M}{\sf T},\quad {\sf C}{\sf \M}^* = \epsilon_{\rm C\M} {\sf \M}{\sf C}.
\end{equation}
By construction, the Hermitized Hamiltonians ${\sf H}$ respect additional chiral symmetry ${\sf \Sigma}$ in Eq.~(\ref{aeq: Hermitized chiral}), which satisfies
\begin{equation}
    {\sf T}{\sf \Sigma}^* = {\sf \Sigma}{\sf T},\quad {\sf C}{\sf \Sigma}^* = -{\sf \Sigma}{\sf C},\quad {\sf \M}{\sf \Sigma} = {\sf \Sigma}{\sf \M}.
\end{equation}
Because of the simultaneous presence of the two independent chiral symmetries specified by $\ii{\sf TC^*}$ and ${\sf \Sigma}$, their combination gives unitary symmetry that commutes with ${\sf H}$:
\begin{equation}
    \left( \ii {\sf TC^*\Sigma}\right) {\sf H} \left( \ii {\sf TC^*\Sigma}\right)^{-1} = {\sf H},\quad \left( \ii {\sf TC^*\Sigma}\right)^2 = 1.
        \label{aeq: iTC-Sigma}
\end{equation}
The algebra between this unitary symmetry and the other symmetries is given by
\begin{align}
    &{\sf T} \left( \ii {\sf TC^*\Sigma}\right)^* = - \left( \ii {\sf TC^*\Sigma}\right) {\sf T}, \nonumber \\
    &{\sf C} \left( \ii {\sf TC^*\Sigma}\right)^* = + \left( \ii {\sf TC^*\Sigma}\right) {\sf C}, \nonumber \\
    &{\sf \Sigma} \left( \ii {\sf TC^*\Sigma}\right) = - \left( \ii {\sf TC^*\Sigma}\right) {\sf \Sigma}, \nonumber \\
    &{\sf \M} \left( \ii {\sf TC^*\Sigma}\right) = \epsilon_{\rm T\M} \epsilon_{\rm C\M} \left( \ii {\sf TC^*\Sigma}\right) {\sf \M}.
        \label{aeq: iTC-Sigma-algebra}
\end{align}
Thus, each block of ${\sf H}$ only respects particle-hole symmetry ${\sf C}$ and belongs to class D.
The presence of reflection symmetry or antisymmetry in the blocks depends on $\epsilon_{\rm T\M} \epsilon_{\rm C\M}$, as follows:
\begin{itemize}
    \item $\epsilon_{\rm T\M} \epsilon_{\rm C\M} = +1$.---Reflection symmetry ${\sf M}$ is respected, and its commutation or anticommutation relation with particle-hole symmetry ${\sf C}$ is specified by $\epsilon_{\rm C\M}$.

    \item $\epsilon_{\rm T\M} \epsilon_{\rm C\M} = -1$.---Reflection antisymmetry ${\sf M\Sigma}$ is respected, and its commutation or anticommutation relation with particle-hole symmetry ${\sf C}$ is specified by $-\epsilon_{\rm C\M}$.
\end{itemize}

%%%%%
\subsubsection{Pseudo-reflection}

We next consider class BDI + $\tilde{\cal \M}_{\pm\pm}$.
Non-Hermitian Hamiltonians $H$ respect time-reversal symmetry in Eq.~(\ref{aeq: TRS}), particle-hole symmetry in Eq.~(\ref{aeq: PHS}), and pseudo-reflection symmetry (or equivalently, reflection symmetry$^{\dag}$) in Eq.~(\ref{aeq: pseudo-reflection}), satisfying
%\begin{align}
%    \mathcal{T} H^{*} \mathcal{T}^{-1} &= H,\quad \mathcal{T}\mathcal{T}^* = + 1, \\
%    \mathcal{C} H^{T} \mathcal{C}^{-1} &= - H,\quad \mathcal{C}\mathcal{C}^* = + 1, \\
%    \mathcal{\M} H^{\dag} \mathcal{\M}^{-1} &= H, \quad \mathcal{\M}^2 = 1.
%\end{align}
%These symmetries are required to respect
\begin{align}
    &\mathcal{T} \mathcal{C}^{*} = \mathcal{C} \mathcal{T}^{*},\quad \mathcal{T} \tilde{\cal \M}^* = \epsilon_{\rm T\M} \tilde{\cal \M} \mathcal{T},\nonumber \\
    &\mathcal{C} \tilde{\cal \M}^* = \epsilon_{\rm C\M} \tilde{\cal \M} \mathcal{C} \quad \left( \epsilon_{\rm T\M}, \epsilon_{\rm C\M} \in \{ \pm 1\} \right).
\end{align}
As a consequence, the Hermitized Hamiltonians ${\sf H}$ in Eq.~(\ref{aeq: Hermitization}) respect Eqs.~(\ref{aeq: TRS - H}), (\ref{aeq: PHS - H}), and (\ref{aeq: pseudo-reflection - H}), 
%\begin{align}
%    {\sf T} {\sf H}^{*} {\sf T}^{-1} &= {\sf H},\quad {\sf T} \coloneqq \begin{pmatrix}
%        {\cal T} & 0 \\
%        0 & {\cal T}
%    \end{pmatrix}, \quad {\sf T}{\sf T}^* = +1, \\
%    {\sf C} {\sf H}^{T} {\sf C}^{-1} &= - {\sf H},\quad {\sf C} \coloneqq \begin{pmatrix}
%        0 & {\cal C} \\
%        {\cal C} & 0
%    \end{pmatrix}, \quad {\sf C}{\sf C}^* = +1, \\
%    {\sf \M} {\sf H} {\sf \M}^{-1} &= {\sf H},\quad {\sf \M} \coloneqq \begin{pmatrix}
%        0 & {\cal \M} \\
%        {\cal \M} & 0
%    \end{pmatrix}, \quad {\sf \M}^2 = 1
%\end{align}
with
\begin{align}
    &{\sf T}{\sf C}^* = {\sf C}{\sf T}^*,\quad {\sf T}{\sf \M}^* = \epsilon_{\rm T\M} {\sf \M}{\sf T}, \nonumber \\
    &{\sf C}{\sf \M}^* = \epsilon_{\rm C\M} {\sf \M}{\sf C}.
\end{align}
The additional chiral symmetry ${\sf \Sigma}$ in Eq.~(\ref{aeq: Hermitized chiral}) satisfies
\begin{equation}
    {\sf T}{\sf \Sigma}^* = {\sf \Sigma}{\sf T},\quad {\sf C}{\sf \Sigma}^* = -{\sf \Sigma}{\sf C},\quad {\sf \M}{\sf \Sigma} = - {\sf \Sigma}{\sf \M}.
\end{equation}
%Because of the simultaneous presence of the two independent chiral symmetries specified by $\ii{\sf TC^*}$ and ${\sf \Sigma}$, their combination gives unitary symmetry that commutes with ${\sf H}$:
%\begin{equation}
%    \left( \ii {\sf TC^*\Sigma}\right) {\sf H} \left( \ii {\sf TC^*\Sigma}\right)^{-1} = {\sf H},\quad \left( \ii {\sf TC^*\Sigma}\right)^2 = 1.
%\end{equation}
%The algebra between this unitary symmetry and the other symmetries is given as
Similarly to Sec.~\ref{subsubsec: reflection - BDI}, ${\sf H}$ is block-diagonalized by the unitary symmetry $\ii {\sf TC^*\Sigma}$ as in Eq.~(\ref{aeq: iTC-Sigma}), satisfying
\begin{align}
    &{\sf T} \left( \ii {\sf TC^*\Sigma}\right)^* = - \left( \ii {\sf TC^*\Sigma}\right) {\sf T}, \nonumber \\
    &{\sf C} \left( \ii {\sf TC^*\Sigma}\right)^* = + \left( \ii {\sf TC^*\Sigma}\right) {\sf C}, \nonumber \\
    &{\sf \Sigma} \left( \ii {\sf TC^*\Sigma}\right) = - \left( \ii {\sf TC^*\Sigma}\right) {\sf \Sigma}, \nonumber \\
    &{\sf \M} \left( \ii {\sf TC^*\Sigma}\right) = -  \epsilon_{\rm T\M} \epsilon_{\rm C\M} \left( \ii {\sf TC^*\Sigma}\right) {\sf \M}.
        \label{aeq: iTC-Sigma-algebra - pseudo}
\end{align}
Thus, each block of ${\sf H}$ only respects particle-hole symmetry ${\sf C}$ and belongs to class D.
The presence of reflection symmetry or antisymmetry in the blocks depends on $\epsilon_{\rm T\M} \epsilon_{\rm C\M}$, as follows:
\begin{itemize}
    \item $\epsilon_{\rm T\M} \epsilon_{\rm C\M} = +1$.---Reflection antisymmetry $i{\sf M\Sigma}$ is respected, and its commutation or anticommutation relation with particle-hole symmetry ${\sf C}$ is specified by $\epsilon_{\rm C\M}$.

    \item $\epsilon_{\rm T\M} \epsilon_{\rm C\M} = -1$.---Reflection symmetry ${\sf M}$ is respected, and its commutation or anticommutation relation with particle-hole symmetry ${\sf C}$ is specified by $\epsilon_{\rm C\M}$.
\end{itemize}

%\clearpage
%%%%%%%%%%
\subsection{Class DIII$^{\dag}$ + ${\cal \M}_{\pm\pm}$/$\tilde{\cal \M}_{\pm\pm}$}

%%%%%
\subsubsection{Reflection}  
    \label{subsubsec: reflection - DIIIdag}

As a prime example of the real Altland-Zirnbauer$^{\dag}$ class, we study class DIII$^{\dag}$ + ${\cal \M}_{\pm\pm}$.
Non-Hermitian Hamiltonians $H$ respect time-reversal symmetry$^{\dag}$ and particle-hole symmetry$^{\dag}$,
\begin{align}
    \mathcal{T} H^{T} \mathcal{T}^{-1} &= H,\quad \mathcal{T}\mathcal{T}^* = - 1, \label{aeq: TRS - dag} \\
    \mathcal{C} H^{*} \mathcal{C}^{-1} &= - H,\quad \mathcal{C}\mathcal{C}^* = + 1, \label{aeq: PHS - dag} %\\
    %\mathcal{\M} H \mathcal{\M}^{-1} &= H, \quad \mathcal{\M}^2 = 1, 
\end{align}
in addition to reflection symmetry in Eq.~(\ref{aeq: reflection}).
These symmetries are required to respect
\begin{align}
    &\mathcal{T} \mathcal{C}^{*} = - \mathcal{C} \mathcal{T}^{*},\quad \mathcal{T} \mathcal{\M}^* = \epsilon_{\rm T\M} \mathcal{\M} \mathcal{T}, \nonumber \\ 
    &\mathcal{C} \mathcal{\M}^* = \epsilon_{\rm C\M} \mathcal{\M} \mathcal{C} \quad \left( \epsilon_{\rm T\M}, \epsilon_{\rm C\M} \in \{ \pm 1\} \right).
\end{align}
As a consequence of these symmetries, the Hermitized Hamiltonians ${\sf H}$ in Eq.~(\ref{aeq: Hermitization}) respect
\begin{align}
    {\sf T} {\sf H}^{T} {\sf T}^{-1} &= {\sf H},\quad {\sf T} \coloneqq \begin{pmatrix}
        0 & {\cal T} \\
        {\cal T} & 0
    \end{pmatrix}, \quad {\sf T}{\sf T}^* = -1, \label{aeq: TRS - dag - H} \\
    {\sf C} {\sf H}^{*} {\sf C}^{-1} &= - {\sf H},\quad {\sf C} \coloneqq \begin{pmatrix}
        {\cal C} & 0 \\
        0 & {\cal C}
    \end{pmatrix}, \quad {\sf C}{\sf C}^* = +1, \label{aeq: PHS - dag - H} %\\
    %{\sf \M} {\sf H} {\sf \M}^{-1} &= {\sf H},\quad {\sf \M} \coloneqq \begin{pmatrix}
    %    {\cal \M} & 0 \\
    %    0 & {\cal \M}
    %\end{pmatrix}, \quad {\sf \M}^2 = 1
\end{align}
and Eq.~(\ref{aeq: reflection - H}), with
\begin{align}
    &{\sf T}{\sf C}^* = - {\sf C}{\sf T}^*,\quad {\sf T}{\sf \M}^* = \epsilon_{\rm T\M} {\sf \M}{\sf T}, \nonumber \\ 
    &{\sf C}{\sf \M}^* = \epsilon_{\rm C\M} {\sf \M}{\sf C}.
\end{align}
By construction, ${\sf H}$ respects additional chiral symmetry ${\sf \Sigma}$ in Eq.~(\ref{aeq: Hermitized chiral}), satisfying
\begin{equation}
    {\sf T}{\sf \Sigma}^* = - {\sf \Sigma}{\sf T},\quad {\sf C}{\sf \Sigma}^* ={\sf \Sigma}{\sf C},\quad {\sf \M}{\sf \Sigma} = {\sf \Sigma}{\sf \M}.
\end{equation}
Because of the simultaneous presence of the two independent chiral symmetries specified by $\ii{\sf TC^*}$ and ${\sf \Sigma}$, their combination gives unitary symmetry that commutes with ${\sf H}$, as in Eq.~(\ref{aeq: iTC-Sigma}). %:
%\begin{equation}
%    \left( \ii {\sf TC^*\Sigma}\right) {\sf H} \left( \ii {\sf TC^*\Sigma}\right)^{-1} = {\sf H},\quad \left( \ii {\sf TC^*\Sigma}\right)^2 = 1.
%\end{equation}
The algebra between this unitary symmetry and the other symmetries also coincides with Eq.~(\ref{aeq: iTC-Sigma-algebra}).
%\begin{align}
%    {\sf T} \left( \ii {\sf TC^*\Sigma}\right)^* &= - \left( \ii {\sf TC^*\Sigma}\right) {\sf T}, \\
%    {\sf C} \left( \ii {\sf TC^*\Sigma}\right)^* &= + \left( \ii {\sf TC^*\Sigma}\right) {\sf C}, \\
%    {\sf \Sigma} \left( \ii {\sf TC^*\Sigma}\right) &= - \left( \ii {\sf TC^*\Sigma}\right) {\sf \Sigma}, \\
%    {\sf \M} \left( \ii {\sf TC^*\Sigma}\right) &= \epsilon_{\rm T\M} \epsilon_{\rm C\M} \left( \ii {\sf TC^*\Sigma}\right) {\sf \M}.
%\end{align}
Thus, each block of ${\sf H}$ only respects particle-hole symmetry ${\sf C}$ and hence belongs to class D.
The presence of reflection symmetry or antisymmetry in the blocks depends on $\epsilon_{\rm T\M} \epsilon_{\rm C\M}$, as follows:
\begin{itemize}
    \item $\epsilon_{\rm T\M} \epsilon_{\rm C\M} = +1$.---Reflection symmetry ${\sf M}$ is respected, and its commutation or anticommutation relation with particle-hole symmetry ${\sf C}$ is specified by $\epsilon_{\rm CM}$.

    \item $\epsilon_{\rm T\M} \epsilon_{\rm C\M} = -1$.---Reflection antisymmetry ${\sf M\Sigma}$ is respected, and its commutation or anticommutation relation with particle-hole symmetry ${\sf C}$ is specified by $\epsilon_{\rm CM}$.
\end{itemize}

%%%%%
\subsubsection{Pseudo-reflection}

We also study class DIII$^{\dag}$ + $\tilde{\cal \M}_{\pm\pm}$.
Non-Hermitian Hamiltonians $H$ respect time-reversal symmetry$^{\dag}$ in Eq.~(\ref{aeq: TRS - dag}), particle-hole symmetry$^{\dag}$ in Eq.~(\ref{aeq: PHS - dag}), and pseudo-reflection symmetry (or equivalently, reflection symmetry$^{\dag}$) in Eq.~(\ref{aeq: pseudo-reflection}), satisfying %, respectively:
%\begin{align}
%    \mathcal{T} H^{T} \mathcal{T}^{-1} &= H,\quad \mathcal{T}\mathcal{T}^* = - 1, \\
%    \mathcal{C} H^{*} \mathcal{C}^{-1} &= - H,\quad \mathcal{C}\mathcal{C}^* = + 1, \\
%    \mathcal{R} H^{\dag} \mathcal{R}^{-1} &= H, \quad \mathcal{R}^2 = 1.
%\end{align}
%These symmetries are required to respect
\begin{align}
    &\mathcal{T} \mathcal{C}^{*} = - \mathcal{C} \mathcal{T}^{*},\quad \mathcal{T} \tilde{\cal \M}^* = \epsilon_{\rm T\M} \tilde{\cal \M} \mathcal{T},\\
    &\mathcal{C} \tilde{\cal \M}^* = \epsilon_{\rm C\M} \tilde{\cal \M} \mathcal{C} \quad \left( \epsilon_{\rm T\M}, \epsilon_{\rm C\M} \in \{ \pm 1\} \right).
\end{align}
The Hermitized Hamiltonian ${\sf H}$ in Eq.~(\ref{aeq: Hermitization}) respects Eqs.~(\ref{aeq: TRS - dag - H}), (\ref{aeq: PHS - dag - H}), and (\ref{aeq: pseudo-reflection - H}),
%\begin{align}
%    {\sf T} {\sf H}^{T} {\sf T}^{-1} &= {\sf H},\quad {\sf T} \coloneqq \begin{pmatrix}
%        0 & {\cal T} \\
%        {\cal T} & 0
%    \end{pmatrix}, \quad {\sf T}{\sf T}^* = -1, \\
%    {\sf C} {\sf H}^{*} {\sf C}^{-1} &= - {\sf H},\quad {\sf C} \coloneqq \begin{pmatrix}
%        {\cal C} & 0 \\
%        0 & {\cal C}
%    \end{pmatrix}, \quad {\sf C}{\sf C}^* = +1, \\
%    {\sf \M} {\sf H} {\sf \M}^{-1} &= {\sf H},\quad {\sf \M} \coloneqq \begin{pmatrix}
%        0 & {\cal \M} \\
%        {\cal \M} & 0
%    \end{pmatrix}, \quad {\sf \M}^2 = 1,
%\end{align}
with
\begin{align}
    &{\sf T}{\sf C}^* = - {\sf C}{\sf T}^*,\quad {\sf T}{\sf \M}^* = \epsilon_{\rm T\M} {\sf \M}{\sf T}, \nonumber \\
    &{\sf C}{\sf \M}^* = \epsilon_{\rm C\M} {\sf \M}{\sf C}.
\end{align}
The additional chiral symmetry ${\sf \Sigma}$ in Eq.~(\ref{aeq: Hermitized chiral}) satisfies
\begin{equation}
    {\sf T}{\sf \Sigma}^* = - {\sf \Sigma}{\sf T},\quad {\sf C}{\sf \Sigma}^* ={\sf \Sigma}{\sf C},\quad {\sf \M}{\sf \Sigma} = - {\sf \Sigma}{\sf \M}.
\end{equation}
%Because of the simultaneous presence of the two independent unitary symmetries specified by $\ii{\sf TC^*}$ and ${\sf \Sigma}$, their combination gives unitary symmetry that commutes with ${\sf H}$:
%\begin{equation}
%    \left( \ii {\sf TC^*\Sigma}\right) {\sf H} \left( \ii {\sf TC^*\Sigma}\right)^{-1} = {\sf H},\quad \left( \ii {\sf TC^*\Sigma}\right)^2 = 1.
%\end{equation}
Similarly to Sec.~\ref{subsubsec: reflection - DIIIdag}, ${\sf H}$ is block diagonalized by the unitary symmetry $\ii {\sf TC^*\Sigma}$, as in Eq.~(\ref{aeq: iTC-Sigma}).
The algebra between this unitary symmetry and the other symmetries also coincides with Eq.~(\ref{aeq: iTC-Sigma-algebra - pseudo}).
%\begin{align}
%    {\sf T} \left( \ii {\sf TC^*\Sigma}\right)^* &= - \left( \ii {\sf TC^*\Sigma}\right) {\sf T}, \\
%    {\sf C} \left( \ii {\sf TC^*\Sigma}\right)^* &= + \left( \ii {\sf TC^*\Sigma}\right) {\sf C}, \\
%    {\sf \Sigma} \left( \ii {\sf TC^*\Sigma}\right) &= - \left( \ii {\sf TC^*\Sigma}\right) {\sf \Sigma}, \\
%    {\sf \M} \left( \ii {\sf TC^*\Sigma}\right) &= - \epsilon_{\rm T\M} \epsilon_{\rm C\M} \left( \ii {\sf TC^*\Sigma}\right) {\sf \M}.
%\end{align}
Thus, each block of ${\sf H}$ only respects particle-hole symmetry ${\sf C}$ and belongs to class D.
The presence of reflection symmetry or antisymmetry in the blocks depends on $\epsilon_{\rm T\M} \epsilon_{\rm C\M}$, as follows:
\begin{itemize}
    \item $\epsilon_{\rm T\M} \epsilon_{\rm C\M} = +1$.---Reflection antisymmetry $i{\sf M\Sigma}$ is respected, and its commutation or anticommutation relation with particle-hole symmetry ${\sf C}$ is specified by $-\epsilon_{\rm C\M}$.

    \item $\epsilon_{\rm T\M} \epsilon_{\rm C\M} = -1$.---Reflection symmetry ${\sf M}$ is respected, and its commutation or anticommutation relation with particle-hole symmetry ${\sf C}$ is specified by $\epsilon_{\rm C\M}$.
\end{itemize}

%%%%%
\subsection{Hermitian boundary classification}

While we have hitherto developed the bulk classification, direct analysis of second-order topological boundary states is also feasible.
As illustrative cases, we here study Hermitian Hamiltonians in classes AIII + $\mathsf{M}_+$, AIII + $\mathsf{U}_{+}$ + $\mathsf{M}_{+-}$, and AIII + $\mathsf{U}_{+}$ + $\mathsf{M}_{+-}$, all of which are included in Table~\ref{tab: complex AZ}.

%%%%%
\subsubsection{Class AIII + $\mathsf{M}_+$}

We investigate second-order topological boundary states of Hermitian systems in class AIII + $\mathsf{M}_+$. 
We consider a Hermitian Hamiltonian $\mathsf{H} \left( k_i, \bm{k}_{\parallel} \right)$ around a reflection-invariant corner or hinge, where $k_i$ denotes momentum perpendicular to the mirror plane, and $\bm{k}_{\parallel}$ momenta within the mirror plane. 
Then, $\mathsf{H} \left( k_i, \bm{k}_{\parallel} \right)$ respects chiral symmetry and reflection symmetry that commute with each other:
\begin{align}
    &\mathsf{\Gamma} \mathsf{H} \left( k_i, \bm{k}_{\parallel} \right) \mathsf{\Gamma}^{-1} = - \mathsf{H} \left( k_i, \bm{k}_{\parallel} \right), \\ 
    &\mathsf{\M}_+ \mathsf{H} \left( k_i, \bm{k}_{\parallel} \right) \mathsf{\M}_+^{-1} = \mathsf{H} \left( - k_i, \bm{k}_{\parallel} \right),
\end{align}
with unitary matrices $\mathsf{\Gamma}$ and $\mathsf{\M}_+$ satisfying
\begin{equation}
    \mathsf{\Gamma}^2 = \mathsf{\M}_+^2 = 1, \quad \left[ \mathsf{\Gamma}, \mathsf{\M}_+ \right] = 0.
\end{equation}
Around a reflection-invariant corner in a 2D system, a gapless state $\mathsf{H} \left( k_i \right) = k_i \sigma_x$ appears, where the symmetry operators are chosen as $\mathsf{\Gamma} = \sigma_z$ and $\mathsf{\M}_+ = \sigma_z$.
On the other hand, away from the reflection-invariant corner, this gapless state is lifted by a perturbation $m\sigma_y$ ($m \in \mathbb{R}$).
This demonstrates the emergence of corner states at zero energy, consistent with Table~\ref{tab: complex AZ}.
By contrast, around a reflection-invariant hinge in a 3D system, no symmetry-preserving gapless states appear, showing the absence of second-order topological boundary states.

%%%%%
\subsubsection{Class AIII + $\mathsf{U}_{+}$ + $\mathsf{M}_{+-}$}

In class AIII + $\mathsf{U}_{+}$ + $\mathsf{M}_{+-}$, a Hermitian Hamiltonian $\mathsf{H} \left( k_i, \bm{k}_{\parallel} \right)$ around a reflection-invariant corner or hinge respects chiral symmetry, onsite unitary symmetry, and reflection symmetry,
\begin{align}
    &\mathsf{\Gamma} \mathsf{H} \left( k_i, \bm{k}_{\parallel} \right) \mathsf{\Gamma}^{-1} = - \mathsf{H} \left( k_i, \bm{k}_{\parallel} \right), \nonumber \\
    &\mathsf{U} \mathsf{H} \left( k_i, \bm{k}_{\parallel} \right) \mathsf{U}^{-1} = \mathsf{H} \left( k_i, \bm{k}_{\parallel} \right), \nonumber  \\ 
    &\mathsf{\M}_{+-} \mathsf{H} \left( k_i, \bm{k}_{\parallel} \right) \mathsf{\M}_{+-}^{-1} = \mathsf{H} \left( - k_i, \bm{k}_{\parallel} \right),
\end{align}
with unitary matrices $\mathsf{\Gamma}$, $\mathsf{U}$, and $\mathsf{\M}_{+-}$ satisfying
\begin{align}
    &\mathsf{\Gamma}^2 = \mathsf{U}^2 = \mathsf{\M}_{+-}^2 = 1, \nonumber \\
    &\left[ \mathsf{\Gamma}, \mathsf{U} \right] = \left[ \mathsf{\Gamma}, \mathsf{\M}_{+-} \right] = \left\{ \mathsf{U}, \mathsf{\M}_{+-} \right\} = 0.
\end{align}
For the choice of $\mathsf{\Gamma} = \tau_z \sigma_0$, $\mathsf{U} = \tau_0 \sigma_z$, $\mathsf{\M}_{+-} = \tau_z \sigma_x$, generic mass terms that preserve both the chiral and onsite unitary symmetries are given as $\tau_x \sigma_0$, $\tau_x \sigma_z$, $\tau_y \sigma_0$, $\tau_y \sigma_z$, two of which (i.e., $\tau_x \sigma_z$, $\tau_y \sigma_z$) are even under reflection, and the other two of which (i.e., $\tau_x \sigma_0$, $\tau_y \sigma_0$) are odd.
Consequently, no gapless states appear around a reflection-invariant corner in a 2D system.
On the other hand, around a reflection-invariant hinge in a 3D system, a first-order gapless surface state $\mathsf{H} \left( k_i, k_\parallel \right) = k_i \tau_x \sigma_0 + k_\parallel \tau_y \sigma_z$ appears, where the symmetry operators are chosen as $\mathsf{\Gamma} = \tau_z \sigma_0$, $\mathsf{U} = \tau_0 \sigma_z$, $\mathsf{\M}_{+-} = \tau_z \sigma_x$.
To eliminate this first-order topology, we stack this surface state and obtain $\mathsf{H} \left( k_i, k_\parallel \right) = \left( k_i \tau_x \sigma_0 + k_\parallel \tau_y \sigma_z \right) \rho_z$ with $\mathsf{\Gamma} = \tau_z \sigma_0 \rho_z$, $\mathsf{U} = \tau_0 \sigma_z \rho_0$, $\mathsf{\M}_{+-} = \tau_z \sigma_x \rho_0$, where $\rho_z$ is the Pauli matrix, and $\rho_0$ is the two-by-two identity matrix.
In such a case, it can be gapped out by a symmetry-preserving mass term $m \rho_{x}$ ($m \in \mathbb{R}$), implying the absence of second-order topological boundary states.

%%%%%
\subsubsection{Class A + $\mathsf{U}$ + $\mathsf{M}_{-}$}

In class A + $\mathsf{U}$ + $\mathsf{M}_{-}$, a Hermitian Hamiltonian $\mathsf{H} \left( k_i, \bm{k}_{\parallel} \right)$ around a reflection-invariant corner or hinge respects onsite unitary symmetry and reflection symmetry,
\begin{align}
    &\mathsf{U} \mathsf{H} \left( k_i, \bm{k}_{\parallel} \right) \mathsf{U}^{-1} = \mathsf{H} \left( k_i, \bm{k}_{\parallel} \right), \nonumber \\ 
    &\mathsf{\M}_{-} \mathsf{H} \left( k_i, \bm{k}_{\parallel} \right) \mathsf{\M}_{-}^{-1} = \mathsf{H} \left( - k_i, \bm{k}_{\parallel} \right),
\end{align}
with unitary matrices $\mathsf{U}$ and $\mathsf{\M}_{-}$ satisfying
\begin{equation}
    \mathsf{U}^2 = \mathsf{\M}_{-}^2 = 1, \quad \left\{ \mathsf{U}, \mathsf{\M}_{-} \right\} = 0.
\end{equation}
Around a reflection-invariant corner in a 2D system, a first-order gapless edge state $\mathsf{H} \left( k_i \right) = k_i \sigma_z$ appears, where the symmetry operators are chosen as $\mathsf{U} = \sigma_z$ and $\mathsf{\M}_{-} = \sigma_x$.
To eliminate this first-order topology, we stack this edge state, yielding $\mathsf{H} \left( k_i \right) = k_i \tau_z \sigma_z$ with $\mathsf{U} = \tau_0 \sigma_z$ and $\mathsf{\M}_{-} = \tau_z \sigma_x$.
In such a case, it can be gapped out by a symmetry-preserving mass term $m\tau_x \sigma_z$ ($m \in \mathbb{R}$).
This implies the absence of second-order topological boundary states, consistent with Table~\ref{tab: complex AZ}.
Additionally, around a reflection-invariant hinge in a 3D system, no symmetry-preserving gapless states appear, showing the absence of second-order topological boundary states.

\subsection{Non-Hermitian boundary classification}\label{sec:nh_boundary}
Here, we develop the classification of non-Hermitian boundary states to reveal the symmetry classes in which second-order boundary states appear. 
We consider whether the second-order boundary states appear based on the analysis of boundary Hamiltonians $H(k_j, \boldsymbol{k}_\parallel)$ around a reflection-invariant corner or hinge, where $k_j$ is a momentum perpendicular to the mirror plane, and $\boldsymbol{k}_\parallel$ is momenta within the mirror plane. 
The results are summarized in the entries specified by ``$*$" in Tables~\ref{tab: complex AZ},~\ref{tab: real AZ}, and~\ref{tab: real AZ dag}.
Below, we explicitly show the presence and absence of second-order boundary states in the effective boundary Hamiltonians.

\subsubsection{Class A + $\mathcal{\M}$/$\tilde{{\cal \M}}$}\label{classA+M/tildeM}
%\subsubsection{Reflection}
{\it Reflection symmetry.---}In this symmetry class, non-Hermitian Hamiltonians respect reflection symmetry
\begin{align}\label{eq:boundary_reflection}
    {\cal \M}H(k_j, \boldsymbol{k}_\parallel){\cal \M}^{-1}=H(-k_j, \boldsymbol{k}_\parallel),\quad {\cal \M}^2=1,
\end{align}
where ${\cal \M}$ is a unitary matrix.
When the symmetry operator is chosen as $\mathcal{\M}=\sigma_x$, symmetry-preserving mass terms are given by $\sigma_0$, $i\sigma_0$, $\sigma_x$, and $i \sigma_x$.
Around a reflection-invariant corner of a 2D system, two generic boundary states are given by $k_j \sigma_y$ and $ik_j \sigma_y$. The former and the latter can be gapped out by symmetry-preserving mass terms $\delta \sigma_x$ and $i\delta \sigma_x$ ($\delta \in \mathbb{R}$), respectively. 
Thus, second-order boundary states do not appear in this symmetry class. 
%Furthermore, we show the absence of second-order boundary states in another way. 
Instead, as discussed below, the non-Hermitian Hamiltonian exhibits second-order skin effects.  
A Hamiltonian block-diagonalized by $\mathcal{\M}$ reads $H(k_j, \boldsymbol{k}_\parallel) = H_{+}(k_j, \boldsymbol{k}_\parallel) \oplus H_{-}(k_j, \boldsymbol{k}_\parallel)$ with $k_j =0, \pi$, where $H_{\pm}(k_j, \boldsymbol{k}_\parallel)$ is the matrix labeled by the reflection-symmetry eigenvalues $\pm1$.
Each matrix belongs to class A and is characterized by the winding number $W_{\pm}$ for $H_{\pm}(k_j, \boldsymbol{k}_\parallel)$. 
When the first-order topology is trivial, the sum of the winding numbers is zero, $W_{+}=-W_{-}$. When the winding numbers $W_{\pm}$ of $H_{\pm}(k_j, \boldsymbol{k}_\parallel)$ in the mirror-invariant plane are nonzero, the system exhibits non-Hermitian skin effects, showing that the $\mathbb{Z}$ classification in this class corresponds to the second-order skin effects protected by reflection symmetry. 

%\subsubsection{Pseudo-reflection}
{\it Pseudo-reflection symmetry.---}In this symmetry class, non-Hermitian Hamiltonians respect pseudo-reflection symmetry
\begin{align}\label{eq:boundary_preflection}
    \tilde{{\cal \M}}H^{\dagger}(k_j, \boldsymbol{k}_\parallel)\tilde{{\cal \M}}^{-1}=H(-k_j, \boldsymbol{k}_\parallel),\quad \tilde{{\cal \M}}^2=1,
\end{align}
where $\tilde{{\cal \M}}$ is a unitary matrix. 
Around a reflection-symmetric hinge of a 3D system, a gapless boundary state is given by
\begin{align}
    H(k_j,k_{\parallel})= ik_j\sigma_x + k_{\parallel}\sigma_0,
\end{align}
where the symmetry operator is chosen as $\tilde{\cal \M}=\sigma_0$. 
This boundary state is point-gapless in the presence of pseudo-reflection symmetry. Conversely, a symmetry-breaking mass term $i\delta \sigma_y$ ($\delta \in \mathbb{R}$) induces a point gap. This indicates the emergence of point-gapless boundary states around pseudo-reflection-invariant hinges. Thus, the $\mathbb{Z}$ classification in this symmetry class corresponds to the emergence of second-order boundary states.

\subsubsection{Class AIII + $\mathcal{\M}_{\pm}/\tilde{\mathcal{\M}}_{\pm}$}\label{classAIII+M/tildeM}
In this symmetry class, non-Hermitian Hamiltonians $H(k_j, \boldsymbol{k}_\parallel)$ respect chiral symmetry
\begin{align}\label{eq:boundary_chiral}
    \Gamma H^{\dagger}(k_j, \boldsymbol{k}_\parallel)\Gamma^{-1}=- H(k_j, \boldsymbol{k}_\parallel), \quad \Gamma^2=1,
\end{align}
with a unitary matrix $\Gamma$.
Additionally, they satisfy reflection symmetry in Eq.~\eqref{eq:boundary_reflection} or pseudo-reflection symmetry in Eq.~\eqref{eq:boundary_preflection}.
%\subsubsection{Class AIII + $\mathcal{\M}_{+}$}

{\it Reflection symmetry} $\mathcal{\M}_{+}$.---Around reflection-invariant hinges of 3D systems, a boundary state is given by
\begin{align}
    H(k_j,k_{\parallel})= k_j\sigma_x + ik_{\parallel}\sigma_0,
\end{align}
where the symmetry operations are chosen as $\Gamma=\sigma_z$ and $\mathcal{M}=\sigma_z$.
This boundary state is point-gapless in the simultaneous presence of chiral and reflection symmetries.
Conversely, a reflection-symmetry-breaking mass term $\delta \sigma_y$ ($\delta \in \mathbb{R}$) induces a point gap. Thus, this boundary state is point-gapless at the reflection-invariant hinges, showing that the $\mathbb{Z}$ classification %in 3D systems in this symmetry class 
includes the second-order boundary states.

%\subsubsection{Class AIII + $\mathcal{\M}_{-}$}
{\it Reflection symmetry} $\mathcal{\M}_{-}$.---Around a reflection-invariant corner of a 2D system, a first-order point-gapless boundary state $H(k_j) = ik_j \sigma_z$ appears, where the symmetry operators are chosen as $\Gamma = \sigma_z$ and $\mathcal{\M} = \sigma_x$. 
To eliminate the first-order topology, we stack this boundary state, yielding $H(k_j) = ik_j\tau_z \sigma_z$ with $\Gamma = \sigma_z$ and $\mathcal{\M} = \sigma_x$. It can be gapped out by a symmetry-preserving mass term $i\delta \tau_x \sigma_0$ ($\delta \in \mathbb{R}$), showing the absence of second-order boundary states. 
Thus, the $\mathbb{Z}$ classification %in this symmetry class
does not include exceptional second-order topological insulators.

%\subsubsection{Class AIII + $\tilde{\mathcal{\M}}_{+}$}
{\it Pseudo-reflection symmetry} $\tilde{\mathcal{\M}}_{+}$.---%In this symmetry class, non-Hermitian Hamiltonians $H(k_j, \boldsymbol{k}_\parallel)$ respect chiral symmetry [Eq.~\eqref{eq:boundary_chiral}] and pseudo-reflection symmetry [Eq.~\eqref{eq:boundary_preflection}]. 
Around a reflection-invariant corner of a 2D system, a point-gapless boundary state 
\begin{align}
    H(k_j)=k_j \sigma_x,
\end{align}
 appears, where the symmetry operators are chosen as $\Gamma = \sigma_z$ and $\tilde{\mathcal{M}} = \sigma_z$. This boundary state is gapless in the simultaneous presence of pseudo-reflection and chiral symmetries. 
This gapless state can be gapped out by a symmetry-breaking mass term $\delta \sigma_y$ ($\delta \in \mathbb{R}$). Thus, the boundary state is point-gapless at the reflection-invariant corner. 

%\subsubsection{Class AIII + $\tilde{\mathcal{\M}}_{-}$}
{\it Pseudo-reflection symmetry} $\tilde{\mathcal{\M}}_{-}$.---%In this symmetry class, non-Hermitian Hamiltonians $H(k_j, \boldsymbol{k}_\parallel)$ respect chiral symmetry [Eq.~\eqref{eq:boundary_chiral}] and pseudo-reflection symmetry [Eq.~\eqref{eq:boundary_preflection}]. 
Around a reflection-invariant hinge of a 3D system, a point-gapless boundary state is given by
\begin{align}
    H(k_j)=k_j \sigma_y + k_{\parallel}\sigma_x,
\end{align}
 where the symmetry operators are chosen as $\Gamma = \sigma_z$ and $\tilde{\mathcal{M}} = \sigma_x$.
This boundary state is gapless in the simultaneous presence of pseudo-reflection and chiral symmetries. 
It can be gapped out by a symmetry-breaking mass term $i \delta \sigma_0$ ($\delta \in \mathbb{R}$). Thus, the boundary state is point-gapless at the reflection-invariant hinge. 

%\subsubsection{Class AIII + $\mathcal{S}_{+}$ + ${\mathcal{\M}}_{\pm \pm}$/${\mathcal{\M}}_{\pm \mp}$/$\tilde{\mathcal{\M}}_{\pm \pm}$/$\tilde{\mathcal{\M}}_{\pm \mp}$}
\subsubsection{Class AIII + $\mathcal{S}_{+}$ + ${\mathcal{\M}}_{\pm \pm}$/${\mathcal{\M}}_{\pm \mp}$\label{classAIII+S+M/tildeM}}
In this symmetry class, non-Hermitian Hamiltonians $H(k_j, \boldsymbol{k}_\parallel)$ respect chiral symmetry in Eq.~\eqref{eq:boundary_chiral}, sublattice symmetry,
\begin{align}\label{eq:boundary_sublattice}
    \mathcal{S}H(k_j, \boldsymbol{k}_\parallel)\mathcal{S}^{-1}=-H(k_j, \boldsymbol{k}_\parallel), \quad \mathcal{S}^2=1,
\end{align}
with a unitary matrix $\mathcal{S}$, and reflection symmetry in Eq.~\eqref{eq:boundary_reflection} or pseudo-reflection symmetry in Eq.~\eqref{eq:boundary_preflection}. 

%\subsubsection{Class AIII + $\mathcal{S}_{+}$ + ${\mathcal{\M}}_{++}$}
{\it Reflection symmetry} ${\mathcal{\M}}_{++}$.---The symmetry operators satisfy $[\Gamma, \mathcal{S}]=[\Gamma, \mathcal{\M}]=[\mathcal{S}, \mathcal{\M}]=0$. 
Around a reflection-invariant corner of a 2D system, a boundary state reads 
\begin{align}
    H(k_j) = ik_j\sigma_x,
\end{align}
where the symmetry operators are chosen as $\Gamma=\sigma_0$, $\mathcal{S}=\sigma_z$, and $\mathcal{\M}=\sigma_z$. 
This boundary state is point-gapless in the simultaneous presence of chiral, sublattice, and reflection symmetries.
Conversely, a reflection-symmetry-breaking mass term $i\delta \sigma_y$ ($\delta \in \mathbb{R}$) induces a point gap. Thus, the boundary state is point-gapless at the reflection-invariant corner, showing the presence of second-order boundary states in the $\mathbb{Z}$ classification. 

{\it Reflection symmetry} ${\mathcal{\M}}_{+-}$.---The symmetry operators satisfy $[\Gamma, \mathcal{S}] = [\Gamma, \mathcal{\M}]=\{ \mathcal{S}, \mathcal{M}\}=0$. 
Around a reflection-invariant hinge of a 3D system, a boundary state is given by
\begin{align}
    H(k_j, k_{\parallel})=ik_j\tau_y \sigma_y + ik_{\parallel}\sigma_x,
\end{align}
where the symmetry operators are chosen as $\Gamma = \sigma_0$, $\mathcal{S}=\sigma_z$, and $\mathcal{\M}=\sigma_x$.
This boundary state is point-gapless in the simultaneous presence of chiral, sublattice, and reflection symmetries.
A mass term $i\delta \tau_x \sigma_y$ ($\delta \in \mathbb{R}$) breaking reflection symmetry induces a point gap. Therefore, the boundary state is point-gapless at the reflection-invariant hinge, showing that the $\mathbb{Z}$ classification %in 3D systems in this symmetry class 
includes the second-order boundary states.

%\subsubsection{Class AIII + $\mathcal{S}_{+}$ + ${\mathcal{\M}}_{-+}$}
{\it Reflection symmetry} ${\mathcal{\M}}_{-+}$.---The symmetry operators satisfy $[\Gamma, \mathcal{S}] = \{ \Gamma, \mathcal{\M} \}=[ \mathcal{S}, \mathcal{M}]=0$.
Symmetry-preserving mass terms are $\tau_y\sigma_x$, $\tau_y\sigma_y$, $i\tau_z\sigma_x$, and $i\tau_z\sigma_y$, where the symmetry operators are chosen as $\Gamma = \tau_z$, $\mathcal{S}=\sigma_z$, and $\mathcal{\M}=\tau_x \sigma_z$.
Around a reflection-invariant corner of a 2D system, boundary states are given by $H(k_j)=i k_j \sigma_x$, $i k_j \sigma_y$, $k_{j} \tau_x \sigma_x$, and $k_{j} \tau_x \sigma_y$. 
All the boundary states can be gapped out by the symmetry-preserving mass terms. 
Thus, second-order boundary states do not appear in 2D systems in this symmetry class. 

%\subsubsection{Class AIII + $\mathcal{S}_{+}$ + ${\mathcal{\M}}_{--}$}
{\it Reflection symmetry} ${\mathcal{\M}}_{--}$.---The symmetry operators satisfy $[\Gamma, \mathcal{S}] = \{ \Gamma, {\mathcal{\M}} \}=\{ \mathcal{S}, {\mathcal{M}}\}=0$.
Around a reflection-invariant hinge of a 3D system, a boundary state is given by
\begin{align}
    H(k_j, k_{\parallel}) = k_j \tau_x  \sigma_y + k_{\parallel} \tau_y \sigma_y,
\end{align}
where the symmetry operators are chosen as $\Gamma = \tau_z$, $\mathcal{S} = \sigma_z$, and ${\mathcal{M}} = \tau_x \sigma_x$.
This boundary state is point-gapless in the simultaneous presence of chiral, sublattice, and reflection symmetries.
A mass term $i\delta \sigma_y$ $(\delta \in \mathbb{R})$ breaking reflection symmetry induces a point gap.
Therefore, the boundary state is point-gapless at the reflection-invariant hinge, showing that the $\mathbb{Z}$ classification %in 3D systems in this symmetry class 
includes second-order boundary states. 

\subsubsection{Class AII$^\dagger$ + ${\mathcal{\M}}_{\pm}$/$\tilde{\mathcal{\M}}_{\pm}$}\label{classAIIdag+M/tildeM}
In this symmetry class, non-Hermitian Hamiltonians $H(k_j, \boldsymbol{k}_\parallel)$ respect time-reversal symmetry$^\dagger$,
\begin{align}\label{eq:boundary_TRSdag}
    \mathcal{T}H^{T}(k_j, \boldsymbol{k}_\parallel)\mathcal{T}^{-1}=H(-k_j, -\boldsymbol{k}_\parallel), \quad \mathcal{T}\mathcal{T}^{*}=-1,
\end{align}
with a unitary matrix $\mathcal{T}$.
Additionally, they satisfy reflection symmetry in Eq.~\eqref{eq:boundary_reflection} or pseudo-reflection symmetry in Eq.~\eqref{eq:boundary_preflection}.

%\subsubsection{Class AII$^\dagger$ + ${\mathcal{\M}}_{+}$}
{\it Reflection symmetry} $\mathcal{\M}_{+}$.---The symmetry operators satisfy $\mathcal{T} \mathcal{\M}=\mathcal{\M}^{*}\mathcal{T}$. Symmetry-preserving mass terms are given by $\sigma_0$, $i\sigma_0$, $\tau_z$, and $i\tau_z$, where the symmetry operators are chosen as $\mathcal{T}=\sigma_y$ and $\mathcal{M}=\tau_z$. Around a reflection-invariant corner of a 2D system, mass terms $\sigma_0$ and $i\sigma_0$ induce a point gap for any boundary states.  
Around a reflection-invariant hinge of a 3D system, boundary states that remain point-gapless in the presence of the mass terms $\sigma_0$ and $i\sigma_0$ are given by $H(k_j,k_{\parallel})=k_j\tau_x \sigma_l+ik_{\parallel}\sigma_l$ ($l=x,y,z$), $ik_j\tau_x \sigma_l+k_{\parallel}\sigma_l$, $k_j\tau_y +i k_{\parallel}\sigma_l$, $ik_j\tau_y + k_{\parallel}\sigma_l$,  $k_j\tau_x\sigma_l +i k_{\parallel}\tau_z\sigma_m$ ($m=x,y,z$, $l\neq m$), $ik_j\tau_x\sigma_l + k_{\parallel}\tau_z\sigma_m$. 
All the above boundary states can be gapped out by adding $\tau_z$ or $i\tau_z$, showing the absence of second-order boundary states. Thus, second-order boundary states do not appear in the $\mathbb{Z}_2$ classifications for 2D and 3D classes AII$^\dagger$ + ${\mathcal{\M}}_{+}$. 

%\subsubsection{Class AII$^\dagger$ + ${\mathcal{\M}}_{-}$}
{\it Reflection symmetry} $\mathcal{\M}_{-}$.---The symmetry operators satisfy $\mathcal{T} \mathcal{\M}=-\mathcal{\M}^{*}\mathcal{T}$.
Symmetry-preserving mass terms include $\sigma_0$ and $i\sigma_0$, where the symmetry operators are chosen as $\mathcal{T}=\sigma_y$ and $\mathcal{M}=\tau_x \sigma_x$.
Around a reflection-invariant corner of a 2D system, symmetry-preserving mass terms $\sigma_0$ and $i\sigma_0$ induce a point gap for any boundary states.
Thus, second-order boundary states do not appear in the $2\mathbb{Z}$ classification for 2D class AII$^\dagger$ + ${\mathcal{\M}}_{-}$. 

%\subsubsection{Class AII$^\dagger$ + ${\tilde{\mathcal{\M}}}_{-}$}
{\it Pseudo-reflection symmetry} $\tilde{\mathcal{\M}}_{-}$.---The symmetry operators satisfy $\mathcal{T}\tilde{\mathcal{M}}=-\tilde{\mathcal{M}}^{*}\mathcal{T}$. 
We here choose them as $\mathcal{T}=\sigma_y$ and $\tilde{\mathcal{\M}}=\sigma_x$.
Around a reflection-invariant corner of a 2D system, a boundary state is given by
\begin{align}
    H(k_j) = k_j \sigma_y, 
\end{align}
which is point-gapless in the simultaneous presence of time-reversal and pseudo-reflection symmetries in two-band systems. 
%A mass term $i\delta_{0}$ ($\delta \in \mathbb{R}$) breaking pseudo-reflection symmetry induces a point gap. 
Because a mass term $\delta\sigma_{0}$ ($\delta \in \mathbb{R}$) shifts this point-gap-closing point, the boundary state is unstable.
%Therefore, the boundary state is point-gapless at the reflection-invariant corner, showing the emergence of second-order boundary states.
%On the other hand, doubling the boundary states, we obtain a boundary state $H(k_j) = k_j \tau_0\sigma_y$. This doubled boundary state can be gapped out by adding a symmetry-preserving mass term $\delta \tau_y\sigma_x$ ($\delta \in \mathbb{R}$). Therefore, %the boundary state is $\mathbb{Z}_2$ classified, showing that 
%the $\mathbb{Z}_2$ classification for 2D class AII$^\dagger$ + ${\tilde{\mathcal{\M}}}_{-}$ includes the second-order boundary states. 

Around a reflection-invariant hinge of a 3D system, a boundary state is given by
\begin{align}\label{eq:boundary_AIIdag_preflection}
    H(k_j, k_{\parallel}) = k_j \sigma_y + k_{\parallel}\sigma_x,
\end{align}
where the symmetry operators are chosen as $\mathcal{T}=\sigma_y$ and $\tilde{\mathcal{\M}}=\sigma_x$.
This boundary state is point-gapless in the simultaneous presence of time-reversal and pseudo-reflection symmetries. 
A mass term $i\delta_{0}$ ($\delta \in \mathbb{R}$) breaking pseudo-reflection symmetry induces a point gap.
In addition, a boundary state $H(k_j, k_{\parallel}) = k_j \tau_0\sigma_y + k_{\parallel} \tau_0\sigma_x$ obtained by doubling Eq.~\eqref{eq:boundary_AIIdag_preflection} cannot be gapped out by adding symmetry-preserving mass terms, showing that this boundary state is $\mathbb{Z}$ classified. 
Thus, the second-order boundary states appear in the $\mathbb{Z}$ classification for 3D class AII$^\dagger$ + ${\tilde{\mathcal{\M}}}_{-}$.

Hereafter, for the other symmetry classes, only the results are presented.

\subsubsection{Class AIII + $\mathcal{S}_+$ + $\mathcal{\tilde{\M}}_{\pm \pm }/\tilde{\mathcal{\M}}_{\pm \mp}$}\label{classAIII+S+tildeM}

{\it Pseudo-reflection symmetry} ${\tilde{\mathcal{\M}}}_{++}$.---Around a reflection-invariant corner of a 2D system, a second-order boundary state is given by
\begin{align}
    H(k_j) = ik_j\sigma_x,
\end{align}
where the symmetry operators are chosen as $\Gamma=\sigma_0$, $\mathcal{S}=\sigma_z$, and $\tilde{\mathcal{\M}}=\sigma_0$. 
A mass term $i\delta \sigma_y$ ($\delta \in \mathbb{R}$) breaking pseudo-reflection symmetry induces a point gap.

{\it Pseudo-reflection symmetry} $\tilde{{\mathcal{\M}}}_{+-}$.---Around a reflection-invariant hinge of a 3D system, a second-order boundary state is given by
\begin{align}
    H(k_j, k_{\parallel})=ik_j\tau_y \sigma_y + ik_{\parallel}\sigma_x,
\end{align}
where the symmetry operators are chosen as $\Gamma = \sigma_0$, $\mathcal{S}=\sigma_z$, and $\tilde{\mathcal{\M}}=\sigma_y$.
A mass term $i\delta \tau_x \sigma_y$ ($\delta \in \mathbb{R}$) breaking pseudo-reflection symmetry induces a point gap. 

{\it Pseudo-reflection symmetry} ${\tilde{\mathcal{\M}}}_{--}$.---Around a reflection-invariant hinge of a 3D system, a second-order boundary state is given by
\begin{align}
    H(k_j, k_{\parallel}) = k_j \tau_x  \sigma_y + k_{\parallel} \tau_y \sigma_y,
\end{align}
where the symmetry operators are chosen as $\Gamma = \tau_z$, $\mathcal{S} = \sigma_z$, and $\tilde{\mathcal{M}} = \tau_y \sigma_y$.
A mass term $i\delta \sigma_y$ $(\delta \in \mathbb{R})$ breaking pseudo-reflection symmetry induces a point gap. 

\subsubsection{Class A + $\mathcal{S}$ + $\mathcal{{\M}}_{\pm}$/$\mathcal{\tilde{\M}}_{\pm}$}\label{classA+S+M/tildeM}

{\it Reflection symmetry} ${{\mathcal{\M}}}_{+}$.---Around a reflection-invariant corner of a 2D system, a second-order boundary state is given by
\begin{align}
    H(k_j) = k_j\sigma_x,
\end{align}
where the symmetry operators are chosen as $\mathcal{S}=\sigma_z$ and ${\mathcal{\M}}=\sigma_z$. 
A mass term $\delta \sigma_y$ ($\delta \in \mathbb{R}$) breaking reflection symmetry induces a point gap.

{\it Pseudo-reflection symmetry} ${\tilde{{\mathcal{\M}}}}_{-}$.---Around a reflection-invariant hinge of a 3D system, a second-order boundary state is given by 
\begin{align}
    H(k_j,k_{\parallel}) = k_j\tau_x\sigma_y+k_{\parallel}\sigma_x,
\end{align}
where the symmetry operators are chosen as $\mathcal{S}=\sigma_z$ and $\tilde{\mathcal{\M}}=\sigma_x$. 
A mass term $\delta \tau_y \sigma_y$ ($\delta \in \mathbb{R}$) breaking pseudo-reflection symmetry induces a point gap.

\subsubsection{Class AIII + $\mathcal{S}_-$ + $\mathcal{{\M}}_{\pm \pm }/{\mathcal{\M}}_{\pm \mp}/\mathcal{\tilde{\M}}_{\pm \pm }/\tilde{\mathcal{\M}}_{\pm \mp}$}\label{classAIII+S-M/pseudoM}

{\it Reflection symmetry} ${{\mathcal{\M}}}_{++}$.---Around a reflection-invariant hinge of a 3D system, a second-order boundary state is given by
\begin{align}
    H(k_j,k_\parallel) = k_j\sigma_y+ik_{\parallel}\tau_x,
\end{align}
where the symmetry operators are chosen as $\Gamma=\sigma_z$, $\mathcal{S}=\tau_z \sigma_x$, and ${\mathcal{\M}}=\tau_x \sigma_z$. 
A mass term $\delta \tau_x \sigma_x$ ($\delta \in \mathbb{R}$) breaking reflection symmetry induces a point gap.

{\it Reflection symmetry} ${{\mathcal{\M}}}_{-+}$.---Around a reflection-invariant corner of a 2D system, a second-order boundary state is given by
\begin{align}
    H(k_j) = k_j\sigma_y,
\end{align}
where the symmetry operators are chosen as $\Gamma=\sigma_z$, $\mathcal{S}=\tau_z \sigma_x$, and ${\mathcal{\M}}=\tau_z \sigma_x$. 
A mass term $\delta \tau_x \sigma_x$ ($\delta \in \mathbb{R}$) breaking reflection symmetry induces a point gap.

{\it Pseudo-reflection symmetry} ${\tilde{\mathcal{\M}}}_{+-}$.---Around a reflection-invariant corner of a 2D system, a second-order boundary state is given by
\begin{align}
    H(k_j) = k_j\sigma_y,
\end{align}
where the symmetry operators are chosen as $\Gamma=\sigma_z$, $\mathcal{S}=\tau_z \sigma_x$, and $\tilde{\mathcal{\M}}=\sigma_z$. 
A mass term $\delta \tau_x \sigma_x$ ($\delta \in \mathbb{R}$) breaking pseudo-reflection symmetry induces a point gap.

{\it Pseudo-reflection symmetry} ${\tilde{\mathcal{\M}}}_{--}$.---Around a reflection-invariant hinge of a 3D system, a second-order boundary state is given by
\begin{align}
    H(k_j,k_\parallel) = ik_j\tau_x+k_{\parallel}\tau_x\sigma_x,
\end{align}
where the symmetry operators are chosen as $\Gamma=\sigma_z$, $\mathcal{S}=\tau_z \sigma_x$, and $\tilde{\mathcal{\M}}=\tau_x \sigma_x$. 
A mass term $\delta \sigma_y$ ($\delta \in \mathbb{R}$) breaking pseudo-reflection symmetry induces a point gap.

\subsubsection{Class AI + $\mathcal{{\M}}_{\pm }/\mathcal{\tilde{\M}}_{\pm  }$}\label{classAI+-M/pseudoM}

{\it Pseudo-reflection symmetry} ${\tilde{\mathcal{\M}}}_{-}$.---Around a reflection-invariant hinge of a 3D system, a second-order boundary state is given by
\begin{align}
    H(k_j,k_\parallel) = ik_j+k_{\parallel}\sigma_z,
\end{align}
where the symmetry operators are chosen as $\mathcal{T}=\sigma_x$ and $\tilde{\mathcal{\M}}= \sigma_z$. 
A mass term $\delta \sigma_x$ ($\delta \in \mathbb{R}$) breaking pseudo-reflection symmetry induces a point gap.

\subsubsection{Class BDI + $\mathcal{{\M}}_{\pm \pm }/{\mathcal{\M}}_{\pm \mp}/\mathcal{\tilde{\M}}_{\pm \pm }/\tilde{\mathcal{\M}}_{\pm \mp}$}\label{classBDI+-M/pseudoM}

{\it Reflection symmetry} ${{\mathcal{\M}}}_{++}$.---Around a reflection-invariant corner of a 2D system, a second-order boundary state is given by
\begin{align}
    H(k_j) = k_j\sigma_z,
\end{align}
where the symmetry operators are chosen as $\mathcal{T}=\sigma_x$, $\mathcal{C}=\sigma_0$, and ${\mathcal{\M}}=\sigma_x$. 
A mass term $\delta \sigma_y$ ($\delta \in \mathbb{R}$) breaking reflection symmetry induces a point gap. Furthermore, around a reflection-invariant hinge of a 3D system, a second-order boundary state reads 
\begin{align}
    H(k_j, k_{\parallel}) = k_j \sigma_z + ik_{\parallel},
\end{align}
where the symmetry operators are chosen as $\mathcal{T}=\sigma_x$, $\mathcal{C}=\sigma_0$, and ${\mathcal{\M}}=\sigma_x$. 
The mass term $\delta \sigma_y$  breaking reflection symmetry induces a point gap.

{\it Pseudo-reflection symmetry} ${\tilde{\mathcal{\M}}}_{++}$.---Around a reflection-invariant corner of a 2D system, a second-order boundary state is given by
\begin{align}
    H(k_j) = k_j\sigma_z,
\end{align}
where the symmetry operators are chosen as $\mathcal{T}=\sigma_x$, $\mathcal{C}=\sigma_0$, and $\tilde{\mathcal{\M}}=\sigma_x$. 
A mass term $\delta \sigma_y$ ($\delta \in \mathbb{R}$) breaking pseudo-reflection symmetry induces a point gap.

{\it Pseudo-reflection symmetry} ${\tilde{\mathcal{\M}}}_{-+}$.---Around a reflection-invariant corner of a 2D system, a second-order boundary state is given by
\begin{align}
    H(k_j) = ik_j,
\end{align}
where the symmetry operators are chosen as $\mathcal{T}=\sigma_x$, $\mathcal{C}=\sigma_0$, and $\tilde{\mathcal{\M}}=\sigma_z$. 
A mass term $\delta \sigma_y$ ($\delta \in \mathbb{R}$) breaking pseudo-reflection symmetry induces a point gap.
Furthermore, around a reflection-invariant hinge of a 3D system, a second-order boundary state reads 
\begin{align}
    H(k_j, k_{\parallel}) = ik_j + k_{\parallel}\sigma_z,
\end{align}
where the symmetry operators are chosen as $\mathcal{T}=\sigma_x$, $\mathcal{C}=\sigma_0$, and $\tilde{\mathcal{\M}}=\sigma_z$. 
The mass term $\delta \sigma_y$  breaking pseudo-reflection symmetry induces a point gap.

\subsubsection{Class D + $\mathcal{{\M}}_{\pm }/\mathcal{\tilde{\M}}_{\pm  }$}\label{classD+-M/pseudoM}

{\it Reflection symmetry} ${{\mathcal{\M}}}_{+}$.---Around a reflection-invariant corner of a 2D system, a second-order boundary state is given by
\begin{align}
    H(k_j) = k_j \sigma_y,
\end{align}
where the symmetry operators are chosen as $\mathcal{C}=\sigma_x$ and ${\mathcal{\M}}= \sigma_x$. 
A mass term $\delta \sigma_z$ ($\delta \in \mathbb{R}$) breaking reflection symmetry induces a point gap.
Furthermore, around a reflection-invariant hinge of a 3D system, a second-order boundary state reads
\begin{align}
    H(k_j k_{\parallel}) = k_j \sigma_y+i k_\parallel ,
\end{align}
where the symmetry operators are chosen as $\mathcal{C}=\sigma_x$ and ${\mathcal{\M}}= \sigma_x$. 
The mass term $\delta \sigma_z$ breaking reflection symmetry induces a point gap.

{\it Pseudo-reflection symmetry} ${\tilde{\mathcal{\M}}}_{+}$.---Around a reflection-invariant corner of a 2D system, a second-order boundary state is given by
\begin{align}
    H(k_j) = ik_j,
\end{align}
where the symmetry operators are chosen as $\mathcal{C}=\sigma_x$ and $\tilde{\mathcal{\M}}= \sigma_x$. 
A mass term $\delta \sigma_z$ ($\delta \in \mathbb{R}$) breaking pseudo-reflection symmetry induces a point gap.
In addition, around a reflection-invariant hinge of a 3D system, a second-order boundary state reads
\begin{align}
    H(k_j,k_\parallel) = ik_j+k_{\parallel}\sigma_x,
\end{align}
where the symmetry operators are chosen as $\mathcal{C}=\sigma_x$ and $\tilde{\mathcal{\M}}= \sigma_x$. 
The mass term $\delta \sigma_z$ breaking pseudo-reflection symmetry induces a point gap.

\subsubsection{Class DIII + $\mathcal{{\M}}_{\pm \pm }/{\mathcal{\M}}_{\pm \mp}/\mathcal{\tilde{\M}}_{\pm \pm }/\tilde{\mathcal{\M}}_{\pm \mp}$}\label{classDIII+-M/pseudoM}

{\it Reflection symmetry} ${{\mathcal{\M}}}_{++}$.---Around a reflection-invariant hinge of a 3D system, a second-order boundary state is given by
\begin{align}
    H(k_j,k_\parallel) = k_j\sigma_x+ik_{\parallel},
\end{align}
where the symmetry operators are chosen as $\mathcal{T}=\tau_x\sigma_y$, $\mathcal{C}=\tau_x\sigma_x$, and ${\mathcal{\M}}=\tau_z \sigma_z$. 
A mass term $\delta \tau_z \sigma_y$ ($\delta \in \mathbb{R}$) breaking reflection symmetry induces a point gap.

{\it Reflection symmetry} ${{\mathcal{\M}}}_{-+}$.---Around a reflection-invariant corner of a 2D system, a second-order boundary state is given by
\begin{align}
    H(k_j) = k_j\sigma_x,
\end{align}
where the symmetry operators are chosen as $\mathcal{T}=\tau_x \sigma_y$, $\mathcal{C}=\tau_x \sigma_x$, and ${\mathcal{\M}}=\tau_x \sigma_y$. 
A mass term $\delta \tau_z \sigma_y$ ($\delta \in \mathbb{R}$) breaking reflection symmetry induces a point gap. Furthermore, around a reflection-invariant hinge of a 3D system, a second-order boundary state reads 
\begin{align}
    H(k_j, k_{\parallel}) = k_j \sigma_x + ik_{\parallel},
\end{align}
where the symmetry operators are chosen as $\mathcal{T}=\tau_x \sigma_y$, $\mathcal{C}=\tau_x \sigma_x$, and ${\mathcal{\M}}=\tau_x \sigma_y$. 
The mass term $\delta \tau_z \sigma_y$  breaking reflection symmetry induces a point gap.

{\it Reflection symmetry} ${{\mathcal{\M}}}_{--}$.---Around a reflection-invariant hinge of a 3D system, a second-order boundary state reads 
\begin{align}
    H(k_j, k_{\parallel}) = k_j \tau_x \sigma_x + ik_{\parallel}\tau_z \sigma_z,
\end{align}
where the symmetry operators are chosen as $\mathcal{T}=\tau_x \sigma_y$, $\mathcal{C}=\tau_x \sigma_x$, and ${\mathcal{\M}}=\tau_z$. 
A mass term $i\delta \tau_y \sigma_z$ ($\delta \in \mathbb{R}$)  breaking reflection symmetry induces a point gap.

{\it Pseudo-reflection symmetry} ${\tilde{\mathcal{\M}}}_{++}$.---Around a reflection-invariant corner of a 2D system, a second-order boundary state is given by
\begin{align}
    H(k_j) = k_j \sigma_x,
\end{align}
where the symmetry operators are chosen as $\mathcal{T}=\tau_x \sigma_y$, $\mathcal{C}=\tau_x \sigma_x$, and $\tilde{\mathcal{\M}}=\tau_z \sigma_z$. 
A mass term $\delta \tau_z \sigma_y$ ($\delta \in \mathbb{R}$) breaking pseudo-reflection symmetry induces a point gap.
Furthermore, around a reflection-invariant hinge of a 3D system, a second-order boundary state reads 
\begin{align}
    H(k_j, k_{\parallel}) = k_j \sigma_x + k_{\parallel} \tau_y \sigma_y,
\end{align}
where the symmetry operators are chosen as $\mathcal{T}=\tau_x \sigma_y$, $\mathcal{C}=\tau_x \sigma_x$, and $\tilde{\mathcal{\M}}=\tau_z \sigma_z$.
The mass term $\delta \tau_z \sigma_y$  breaking pseudo-reflection symmetry induces a point gap.

{\it Pseudo-reflection symmetry} ${\tilde{\mathcal{\M}}}_{+-}$.---Around a reflection-invariant hinge of a 3D system, a second-order boundary state is given by
\begin{align}
    H(k_j, k_{\parallel}) = ik_j + k_{\parallel} \tau_x \sigma_x,
\end{align}
where the symmetry operators are chosen as $\mathcal{T}=\tau_x\sigma_y$, $\mathcal{C}=\tau_y \sigma_y$, and $\tilde{\mathcal{\M}}=\tau_x$. 
A mass term $\delta \tau_y$ ($\delta \in \mathbb{R}$) breaking pseudo-reflection symmetry induces a point gap.

{\it Pseudo-reflection symmetry} ${\tilde{\mathcal{\M}}}_{-+}$.---Around a reflection-invariant hinge of a 3D system, a second-order boundary state is given by
\begin{align}
    H(k_j, k_{\parallel}) = ik_j + k_{\parallel} \tau_x \sigma_y,
\end{align}
where the symmetry operators are chosen as $\mathcal{T}=\tau_x\sigma_y$, $\mathcal{C}=\tau_x \sigma_x$, and $\tilde{\mathcal{\M}}=\tau_x\sigma_y$. 
A mass term $\delta \tau_z \sigma_y$ ($\delta \in \mathbb{R}$) breaking pseudo-reflection symmetry induces a point gap.

{\it Pseudo-reflection symmetry} ${\tilde{\mathcal{\M}}}_{--}$.---Around a reflection-invariant corner of a 2D system, a second-order boundary state is given by
\begin{align}
    H(k_j) = k_j \sigma_x,
\end{align}
where the symmetry operators are chosen as $\mathcal{T}=\tau_x\sigma_y$, $\mathcal{C}=\tau_x \sigma_x$, and $\tilde{\mathcal{\M}}=\sigma_z$. 
A mass term $\delta \tau_z \sigma_y$ ($\delta \in \mathbb{R}$) breaking pseudo-reflection symmetry induces a point gap.

\subsubsection{Class AII + $\mathcal{{\M}}_{\pm }/\mathcal{\tilde{\M}}_{\pm  }$}\label{classAII+-M/pseudoM}

{\it Reflection symmetry} ${{\mathcal{\M}}}_{-}$.---Around a reflection-invariant corner of a 2D system, a second-order boundary state is given by
\begin{align}
    H(k_j) = k_j\sigma_y,
\end{align}
where the symmetry operators are chosen as $\mathcal{T}=\sigma_y$ and ${\mathcal{\M}}= \sigma_x$. Because a mass term $\delta \sigma_0$ ($\delta \in \mathbb{R}$) shifts the point-gap-closing point, this boundary state is unstable.  

{\it Pseudo-reflection symmetry} ${\tilde{\mathcal{\M}}}_{+}$.---Around a reflection-invariant hinge of a 3D system, a second-order boundary state is given by
\begin{align}
    H(k_j,k_\parallel) = ik_j+k_{\parallel}\tau_x\sigma_x,
\end{align}
where the symmetry operators are chosen as $\mathcal{T}=\tau_x \sigma_y$ and $\tilde{\mathcal{\M}}= \tau_x$. 
A mass term $\delta \tau_y$ ($\delta \in \mathbb{R}$) breaking pseudo-reflection symmetry induces a point gap.

{\it Pseudo-reflection symmetry} ${\tilde{\mathcal{\M}}}_{-}$.---Around a reflection-invariant hinge of a 3D system, a second-order boundary state is given by
\begin{align}
    H(k_j,k_\parallel) = ik_j+k_{\parallel}\sigma_x,
\end{align}
where the symmetry operators are chosen as $\mathcal{T}=\sigma_y$ and $\tilde{\mathcal{\M}}= \sigma_x$. 
A mass term $\delta \tau_y \sigma_y$ ($\delta \in \mathbb{R}$) breaking pseudo-reflection symmetry induces a point gap.

\subsubsection{Class CII + $\mathcal{{\M}}_{\pm \pm }/{\mathcal{\M}}_{\pm \mp}/\mathcal{\tilde{\M}}_{\pm \pm }/\tilde{\mathcal{\M}}_{\pm \mp}$}\label{classCII+-M/pseudoM}

{\it Reflection symmetry} ${{\mathcal{\M}}}_{++}$.---Around a reflection-invariant hinge of a 3D system, a second-order boundary state is given by
\begin{align}
    H(k_j,k_\parallel) = k_j \sigma_x+ik_{\parallel},
\end{align}
where the symmetry operators are chosen as $\mathcal{T}=\sigma_y$, $\mathcal{C}=\tau_y$, and ${\mathcal{\M}}=\tau_y \sigma_y$. 
A mass term $\delta \tau_y \sigma_z$ ($\delta \in \mathbb{R}$) breaking reflection symmetry induces a point gap.

{\it Reflection symmetry} ${{\mathcal{\M}}}_{--}$.---Around a reflection-invariant hinge of a 3D system, a second-order boundary state is given by
\begin{align}
    H(k_j,k_\parallel) = k_j \sigma_z +ik_{\parallel},
\end{align}
where the symmetry operators are chosen as $\mathcal{T}=\sigma_y$, $\mathcal{C}=\tau_y$, and ${\mathcal{\M}}=\tau_x \sigma_x$. 
A mass term $\delta \tau_y \sigma_x$ ($\delta \in \mathbb{R}$) breaking reflection symmetry induces a point gap.

{\it Pseudo-reflection symmetry} ${\tilde{\mathcal{\M}}}_{++}$.---Around a reflection-invariant corner of a 2D system, a second-order boundary state is given by
\begin{align}
    H(k_j) = k_j \sigma_x,
\end{align}
where the symmetry operators are chosen as $\mathcal{T}=\sigma_y$, $\mathcal{C}=\tau_y$, and $\tilde{\mathcal{\M}}=\tau_y \sigma_y$. 
A mass term $\delta \tau_y \sigma_z$ ($\delta \in \mathbb{R}$) breaking pseudo-reflection symmetry induces a point gap.

{\it Pseudo-reflection symmetry} ${\tilde{\mathcal{\M}}}_{+-}$.---Around a reflection-invariant hinge of a 3D system, a second-order boundary state is given by
\begin{align}
    H(k_j,k_\parallel) = ik_j +k_{\parallel}\tau_x\sigma_y,
\end{align}
where the symmetry operators are chosen as $\mathcal{T}=\sigma_y$, $\mathcal{C}=\tau_y$, and $\tilde{\mathcal{\M}}=\tau_x$. 
A mass term $\delta \tau_z$ ($\delta \in \mathbb{R}$) breaking pseudo-reflection symmetry induces a point gap.

{\it Pseudo-reflection symmetry} ${\tilde{\mathcal{\M}}}_{-+}$.---Around a reflection-invariant hinge of a 3D system, a second-order boundary state is given by
\begin{align}
    H(k_j,k_\parallel) = ik_j +k_{\parallel}\sigma_x,
\end{align}
where the symmetry operators are chosen as $\mathcal{T}=\sigma_y$, $\mathcal{C}=\tau_y$, and $\tilde{\mathcal{\M}}=\sigma_x$. 
A mass term $\delta \tau_y \sigma_z$ ($\delta \in \mathbb{R}$) breaking pseudo-reflection symmetry induces a point gap.

\subsubsection{Class C + $\mathcal{{\M}}_{\pm }/\mathcal{\tilde{\M}}_{\pm}$}\label{classC+-M/pseudoM}

{\it Pseudo-reflection symmetry} ${\tilde{\mathcal{\M}}}_{+}$.---Around a reflection-invariant hinge of a 3D system, a second-order boundary state is given by
\begin{align}
    H(k_j,k_\parallel) = i k_j \tau_x+k_\parallel,
\end{align}
where the symmetry operators are chosen as $\mathcal{C}=\sigma_y$ and $\tilde{\mathcal{\M}}= \sigma_0$. 
A mass term $i\delta \tau_y$ ($\delta \in \mathbb{R}$) breaking pseudo-reflection symmetry induces a point gap.

\subsubsection{Class CI + $\mathcal{{\M}}_{\pm \pm }/{\mathcal{\M}}_{\pm \mp}/\mathcal{\tilde{\M}}_{\pm \pm }/\tilde{\mathcal{\M}}_{\pm \mp}$}\label{classCI+-M/pseudoM}

{\it Reflection symmetry} ${{\mathcal{\M}}}_{--}$.---Around a reflection-invariant hinge of a 3D system, a second-order boundary state is given by
\begin{align}
    H(k_j,k_\parallel) = ik_j \sigma_x+ik_{\parallel}\tau_x\sigma_y,
\end{align}
where the symmetry operators are chosen as $\mathcal{T}=\sigma_x$, $\mathcal{C}=\tau_y$, and ${\mathcal{\M}}=\tau_z \sigma_z$. 
A mass term $i\delta \tau_x \sigma_z$ ($\delta \in \mathbb{R}$) breaking reflection symmetry induces a point gap.

{\it Pseudo-reflection symmetry} ${\tilde{\mathcal{\M}}}_{-+}$.---Around a reflection-invariant hinge of a 3D system, a second-order boundary state is given by
\begin{align}
    H(k_j,k_\parallel) = ik_j +k_{\parallel}\sigma_z,
\end{align}
where the symmetry operators are chosen as $\mathcal{T}=\sigma_x$, $\mathcal{C}=\tau_y$, and $\tilde{\mathcal{\M}}=\sigma_z$. 
A mass term $\delta \sigma_y$ ($\delta \in \mathbb{R}$) breaking pseudo-reflection symmetry induces a point gap.

{\it Pseudo-reflection symmetry} ${\tilde{\mathcal{\M}}}_{--}$.---Around a reflection-invariant corner of a 2D system, a second-order boundary state is given by
\begin{align}
    H(k_j) = ik_j,
\end{align}
where the symmetry operators are chosen as $\mathcal{T}=\sigma_x$, $\mathcal{C}=\tau_y$, and $\tilde{\mathcal{\M}}=\tau_y \sigma_x$. 
A mass term $\delta \sigma_y$ ($\delta \in \mathbb{R}$) breaking pseudo-reflection symmetry induces a point gap.

\subsubsection{Class AI$^{\dagger}$ + $\mathcal{{\M}}_{\pm }/\mathcal{\tilde{\M}}_{\pm  }$}\label{classAIdag+-M/pseudoM}

{\it Pseudo-reflection symmetry} ${\tilde{\mathcal{\M}}}_{-}$.---Around a reflection-invariant hinge of a 3D system, a second-order boundary state is given by
\begin{align}
    H(k_j,k_\parallel) = ik_j\tau_z\sigma_z+k_{\parallel}\sigma_z,
\end{align}
where the symmetry operator are chosen as $\mathcal{T}=\sigma_x$ and $\tilde{\mathcal{\M}}= \sigma_z$. 
A mass term $\delta \tau_x \sigma_y$ ($\delta \in \mathbb{R}$) breaking pseudo-reflection symmetry induces a point gap.

\subsubsection{Class BDI$^\dagger$ + $\mathcal{{\M}}_{\pm \pm }/{\mathcal{\M}}_{\pm \mp}/\mathcal{\tilde{\M}}_{\pm \pm }/\tilde{\mathcal{\M}}_{\pm \mp}$}\label{classBDIdag+-M/pseudoM}

{\it Reflection symmetry} ${{\mathcal{\M}}}_{--}$.---Around a reflection-invariant hinge of a 3D system, a second-order boundary state is given by
\begin{align}
    H(k_j,k_\parallel) = k_j \tau_y +ik_{\parallel}\tau_y \sigma_y,
\end{align}
where the symmetry operators are chosen as $\mathcal{T}=\sigma_x$, $\mathcal{C}=\tau_x$, and ${\mathcal{\M}}=\tau_z \sigma_z$. 
A mass term $i \delta \sigma_x$ ($\delta \in \mathbb{R}$) breaking reflection symmetry induces a point gap.

{\it Pseudo-reflection symmetry} ${\tilde{\mathcal{\M}}}_{++}$.---Around a reflection-invariant corner of a 2D system, a second-order boundary state is given by
\begin{align}
    H(k_j) = k_j\sigma_z,
\end{align}
where the symmetry operators are chosen as $\mathcal{T}=\sigma_x$, $\mathcal{C}=\tau_x$, and $\tilde{\mathcal{\M}}=\tau_x \sigma_x$. 
A mass term $\delta \sigma_y$ ($\delta \in \mathbb{R}$) breaking pseudo-reflection symmetry induces a point gap.

{\it Pseudo-reflection symmetry} ${\tilde{\mathcal{\M}}}_{-+}$.---Around a reflection-invariant hinge of a 3D system, a second-order boundary state is given by
\begin{align}
    H(k_j,k_\parallel) = k_j\tau_y\sigma_x + k_\parallel \sigma_z,
\end{align}
where the symmetry operators are chosen as $\mathcal{T}=\sigma_x$, $\mathcal{C}=\tau_x$, and $\tilde{\mathcal{\M}}=\sigma_z$. 
A mass term $i\delta \sigma_0$ ($\delta \in \mathbb{R}$) breaking pseudo-reflection symmetry induces a point gap.

\subsubsection{Class D$^{\dagger}$ + $\mathcal{{\M}}_{\pm }/\mathcal{\tilde{\M}}_{\pm}$}\label{classDdag+-M/pseudoM}

{\it Pseudo-reflection symmetry} ${\tilde{\mathcal{\M}}}_{+}$.---Around a reflection-invariant hinge of a 3D system, a second-order boundary state is given by
\begin{align}
    H(k_j,k_\parallel) = k_j \sigma_y+k_\parallel \sigma_x,
\end{align}
where the symmetry operators are chosen as $\mathcal{C}=\sigma_x$ and $\tilde{\mathcal{\M}}= \sigma_x$. 
A mass term $i\delta \sigma_0$ ($\delta \in \mathbb{R}$) breaking pseudo-reflection symmetry induces a point gap.

\subsubsection{Class DIII$^\dagger$ + $\mathcal{{\M}}_{\pm \pm }/{\mathcal{\M}}_{\pm \mp}/\mathcal{\tilde{\M}}_{\pm \pm }/\tilde{\mathcal{\M}}_{\pm \mp}$}\label{classDIIIdag+-M/pseudoM}

{\it Reflection symmetry} ${{\mathcal{\M}}}_{++}$.---Around a reflection-invariant hinge of a 3D system, a second-order boundary state is given by
\begin{align}
    H(k_j,k_\parallel) = k_j \tau_z \sigma_y +ik_{\parallel}\sigma_y,
\end{align}
where the symmetry operators are chosen as $\mathcal{T}=\sigma_y$, $\mathcal{C}=\tau_x$, and ${\mathcal{\M}}=\tau_x$. 
A mass term $i \delta \tau_y \sigma_x$ ($\delta \in \mathbb{R}$) breaking reflection symmetry induces a point gap.

{\it Pseudo-reflection symmetry} ${\tilde{\mathcal{\M}}}_{-+}$.---Around a reflection-invariant corner of a 2D system, a second-order boundary state is given by
\begin{align}
    H(k_j) = k_j \sigma_x,
\end{align}
where the symmetry operators are chosen as $\mathcal{T}=\sigma_y$, $\mathcal{C}=\sigma_0 $, and $\tilde{\mathcal{\M}}=\sigma_z$. 
A mass term $i \delta \sigma_0$ ($\delta \in \mathbb{R}$) breaking pseudo-reflection symmetry induces a point gap.
Furthermore, around a reflection-invariant hinge of a 3D system, a second-order boundary state reads 
\begin{align}
    H(k_j, k_{\parallel}) = k_j \sigma_x + k_{\parallel}\sigma_z,
\end{align}
where the symmetry operators are chosen as $\mathcal{T}=\sigma_y$, $\mathcal{C}=\tau_x$, and $\tilde{\mathcal{\M}}=\sigma_z$.
The mass term $\delta \tau_y \sigma_y$  breaking pseudo-reflection symmetry induces a point gap.

{\it Pseudo-reflection symmetry} ${\tilde{\mathcal{\M}}}_{--}$.---Around a reflection-invariant corner of a 2D system, a second-order boundary state is given by
\begin{align}
    H(k_j) = k_j \sigma_x,
\end{align}
where the symmetry operators are chosen as $\mathcal{T}=\sigma_y$, $\mathcal{C}=\tau_x$, and $\tilde{\mathcal{\M}}=\tau_x\sigma_y$. 
A mass term $\delta \tau_y \sigma_y$ ($\delta \in \mathbb{R}$) breaking pseudo-reflection symmetry induces a point gap.

\subsubsection{Class CII$^\dagger$ + $\mathcal{{\M}}_{\pm \pm }/{\mathcal{\M}}_{\pm \mp}/\mathcal{\tilde{\M}}_{\pm \pm }/\tilde{\mathcal{\M}}_{\pm \mp}$}\label{classCIIdag+-M/pseudoM}

{\it Reflection symmetry} ${{\mathcal{\M}}}_{++}$.---Around a reflection-invariant hinge of a 3D system, a second-order boundary state is given by
\begin{align}
    H(k_j,k_\parallel) = k_j \sigma_x +ik_{\parallel}\tau_y,
\end{align}
where the symmetry operators are chosen as $\mathcal{T}=\sigma_y$, $\mathcal{C}=\tau_y$, and ${\mathcal{\M}}=\tau_y \sigma_y$. 
A mass term $\delta \tau_y \sigma_z$ ($\delta \in \mathbb{R}$) breaking reflection symmetry induces a point gap.

{\it Reflection symmetry} ${{\mathcal{\M}}}_{+-}$.---Around a reflection-invariant corner of a 2D system, a second-order boundary state is given by
\begin{align}
    H(k_j) = ik_j \tau_y,
\end{align}
where the symmetry operators are chosen as $\mathcal{T}=\sigma_y$, $\mathcal{C}=\tau_y$, and ${\mathcal{\M}}=\tau_z$. 
%A mass term $\delta \tau_y \sigma_x$ ($\delta \in \mathbb{R}$) breaking reflection symmetry induces a point gap.
Because a mass term $i\delta \sigma_0$ ($\delta \in \mathbb{R}$) shifts the point-gap-closing point, this boundary state is unstable.
Around a reflection-invariant hinge of a 3D system, a second-order boundary state is given by
\begin{align}
    H(k_j,k_\parallel) = ik_j \tau_y +k_{\parallel}\sigma_z,
\end{align}
where the symmetry operators are chosen as $\mathcal{T}=\sigma_y$, $\mathcal{C}=\tau_y$, and ${\mathcal{\M}}=\tau_z$. 
The mass term $\delta \tau_y \sigma_x$ breaking reflection symmetry induces a point gap.

{\it Reflection symmetry} ${{\mathcal{\M}}}_{--}$.---Around a reflection-invariant hinge of a 3D system, a second-order boundary state is given by
\begin{align}
    H(k_j,k_\parallel) = ik_j \sigma_y +ik_{\parallel}\tau_z \sigma_z,
\end{align}
where the symmetry operators are chosen as $\mathcal{T}=\sigma_y$, $\mathcal{C}=\tau_y$, and ${\mathcal{\M}}=\tau_x \sigma_x$. 
A mass term $\delta \tau_z$ ($\delta \in \mathbb{R}$) breaking reflection symmetry induces a point gap.

{\it Pseudo-reflection symmetry} ${\tilde{\mathcal{\M}}}_{++}$.---Around a reflection-invariant corner of a 2D system, a second-order boundary state is given by
\begin{align}
    H(k_j) = k_j \sigma_x,
\end{align}
where the symmetry operators are chosen as $\mathcal{T}=\sigma_y$, $\mathcal{C}=\tau_y$, and $\tilde{\mathcal{\M}}=\tau_y \sigma_y$. 
A mass term $i\delta \sigma_0$ ($\delta \in \mathbb{R}$) breaking pseudo-reflection symmetry induces a point gap.

{\it Pseudo-reflection symmetry} ${\tilde{\mathcal{\M}}}_{+-}$.---Around a reflection-invariant hinge of a 3D system, a second-order boundary state is given by
\begin{align}
    H(k_j,k_\parallel) = k_j \tau_x\sigma_y + k_\parallel \sigma_x,
\end{align}
where the symmetry operators are chosen as $\mathcal{T}=\sigma_y$, $\mathcal{C}=\tau_y$, and $\tilde{\mathcal{\M}}=\tau_z$. 
A mass term $i\delta \sigma_0$ ($\delta \in \mathbb{R}$) breaking pseudo-reflection symmetry induces a point gap.

{\it Pseudo-reflection symmetry} ${\tilde{\mathcal{\M}}}_{-+}$.---Around a reflection-invariant hinge of a 3D system, a second-order boundary state is given by
\begin{align}
    H(k_j,k_\parallel) = k_j \sigma_x + k_\parallel \sigma_z,
\end{align}
where the symmetry operators are chosen as $\mathcal{T}=\sigma_y$, $\mathcal{C}=\tau_y$, and $\tilde{\mathcal{\M}}=\sigma_z$. 
A mass term $i\delta \sigma_0$ ($\delta \in \mathbb{R}$) breaking pseudo-reflection symmetry induces a point gap.

{\it Pseudo-reflection symmetry} ${\tilde{\mathcal{\M}}}_{--}$.---Around a reflection-invariant corner of a 2D system, a second-order boundary state is given by
\begin{align}
    H(k_j) = k_j \sigma_z,
\end{align}
where the symmetry operators are chosen as $\mathcal{T}=\sigma_y$, $\mathcal{C}=\tau_y$, and $\tilde{\mathcal{\M}}=\tau_x \sigma_x$. 
A mass term $i\delta \sigma_0$ ($\delta \in \mathbb{R}$) breaking pseudo-reflection symmetry induces a point gap.
Around a reflection-invariant hinge of a 3D system, a second-order boundary state reads
\begin{align}
    H(k_j,k_\parallel) = k_j \sigma_z + k_\parallel \sigma_x,
\end{align}
where the symmetry operators are chosen as $\mathcal{T}=\sigma_y$, $\mathcal{C}=\tau_y$, and $\tilde{\mathcal{\M}}=\tau_x\sigma_x$. 
The mass term $i\delta \sigma_0$ breaking pseudo-reflection symmetry induces a point gap.

\subsubsection{Class C$^\dagger$ + $\mathcal{{\M}}_{\pm }/\mathcal{\tilde{\M}}_{\pm  }$}\label{classCdag+-M/pseudoM}

{\it Reflection symmetry} ${{\mathcal{\M}}}_{-}$.---Around a reflection-invariant corner of a 2D system, a second-order boundary state is given by
\begin{align}
    H(k_j) = ik_j\sigma_x,
\end{align}
where the symmetry operators are chosen as $\mathcal{C}=\sigma_y$ and ${\mathcal{\M}}= \sigma_y$. 
Because a mass term $i\delta \sigma_0$ ($\delta \in \mathbb{R}$) shifts the point-gap-closing point, this boundary state is unstable.
%A mass term $\delta \sigma_x$ ($\delta \in \mathbb{R}$) breaking reflection symmetry induces a point gap.

{\it Pseudo-reflection symmetry} ${\tilde{\mathcal{\M}}}_{+}$.---Around a reflection-invariant hinge of a 3D system, a second-order boundary state is given by
\begin{align}
    H(k_j,k_\parallel) = k_j \tau_x+k_\parallel \tau_z,
\end{align}
where the symmetry operators are chosen as $\mathcal{C}=\sigma_y$ and $\tilde{\mathcal{\M}}= \tau_z$. 
A mass term $i\delta \sigma_0$ ($\delta \in \mathbb{R}$) breaking pseudo-reflection symmetry induces a point gap.

{\it Pseudo-reflection symmetry} ${\tilde{\mathcal{\M}}}_{-}$.---Around a reflection-invariant hinge of a 3D system, a second-order boundary state is given by
\begin{align}
    H(k_j,k_\parallel) = k_j \tau_y\sigma_x+ik_\parallel \sigma_x,
\end{align}
where the symmetry operators are chosen as $\mathcal{C}=\sigma_y$ and $\tilde{\mathcal{\M}}= \sigma_y$. 
A mass term $\delta \tau_x \sigma_x$ ($\delta \in \mathbb{R}$) breaking pseudo-reflection symmetry induces a point gap.

\subsubsection{Class CI$^\dagger$ + $\mathcal{{\M}}_{\pm \pm }/{\mathcal{\M}}_{\pm \mp}/\mathcal{\tilde{\M}}_{\pm \pm }/\tilde{\mathcal{\M}}_{\pm \mp}$}\label{classCIdag+-M/pseudoM}

{\it Reflection symmetry} ${{\mathcal{\M}}}_{++}$.---Around a reflection-invariant hinge of a 3D system, a second-order boundary state is given by
\begin{align}
    H(k_j,k_\parallel) = k_j \sigma_z +ik_{\parallel}\tau_y,
\end{align}
where the symmetry operators are chosen as $\mathcal{T}=\sigma_x$, $\mathcal{C}=\tau_y$, and ${\mathcal{\M}}=\sigma_x$. 
A mass term $\delta \sigma_y$ ($\delta \in \mathbb{R}$) breaking reflection symmetry induces a point gap.

{\it Reflection symmetry} ${{\mathcal{\M}}}_{--}$.---Around a reflection-invariant hinge of a 3D system, a second-order boundary state is given by
\begin{align}
    H(k_j,k_\parallel) = ik_j \tau_y +ik_{\parallel}\tau_z \sigma_z,
\end{align}
where the symmetry operators are chosen as $\mathcal{T}=\sigma_x$, $\mathcal{C}=\tau_y$, and ${\mathcal{\M}}=\tau_z\sigma_z$. 
A mass term $i\delta \tau_z \sigma_y$ ($\delta \in \mathbb{R}$) breaking reflection symmetry induces a point gap.

{\it Pseudo-reflection symmetry} ${\tilde{\mathcal{\M}}}_{+-}$.---Around a reflection-invariant hinge of a 3D system, a second-order boundary state is given by
\begin{align}
    H(k_j,k_\parallel) = ik_j \tau_z\sigma_z + k_\parallel \sigma_z,
\end{align}
where the symmetry operators are chosen as $\mathcal{T}=\sigma_x$, $\mathcal{C}=\tau_y$, and $\tilde{\mathcal{\M}}=\tau_z$. 
A mass term $\delta \tau_x \sigma_x$ ($\delta \in \mathbb{R}$) breaking pseudo-reflection symmetry induces a point gap.

{\it Pseudo-reflection symmetry} ${\tilde{\mathcal{\M}}}_{-+}$.---Around a reflection-invariant hinge of a 3D system, a second-order boundary state is given by
\begin{align}
    H(k_j,k_\parallel) = ik_j \tau_z\sigma_z + k_\parallel \sigma_z,
\end{align}
where the symmetry operators are chosen as $\mathcal{T}=\sigma_x$, $\mathcal{C}=\tau_y$, and $\tilde{\mathcal{\M}}=\sigma_z$. 
A mass term $\delta \tau_x \sigma_x$ ($\delta \in \mathbb{R}$) breaking pseudo-reflection symmetry induces a point gap.

{\it Pseudo-reflection symmetry} ${\tilde{\mathcal{\M}}}_{--}$.---Around a reflection-invariant corner of a 2D system, a second-order boundary state is given by
\begin{align}
    H(k_j) = k_j \sigma_z,
\end{align}
where the symmetry operators are chosen as $\mathcal{T}=\sigma_x$, $\mathcal{C}=\tau_y$, and $\tilde{\mathcal{\M}}=\tau_y\sigma_x$. 
A mass term $i\delta \sigma_0$ ($\delta \in \mathbb{R}$) breaking pseudo-reflection symmetry induces a point gap.

%\clearpage

\begin{table*}
	\centering
	\caption{Classification of second-order point-gap topology in the complex Altland-Zirnbauer symmetry classes with reflection symmetry.
    The subscript of $\mathcal{S}$ specifies the commutation ($+$) or anticommutation ($-$) relation with chiral symmetry.
    ${\cal \M}$ and $\tilde{\cal \M}$ denote reflection symmetry and pseudo-reflection symmetry (or equivalently, reflection symmetry$^{\dag}$), respectively.
    In classes AIII and A + ${\cal S}$, the subscript of ${\cal \M}_\pm$ or $\tilde{\cal \M}_\pm$ specifies the commutation ($+$) or anticommutation ($-$) relation with chiral and sublattice symmetries, respectively.
    In classes AIII + ${\cal S}_{\pm}$, the first subscript of ${\cal \M}_{\pm\pm}$ or $\tilde{\cal \M}_{\pm\pm}$ specifies the relation to chiral symmetry and the second one to sublattice symmetry. For the classifications specified by ``$*$", second-order boundary states appear, as determined by the non-Hermitian boundary classification in Sec.~\ref{sec:nh_boundary}.
    }\label{tab: complex AZ}
     \begin{tabular}{cccc} \hline \hline
    Class & Hermitization & $d=2$ & $d=3$ \\ \hline    
    A + ${\cal \M}$ & AIII + ${\sf \M}_+$ & $\mathbb{Z}$  & $0$ \\
    A + $\tilde{\cal \M}$ & AIII + ${\sf \M}_-$ & $0$ &  $\mathbb{Z}^{\textcolor{purple}{*}}$ [Sec.~\ref{classA+M/tildeM}]  \\ \hline
    AIII + ${\cal \M}_+$ & A + ${\mathsf \M}$ & $0$ & $\mathbb{Z}^{\textcolor{purple}{*}}$ [Sec.~\ref{classAIII+M/tildeM}]  \\
    AIII + ${\cal \M}_-$ & A + $\bar{\mathsf \M}$ &  $\mathbb{Z}$ & $0$ \\
    AIII + $\tilde{\cal \M}_+$ & A + $\bar{\mathsf \M}$ &  $\mathbb{Z}^{\textcolor{purple}{*}}$ [Sec.~\ref{classAIII+M/tildeM}] & $0$ \\
    AIII + $\tilde{\cal \M}_-$ & A + ${\mathsf \M}$ & $0$ & $\mathbb{Z}^{\textcolor{purple}{*}}$ [Sec.~\ref{classAIII+M/tildeM}] \\ \hline
    AIII + ${\cal S}_+$ + ${\cal \M}_{++}$ & AIII + ${\mathsf \M}_+$ & $\mathbb{Z}^{\textcolor{purple}{*}}$ [Sec.~\ref{classAIII+S+M/tildeM}] & $0$ \\
    AIII + ${\cal S}_+$ + ${\cal \M}_{+-}$ & %AIII + $\bar{\sf M}_-$ = 
    AIII + ${\sf M}_-$ & $0$ &  $\mathbb{Z}^{\textcolor{purple}{*}}$ [Sec.~\ref{classAIII+S+M/tildeM}]  \\
    AIII + ${\cal S}_+$ + ${\cal \M}_{-+}$ & AIII + ${\mathsf \M}_+$ & $\mathbb{Z}$ & $0$ \\
    AIII + ${\cal S}_+$ + ${\cal \M}_{--}$ & %AIII + $\bar{\sf M}_-$ = 
    AIII + ${\sf M}_-$ & $0$ &  $\mathbb{Z}^{\textcolor{purple}{*}}$ [Sec.~\ref{classAIII+S+M/tildeM}] \\
    AIII + ${\cal S}_+$ + $\tilde{\cal \M}_{++}$ & %AIII + $\bar{\sf M}_+$ = 
    AIII + ${\sf M}_+$ & $\mathbb{Z}^{\textcolor{purple}{*}}$ [Sec.~\ref{classAIII+S+tildeM}]  & $0$ \\
    AIII + ${\cal S}_+$ + $\tilde{\cal \M}_{+-}$ & AIII + ${\mathsf \M}_-$ & $0$ &  $\mathbb{Z}^{\textcolor{purple}{*}}$ [Sec.~\ref{classAIII+S+tildeM}] \\
    AIII + ${\cal S}_+$ + $\tilde{\cal \M}_{-+}$ & %AIII + $\bar{\sf M}_+$ = 
    AIII + ${\sf M}_+$ & $\mathbb{Z}$  & $0$ \\
    AIII + ${\cal S}_+$ + $\tilde{\cal \M}_{--}$ & AIII + ${\mathsf \M}_-$ & $0$ &  $\mathbb{Z}^{\textcolor{purple}{*}}$ [Sec.~\ref{classAIII+S+tildeM}] \\ \hline
    A + ${\cal S}$ + ${\cal \M}_+$ & ~~$\left( \text{AIII~+~}{\sf \M}_+ \right)$ $\times$ $\left( \text{AIII~+~}{\sf \M}_+ \right)$~~ & ~~$\mathbb{Z} \oplus \mathbb{Z}^{\textcolor{purple}{*}}$ [Sec.~\ref{classA+S+M/tildeM}]~~ & $0$ \\
    A + ${\cal S}$ + ${\cal \M}_-$ & %AIII + $\mathsf{S}_{+}$ + $\mathsf{M}_{+-}$ = 
    AIII + $\mathsf{U}_{+}$ + $\mathsf{M}_{+-}$ & $0$ & $0$ \\
    A + ${\cal S}$ + $\tilde{\cal \M}_+$ & %AIII + $\mathsf{S}_{+}$ + $\mathsf{M}_{+-}$ = 
    AIII + $\mathsf{U}_{+}$ + $\mathsf{M}_{+-}$ & $0$ & $0$ \\
    A + ${\cal S}$ + $\tilde{\cal \M}_-$ & $\left( \text{AIII~+~}{\sf \M}_- \right)$ $\times$ $\left( \text{AIII~+~}{\sf \M}_- \right)$ & $0$ &  $\mathbb{Z}\oplus\mathbb{Z}^{\textcolor{purple}{*}}$ [Sec.~\ref{classA+S+M/tildeM}] \\ \hline
    AIII + ${\cal S}_-$ + ${\cal \M}_{++}$ & $\left( \text{A~+~}{\sf \M} \right)$ $\times$ $\left( \text{A~+~}{\sf \M} \right)$ & $0$ & $\mathbb{Z} \oplus \mathbb{Z}^{\textcolor{purple}{*}}$ [Sec.~\ref{classAIII+S-M/pseudoM}] \\
    AIII + ${\cal S}_-$ + ${\cal \M}_{+-}$ & A + $\mathsf{U}$ + $\mathsf{\M}_-$ & $0$ & $0$ \\
    AIII + ${\cal S}_-$ + ${\cal \M}_{-+}$ & $\left( \text{A~+~}\bar{{\sf \M}} \right)$ $\times$ $\left( \text{A~+~}\bar{{\sf \M}} \right)$ & $\mathbb{Z}\oplus\mathbb{Z}^{\textcolor{purple}{*}}$ [Sec.~\ref{classAIII+S-M/pseudoM}] & $0$ \\
    AIII + ${\cal S}_-$ + ${\cal \M}_{--}$ & A + $\mathsf{U}$ + $\mathsf{\M}_-$ & $0$ & $0$ \\ 
    AIII + ${\cal S}_-$ + $\tilde{\cal \M}_{++}$ & A + $\mathsf{U}$ + $\mathsf{\M}_-$ & $0$ & $0$ \\
    AIII + ${\cal S}_-$ + $\tilde{\cal \M}_{+-}$ & $\left( \text{A~+~}\bar{{\sf \M}} \right)$ $\times$ $\left( \text{A~+~}\bar{{\sf \M}} \right)$ & $\mathbb{Z}\oplus\mathbb{Z}^{\textcolor{purple}{*}}$ [Sec.~\ref{classAIII+S-M/pseudoM}] & $0$ \\
    AIII + ${\cal S}_-$ + $\tilde{\cal \M}_{-+}$ & A + $\mathsf{U}$ + $\mathsf{\M}_-$ & $0$ & $0$ \\
    ~~AIII + ${\cal S}_-$ + $\tilde{\cal \M}_{--}$~~ & $\left( \text{A~+~}{\sf \M} \right)$ $\times$ $\left( \text{A~+~}{\sf \M} \right)$ & $0$ & ~~$\mathbb{Z} \oplus \mathbb{Z}^{\textcolor{purple}{*}}$ [Sec.~\ref{classAIII+S-M/pseudoM}]~~ \\ \hline \hline
  \end{tabular}
\end{table*}

\begin{table*}
	\centering
	\caption{Classification of second-order point-gap topology in the real Altland-Zirnbauer symmetry classes with reflection symmetry.
    ${\cal \M}$ and $\tilde{\cal \M}$ denote reflection symmetry and pseudo-reflection symmetry (or equivalently, reflection symmetry$^{\dag}$), respectively.
    The subscript of ${\cal \M}_\pm$ or $\tilde{\cal \M}_\pm$ specifies the commutation ($+$) or anticommutation ($-$) relation to time-reversal symmetry or particle-hole symmetry.
    For the symmetry classes involving both time-reversal symmetry and particle-hole symmetry (i.e., classes BDI, DIII, CII, and CI), the first subscript of ${\cal \M}_{\pm\pm}$ or $\tilde{\cal \M}_{\pm\pm}$ specifies the relation to time-reversal symmetry and the second one to particle-hole symmetry.
    For the classifications specified by ``$*$", second-order boundary states appear, as %. The presence and absence of second-order boundary states are 
    determined by the non-Hermitian boundary classification in Sec.~\ref{sec:nh_boundary}.
    }
	\label{tab: real AZ}
     \begin{tabular}{cccc} \hline \hline
    Class & Hermitization & $d=2$ & $d=3$ \\ \hline    
    AI + ${\cal \M}_+$ & ~~BDI + ${\sf \M}_{++}$~~ & $\mathbb{Z}$ & $0$ \\
    AI + ${\cal \M}_-$ & BDI + ${\sf \M}_{--}$ & 0 & $0$ \\
    AI + $\tilde{\cal \M}_+$ & BDI + ${\sf \M}_{+-}$ & $0$ & $0$ \\
    AI + $\tilde{\cal \M}_-$ & BDI + ${\sf \M}_{-+}$ & $0$ &  $\mathbb{Z}^{\textcolor{purple}{*}}$ [Sec.~\ref{classAI+-M/pseudoM}] \\ \hline
    BDI + ${\cal \M}_{++}$ & D + ${\sf \M}_+$ & $\mathbb{Z}_2^{\textcolor{purple}{*}}$ [Sec.~\ref{classBDI+-M/pseudoM}]& $\mathbb{Z}^{\textcolor{purple}{*}}$ [Sec.~\ref{classBDI+-M/pseudoM}] \\
    BDI + ${\cal \M}_{+-}$ & D + $\bar{\sf \M}_+$ &  $\mathbb{Z}$ & $0$ \\
    BDI + ${\cal \M}_{-+}$ & D + $\bar{\sf \M}_-$ & $0$ & $0$ \\
    BDI + ${\cal \M}_{--}$ & D + ${\sf \M}_-$ & $0$ &  $0$ \\ 
    BDI + $\tilde{\cal \M}_{++}$ & D + $\bar{\sf \M}_+$ & $\mathbb{Z}^{\textcolor{purple}{*}}$ [Sec.~\ref{classBDI+-M/pseudoM}] & $0$ \\
    BDI + $\tilde{\cal \M}_{+-}$ & D + ${\sf \M}_-$ & $0$ &  $0$ \\
    BDI + $\tilde{\cal \M}_{-+}$ & D + ${\sf \M}_+$ & $\mathbb{Z}_2^{\textcolor{purple}{*}}$ [Sec.~\ref{classBDI+-M/pseudoM}] & $\mathbb{Z}^{\textcolor{purple}{*}}$ [Sec.~\ref{classBDI+-M/pseudoM}] \\
    BDI + $\tilde{\cal \M}_{--}$ & D + $\bar{\sf \M}_-$ & $0$ & $0$ \\ \hline
    D + ${\cal \M}_+$ & ~~DIII + ${\mathsf \M}_{++}$~~ & $\mathbb{Z}_2^{\textcolor{purple}{*}}$ [Sec.~\ref{classD+-M/pseudoM}] & $\mathbb{Z}_2^{\textcolor{purple}{*}}$ [Sec.~\ref{classD+-M/pseudoM}] \\
    D + ${\cal \M}_-$ & DIII + ${\mathsf \M}_{--}$ & $2\mathbb{Z}$ & $0$ \\
    D + $\tilde{\cal \M}_+$ & DIII + ${\mathsf \M}_{-+}$ & $\mathbb{Z}_2^{\textcolor{purple}{*}}$ [Sec.~\ref{classD+-M/pseudoM}] & $\mathbb{Z}^{\textcolor{purple}{*}}$ [Sec.~\ref{classD+-M/pseudoM}] \\
    D + $\tilde{\cal \M}_-$ & DIII + ${\mathsf \M}_{+-}$ & $0$ & $0$ \\ \hline
    ~~DIII + ${\cal \M}_{++}$~~ & AII + ${\sf \M}_{+}$ & $0$ & $\mathbb{Z}_2^{\textcolor{purple}{*}}$ [Sec.~\ref{classDIII+-M/pseudoM}] \\
    DIII + ${\cal \M}_{+-}$ & AII + $\bar{\sf M}_+$ & $2\mathbb{Z}$ & $0$ \\
    DIII + ${\cal \M}_{-+}$ & AII + $\bar{\sf M}_-$ &  $\mathbb{Z}_2^{\textcolor{purple}{*}}$ [Sec.~\ref{classDIII+-M/pseudoM}] & $\mathbb{Z}_2^{\textcolor{purple}{*}}$ [Sec.~\ref{classDIII+-M/pseudoM}] \\
    DIII + ${\cal \M}_{--}$ & AII + ${\sf \M}_{-}$ & $0$ & $2\mathbb{Z}^{\textcolor{purple}{*}}$ [Sec.~\ref{classDIII+-M/pseudoM}] \\ 
    DIII + $\tilde{\cal \M}_{++}$ & AII + $\bar{\sf M}_-$ &  $\mathbb{Z}_2^{\textcolor{purple}{*}}$ [Sec.~\ref{classDIII+-M/pseudoM}] & $\mathbb{Z}_2^{\textcolor{purple}{*}}$ [Sec.~\ref{classDIII+-M/pseudoM}] \\
    DIII + $\tilde{\cal \M}_{+-}$ & AII + ${\sf \M}_{+}$ & $0$ & $\mathbb{Z}_2^{\textcolor{purple}{*}}$ [Sec.~\ref{classDIII+-M/pseudoM}] \\
    DIII + $\tilde{\cal \M}_{-+}$ & AII + ${\sf \M}_{-}$ & $0$ & $2\mathbb{Z}^{\textcolor{purple}{*}}$ [Sec.~\ref{classDIII+-M/pseudoM}] \\
    DIII + $\tilde{\cal \M}_{--}$ & AII + $\bar{\sf M}_+$ & $2\mathbb{Z}^{\textcolor{purple}{*}}$ [Sec.~\ref{classDIII+-M/pseudoM}] & $0$ \\ \hline
    AII + ${\cal \M}_+$ & CII + ${\sf \M}_{++}$ & ~~$2\mathbb{Z}$~~ & $0$ \\
    AII + ${\cal \M}_-$ & CII + ${\sf \M}_{--}$ &  $\mathbb{Z}_2$ & $0$ \\
    AII + $\tilde{\cal \M}_+$ & CII + ${\sf \M}_{+-}$ & $0$ &  $\mathbb{Z}_2^{\textcolor{purple}{*}}$ [Sec.~\ref{classAII+-M/pseudoM}] \\
    AII + $\tilde{\cal \M}_-$ & CII + ${\sf \M}_{-+}$ & $0$ &$2\mathbb{Z}^{\textcolor{purple}{*}}$ [Sec.~\ref{classAII+-M/pseudoM}] \\ \hline
    CII + ${\cal \M}_{++}$ & C + ${\sf \M}_+$ & $0$ & ~~$2\mathbb{Z}^{\textcolor{purple}{*}}$ [Sec.~\ref{classCII+-M/pseudoM}]~~ \\
    CII + ${\cal \M}_{+-}$ & C + $\bar{\sf M}_+$ &  $2\mathbb{Z}$ & $0$ \\
    CII + ${\cal \M}_{-+}$ & C + $\bar{\sf M}_-$ & $0$ & $0$ \\
    CII + ${\cal \M}_{--}$ & C + ${\sf \M}_-$ & $0$ &  $\mathbb{Z}_2^{\textcolor{purple}{*}}$ [Sec.~\ref{classCII+-M/pseudoM}] \\ 
    CII + $\tilde{\cal \M}_{++}$ & C + $\bar{\sf M}_+$ &  $2\mathbb{Z}^{\textcolor{purple}{*}}$ [Sec.~\ref{classCII+-M/pseudoM}] & $0$\\
    CII + $\tilde{\cal \M}_{+-}$ & C + ${\sf \M}_-$ & $0$ &  {$\mathbb{Z}_2^{\textcolor{purple}{*}}$} [Sec.~\ref{classCII+-M/pseudoM}] \\
    CII + $\tilde{\cal \M}_{-+}$ & C + ${\sf \M}_+$ & $0$ & $2\mathbb{Z}^{\textcolor{purple}{*}}$ [Sec.~\ref{classCII+-M/pseudoM}] \\
    CII + $\tilde{\cal \M}_{--}$ & C + $\bar{\sf M}_-$ & $0$ & $0$ \\ \hline
    C + ${\cal \M}_+$ & CI + ${\sf \M}_{++}$ & $0$ & $0$ \\
    C + ${\cal \M}_-$ & CI + ${\sf \M}_{--}$ &  $2\mathbb{Z}$ & $0$ \\
    C + $\tilde{\cal \M}_+$ & CI + ${\sf \M}_{-+}$ & $0$ &  $2\mathbb{Z}^{\textcolor{purple}{*}}$ [Sec.~\ref{classC+-M/pseudoM}] \\
    C + $\tilde{\cal \M}_-$ & CI + ${\sf \M}_{+-}$ & $0$ & $0$ \\ \hline
    CI + ${\cal \M}_{++}$ & AI + ${\sf \M}_+$ & $0$ & $0$ \\
    CI + ${\cal \M}_{+-}$ & AI + $\bar{\sf M}_+$ & $\mathbb{Z}$ & $0$ \\
    CI + ${\cal \M}_{-+}$ & AI + $\bar{\sf M}_-$ & $0$ & $0$ \\
    CI + ${\cal \M}_{--}$ & AI + ${\sf \M}_-$ & $0$ & $2\mathbb{Z}^{\textcolor{purple}{*}}$ [Sec.~\ref{classCI+-M/pseudoM}]\\ 
    CI + $\tilde{\cal \M}_{++}$ & AI + $\bar{\sf M}_-$ & $0$ & $0$ \\
    CI + $\tilde{\cal \M}_{+-}$ & AI + ${\sf \M}_+$ & $0$ & $0$ \\
    CI + $\tilde{\cal \M}_{-+}$ & AI + ${\sf \M}_-$ & $0$ & {$2\mathbb{Z}^{\textcolor{purple}{*}}$} [Sec.~\ref{classCI+-M/pseudoM}] \\
    CI + $\tilde{\cal \M}_{--}$ & AI + $\bar{\sf M}_+$ &  $\mathbb{Z}^{\textcolor{purple}{*}}$ [Sec.~\ref{classCI+-M/pseudoM}] & $0$ \\ \hline \hline
  \end{tabular}
\end{table*}

%\clearpage
\begin{table*}
	\centering
	\caption{Classification of second-order point-gap topology in the real Altland-Zirnbauer$^{\dag}$ symmetry classes with reflection symmetry.
    ${\cal \M}$ and $\tilde{\cal \M}$ denote reflection symmetry and pseudo-reflection symmetry (or equivalently, reflection symmetry$^{\dag}$), respectively.
    The subscript of ${\cal \M}_\pm$ or $\tilde{\cal \M}_\pm$ specifies the commutation ($+$) or anticommutation ($-$) relation to time-reversal symmetry$^{\dag}$ or particle-hole symmetry$^{\dag}$.
    For the symmetry classes involving both time-reversal symmetry$^{\dag}$ and particle-hole symmetry$^{\dag}$ (i.e., classes BDI$^{\dag}$, DIII$^{\dag}$, CII$^{\dag}$, and CI$^{\dag}$), the first subscript of ${\cal \M}_{\pm\pm}$ or $\tilde{\cal \M}_{\pm\pm}$ specifies the relation to time-reversal symmetry$^{\dag}$ and the second one to particle-hole symmetry$^{\dag}$.
    For the classifications specified by ``$*$", second-order boundary states appear, as determined by the non-Hermitian boundary classification in Sec.~\ref{sec:nh_boundary}.
    }
	\label{tab: real AZ dag}
     \begin{tabular}{cccc} \hline \hline
    Class & Hermitization & $d=2$ & $d=3$ \\ \hline    
    AI$^{\dag}$ + ${\cal \M}_+$ & CI + ${\sf \M}_{++}$ & $0$ & $0$ \\
    AI$^{\dag}$ + ${\cal \M}_-$ & CI + ${\sf \M}_{--}$ & $2\mathbb{Z}$ & $0$ \\
    AI$^{\dag}$ + $\tilde{\cal \M}_+$ & CI + ${\sf \M}_{+-}$ & $0$ & $0$ \\
    AI$^{\dag}$ + $\tilde{\cal \M}_-$ & CI + ${\sf \M}_{-+}$ & $0$ &  $2\mathbb{Z}^{\textcolor{purple}{*}}$ [Sec.~\ref{classAIdag+-M/pseudoM}] \\ \hline
    BDI$^{\dag}$ + ${\cal \M}_{++}$ & AI + ${\sf \M}_+$ & $0$ & $0$ \\
    BDI$^{\dag}$ + ${\cal \M}_{+-}$ & AI + $\bar{\sf M}_-$ & $0$ & $0$ \\
    BDI$^{\dag}$ + ${\cal \M}_{-+}$ & AI + $\bar{\sf M}_+$ & $\mathbb{Z}$ & $0$ \\
    BDI$^{\dag}$ + ${\cal \M}_{--}$ & AI + ${\sf \M}_-$ & $0$ &  $2\mathbb{Z}^{\textcolor{purple}{*}}$ [Sec.~\ref{classBDIdag+-M/pseudoM}] \\ 
    BDI$^{\dag}$ + $\tilde{\cal \M}_{++}$ & AI + $\bar{\sf M}_+$ & $\mathbb{Z}^{\textcolor{purple}{*}}$ [Sec.~\ref{classBDIdag+-M/pseudoM}] & $0$ \\
    BDI$^{\dag}$ + $\tilde{\cal \M}_{+-}$ & AI + ${\sf \M}_+$ & $0$ & $0$ \\
    BDI$^{\dag}$ + $\tilde{\cal \M}_{-+}$ & AI + ${\sf \M}_-$ & $0$ & $2\mathbb{Z}^{\textcolor{purple}{*}}$ [Sec.~\ref{classBDIdag+-M/pseudoM}] \\
    BDI$^{\dag}$ + $\tilde{\cal \M}_{--}$ & AI + $\bar{\sf M}_-$ & $0$ & $0$ \\ \hline
    D$^{\dag}$ + ${\cal \M}_+$ & BDI + ${\sf \M}_{++}$ & $\mathbb{Z}$ & $0$ \\
    D$^{\dag}$ + ${\cal \M}_-$ & BDI + ${\sf \M}_{--}$ &  $0$ & $0$ \\
    D$^{\dag}$ + $\tilde{\cal \M}_+$ & BDI + ${\sf \M}_{-+}$ & $0$ &  $\mathbb{Z}^{\textcolor{purple}{*}}$ [Sec.~\ref{classDdag+-M/pseudoM}] \\
    D$^{\dag}$ + $\tilde{\cal \M}_-$ & BDI + ${\sf \M}_{+-}$ & $0$ & $0$ \\ \hline
    ~~DIII$^{\dag}$ + ${\cal \M}_{++}$~~ & D + ${\sf \M}_+$ & $\mathbb{Z}_2$ & $\mathbb{Z}^{\textcolor{purple}{*}}$ [Sec.~\ref{classDIIIdag+-M/pseudoM}] \\
    DIII$^{\dag}$ + ${\cal \M}_{+-}$ & D + $\bar{\sf M}_-$ & $0$ & $0$ \\
    DIII$^{\dag}$ + ${\cal \M}_{-+}$ & D + $\bar{\sf M}_+$ &  $\mathbb{Z}$ & $0$ \\
    DIII$^{\dag}$ + ${\cal \M}_{--}$ & D + ${\sf \M}_-$ & $0$ &  $0$ \\ 
    DIII$^{\dag}$ + $\tilde{\cal \M}_{++}$ & D + $\bar{\sf M}_-$ & $0$ & $0$ \\
    DIII$^{\dag}$ + $\tilde{\cal \M}_{+-}$ & D + ${\sf \M}_-$ & $0$ & $0$ \\
    DIII$^{\dag}$ + $\tilde{\cal \M}_{-+}$ & D + ${\sf \M}_+$ & $\mathbb{Z}_2^{\textcolor{purple}{*}}$ [Sec.~\ref{classDIIIdag+-M/pseudoM}] & $\mathbb{Z}^{\textcolor{purple}{*}}$ [Sec.~\ref{classDIIIdag+-M/pseudoM}] \\
    DIII$^{\dag}$ + $\tilde{\cal \M}_{--}$ & D + $\bar{\sf M}_+$ & $\mathbb{Z}^{\textcolor{purple}{*}}$ [Sec.~\ref{classDIIIdag+-M/pseudoM}] & $0$ \\ \hline
    AII$^{\dag}$ + ${\cal \M}_+$ & ~~DIII + ${\sf \M}_{++}$~~ & $\mathbb{Z}_2$ & $\mathbb{Z}_2$ \\
    AII$^{\dag}$ + ${\cal \M}_-$ & DIII + ${\sf \M}_{--}$ & $2\mathbb{Z}$ & $0$ \\
    AII$^{\dag}$ + $\tilde{\cal \M}_+$ & DIII + ${\sf \M}_{+-}$ & $0$ & $0$ \\
    AII$^{\dag}$ + $\tilde{\cal \M}_-$ & DIII + ${\sf \M}_{-+}$ & $\mathbb{Z}_2$ & $\mathbb{Z}^{\textcolor{purple}{*}}$ [Sec.~\ref{classAIIdag+M/tildeM}] \\ \hline
    CII$^{\dag}$ + ${\cal \M}_{++}$ & AII + ${\sf \M}_+$ & $0$ & $\mathbb{Z}_2^{\textcolor{purple}{*}}$ [Sec.~\ref{classCIIdag+-M/pseudoM}] \\
    CII$^{\dag}$ + ${\cal \M}_{+-}$ & AII + $\bar{\sf M}_-$ &  $\mathbb{Z}_2$ & $\mathbb{Z}_2^{\textcolor{purple}{*}}$ [Sec.~\ref{classCIIdag+-M/pseudoM}] \\
    CII$^{\dag}$ + ${\cal \M}_{-+}$ & AII + $\bar{\sf M}_+$ &  $2\mathbb{Z}$ & $0$ \\
    CII$^{\dag}$ + ${\cal \M}_{--}$ & AII + ${\sf \M}_-$ & $0$ &  $2\mathbb{Z}^{\textcolor{purple}{*}}$ [Sec.~\ref{classCIIdag+-M/pseudoM}] \\ 
    CII$^{\dag}$ + $\tilde{\cal \M}_{++}$ & AII + $\bar{\sf M}_+$ & $2\mathbb{Z}^{\textcolor{purple}{*}}$ [Sec.~\ref{classCIIdag+-M/pseudoM}] & $0$ \\
    CII$^{\dag}$ + $\tilde{\cal \M}_{+-}$ & AII + ${\sf \M}_+$ & $0$ & $\mathbb{Z}_2^{\textcolor{purple}{*}}$ [Sec.~\ref{classCIIdag+-M/pseudoM}] \\
    CII$^{\dag}$ + $\tilde{\cal \M}_{-+}$ & AII + ${\sf \M}_-$ & $0$ &  $2\mathbb{Z}^{\textcolor{purple}{*}}$ [Sec.~\ref{classCIIdag+-M/pseudoM}]\\
    CII$^{\dag}$ + $\tilde{\cal \M}_{--}$ & AII + $\bar{\sf M}_-$ &  $\mathbb{Z}_2^{\textcolor{purple}{*}}$ [Sec.~\ref{classCIIdag+-M/pseudoM}] & $\mathbb{Z}_2^{\textcolor{purple}{*}}$ [Sec.~\ref{classCIIdag+-M/pseudoM}] \\ \hline
    C$^{\dag}$ + ${\cal \M}_+$ & CII + ${\sf \M}_{++}$ & ~~$2\mathbb{Z}$~~ & $0$ \\
    C$^{\dag}$ + ${\cal \M}_-$ & CII + ${\sf \M}_{--}$ &  $\mathbb{Z}_2$ & $0$ \\
    C$^{\dag}$ + $\tilde{\cal \M}_+$ & CII + ${\sf \M}_{-+}$ & $0$ & $2\mathbb{Z}^{\textcolor{purple}{*}}$ [Sec.~\ref{classCdag+-M/pseudoM}] \\
    C$^{\dag}$ + $\tilde{\cal \M}_-$ & CII + ${\sf \M}_{+-}$ & $0$ & $\mathbb{Z}_2^{\textcolor{purple}{*}}$ [Sec.~\ref{classCdag+-M/pseudoM}] \\ \hline
    CI$^{\dag}$ + ${\cal \M}_{++}$ & C + ${\sf \M}_+$ & $0$ & ~~$2\mathbb{Z}^{\textcolor{purple}{*}}$ [Sec.~\ref{classCIdag+-M/pseudoM}]~~ \\
    CI$^{\dag}$ + ${\cal \M}_{+-}$ & C + $\bar{\sf M}_-$ & $0$ & $0$ \\
    CI$^{\dag}$ + ${\cal \M}_{-+}$ & C + $\bar{\sf M}_+$ &  $2\mathbb{Z}$ & $0$ \\
    CI$^{\dag}$ + ${\cal \M}_{--}$ & C + ${\sf \M}_-$ & $0$ &  $\mathbb{Z}_2^{\textcolor{purple}{*}}$ [Sec.~\ref{classCIdag+-M/pseudoM}]\\ 
    CI$^{\dag}$ + $\tilde{\cal \M}_{++}$ & C + $\bar{\sf M}_-$ & $0$ & $0$ \\
    CI$^{\dag}$ + $\tilde{\cal \M}_{+-}$ & C + ${\sf \M}_-$ & $0$ &  $\mathbb{Z}_2^{\textcolor{purple}{*}}$ [Sec.~\ref{classCIdag+-M/pseudoM}]\\
    CI$^{\dag}$ + $\tilde{\cal \M}_{-+}$ & C + ${\sf \M}_+$ & $0$ & $2\mathbb{Z}^{\textcolor{purple}{*}}$ [Sec.~\ref{classCIdag+-M/pseudoM}]\\
    CI$^{\dag}$ + $\tilde{\cal \M}_{--}$ & C + $\bar{\sf M}_+$ &  $2\mathbb{Z}^{\textcolor{purple}{*}}$ [Sec.~\ref{classCIdag+-M/pseudoM}] & $0$ \\ \hline \hline
  \end{tabular}
\end{table*}

%\clearpage

\section{Continuous deformation into line-gap topological phase}\label{subsec: cont_deform}

\begin{figure*}
\includegraphics[width=2.\columnwidth]{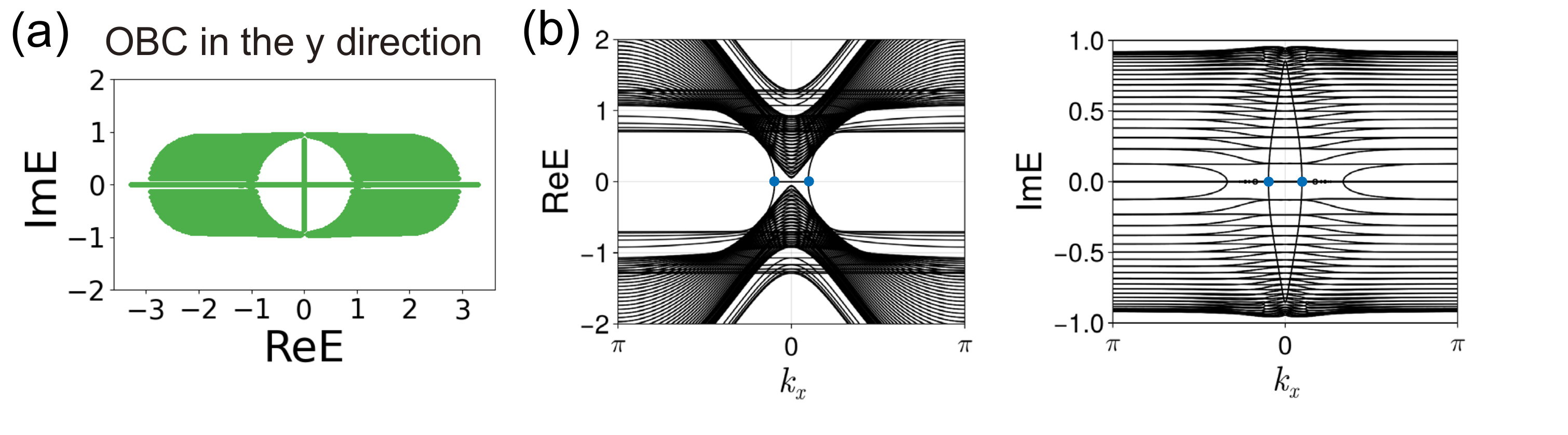}
	\caption{(a)~Complex energy spectrum of the Hamiltonian $H_{\tilde{\mathcal{M}}}(\boldsymbol{k})$ in Eq.~(\ref{eq:classD_2D_pseud}) of Sec.~\ref{sec:2DETI} with $m_1=0.1$ and $m_2=0.3$ under the open boundary conditions (OBC) in the $y$ direction. 
    (b)~Energy dispersion of $H_{\tilde{\mathcal{M}}}(\boldsymbol{k})$. The blue points indicate the exceptional points.
    The system size is 51 in the $y$ direction with the momentum resolution $\Delta k_x=2\pi/10000$. 
 }
	\label{fig:2DclassD_pseudo_yobc}
\end{figure*}

We show that the point-gap topological phase in $H_{\tilde{\mathcal{M}}}(\boldsymbol{k})$ introduced in Eq.~(\ref{eq:classD_2D_pseud}) of Sec.~\ref{sec:2DETI} can be continuously deformed into a line-gap topological phase with respect to ${\rm Re} E=0$. 
We calculate the energy spectrum of the Hamiltonian $H_{\tilde{\mathcal{M}}}(\boldsymbol{k})$
under the open boundary conditions (OBC) along the $y$ direction [Fig.~\ref{fig:2DclassD_pseudo_yobc}(a)]. 
Since the boundaries along the $x$ axis are invariant under the reflection $x\rightarrow -x$, the boundary states are protected by pseudo-reflection symmetry. 
The boundary states support a pair of exceptional points    [Fig.~\ref{fig:2DclassD_pseudo_yobc}(b)], which is consistent with the continuum theory in terms of the effective Hamiltonian given by Eq.~(\ref{eq:boundary}) in Sec.~\ref{sec:2DETI}.

\begin{figure*}
\includegraphics[width=2.\columnwidth]{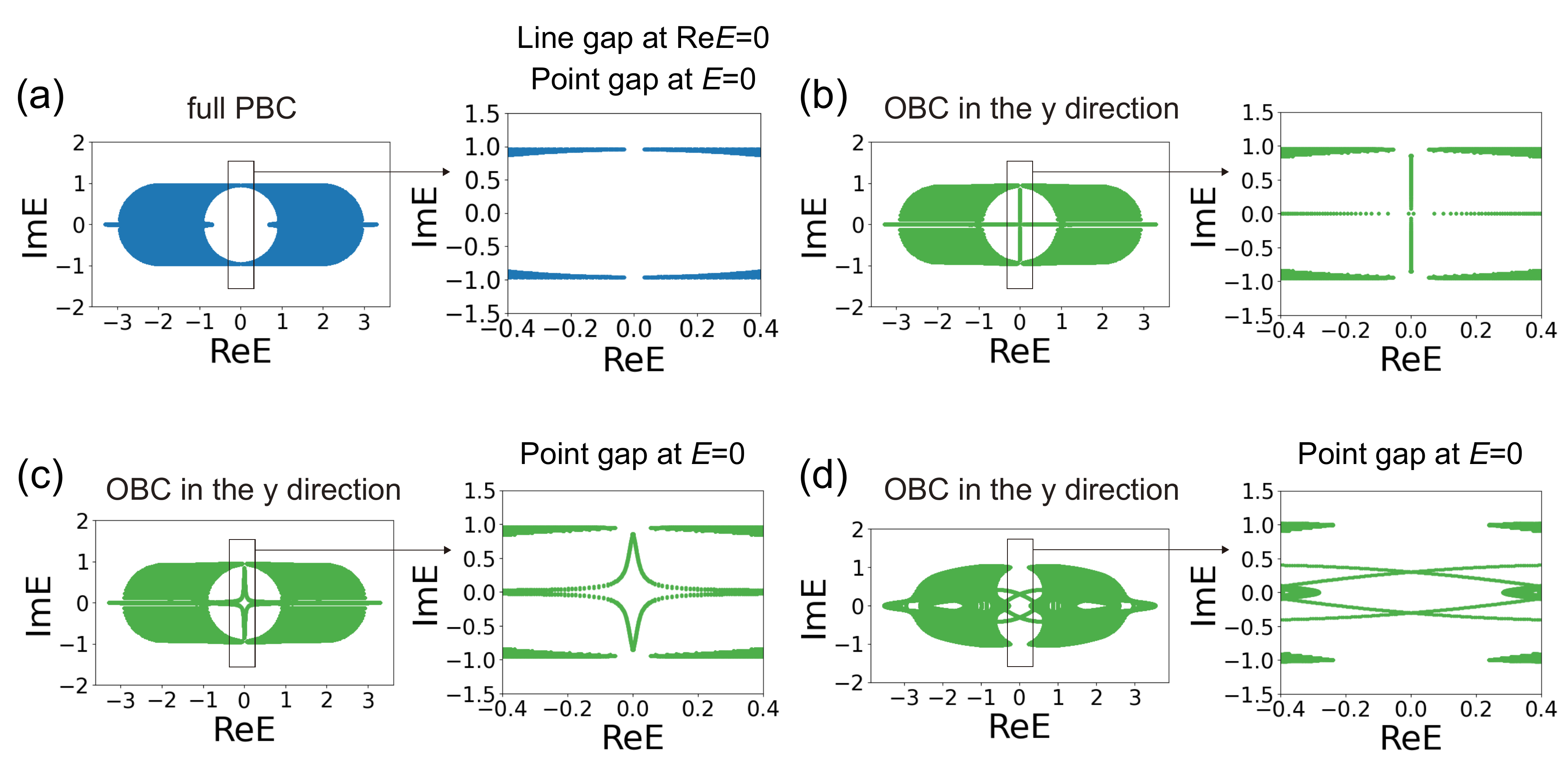}
	\caption{Complex energy spectra of the Hamiltonian $H'_{\tilde{\mathcal{M}}}(\boldsymbol{k})$ in Eq.~(\ref{eq:2DclassD_line}) with $m_1=0.1$, $m_2=0.3$, (a,b)~ $m_3=m_4=0$, (c)~$m_3=m_4=0.01$, and (d)~$m_3=m_4=0.5$ under (a)~the full periodic boundary conditions (PBC) and (b-d)~the open boundary conditions (OBC) in the $y$ direction and PBC in the $x$ direction. 
    The system size is $51$ in the $y$ direction with the momentum resolution $\Delta k_x=2\pi/2000$ in (b-d).
 }
	\label{fig:transition_pgt_lgt}
\end{figure*}

We find that the complex bulk spectrum of $H_{\tilde{\mathcal{M}}}(\boldsymbol{k})$ exhibits a line gap at ${\rm Re}\,E=0$ [Fig.~\ref{fig:transition_pgt_lgt}(a)].
Consequently, the second-order point-gap topological phase of $H_{\tilde{\mathcal{M}}}(\boldsymbol{k})$ is continuously deformable into a line-gap topological phase.
To further clarify this, we add mass terms to $H_{\tilde{\mathcal{M}}}(\boldsymbol{k})$:
\begin{align}\label{eq:2DclassD_line}
    H'_{\tilde{\mathcal{M}}}(\boldsymbol{k}) = H_{\tilde{\mathcal{M}}}(\boldsymbol{k})+im_3 \sigma_y+im_4\tau_z \sigma_y,
\end{align}
which preserve both particle-hole symmetry and pseudo-reflection symmetry.
The original point-gapless boundary states [Fig.~\ref{fig:transition_pgt_lgt}(b)] are deformed into the point-gapped but line-gapless boundary states [Fig.~\ref{fig:transition_pgt_lgt}(c)].
Thus, by incorporating appropriate mass terms, the exceptional second-order topological insulator within this symmetry class can be continuously deformed into the line-gap topological phase [Fig.~\ref{fig:transition_pgt_lgt}(d)].

%\clearpage

\section{Hermitized Hamiltonian in class DIII}\label{appendix: Hermitized_DIII}

We study a 2D Hermitized Hamiltonian in class DIII obtained from a non-Hermitian Hamiltonian in class D or class AII$^\dagger$.
When the Hermitized Hamiltonian in class DIII exhibits the reflection-symmetric second-order topological insulator phase, the corresponding non-Hermitian Hamiltonian exhibits exceptional second-order topological phase or second-order skin effect, depending on whether the symmetry class of the non-Hermitian Hamiltonian is class D or class AII$^\dagger$, as discussed below.

\subsection{Edge theory}

We consider the Hermitized boundary Hamiltonian obtained from the generic non-Hermitian boundary Hamiltonian in Eq.~(\ref{eq:boundary}) of Sec.~\ref{sec:2DETI}.
The Hermitian edge state reads 
\begin{align}
    {\sf H}(k_x) =k_x \tau_x \sigma_z - \delta \tau_y \sigma_y  \quad (\delta \in \mathbb{R}). 
\end{align}
This edge Hamiltonian respects reflection symmetry, time-reversal symmetry, particle-hole symmetry, and chiral symmetry:
\begin{align}
    &{\sf M} {\sf H}(k_x) {\sf M}^{-1}={\sf H}(-k_x), \quad {\sf M}= \tau_x \sigma_x, \label{eq:DIII_mirror} \\
    &{\sf T} {\sf H}^{*}(k_x) {\sf T}^{-1}={\sf H}(-k_x) \quad {\sf T} = i \tau_y, \label{eq:DIII_TRS} \\
    &{\sf C} {\sf H}^{*}(k_x) {\sf C}^{-1}=-{\sf H}(-k_x), \quad {\sf C}=\tau_x, \label{eq:DIII_PHS} \\
    &{\sf \Gamma} {\sf H}(k_x) {\sf \Gamma}^{-1}=-{\sf H}(k_x), \quad {\sf \Gamma}=\tau_z, \label{eq:DIII_chiral}
\end{align}
and belongs to class DIII $+$ ${\sf M}_{-+}$.
The energy dispersion is $E(k_x)= \pm {k_x\pm \delta}$, showing the gapless points at $k_x=\pm \delta$. 
On the other hand, in the absence of reflection symmetry, a mass term $\delta' \tau_x \sigma_y$ ($\delta' \in \mathbb{R}$) opens a gap in this edge state. 
Thus, the gapless boundary states appear only at the reflection-symmetric positions, showing the emergence of second-order topological boundary states protected by reflection symmetry. 

\subsection{Lattice model}

\begin{figure}
\includegraphics[width=0.8\columnwidth]{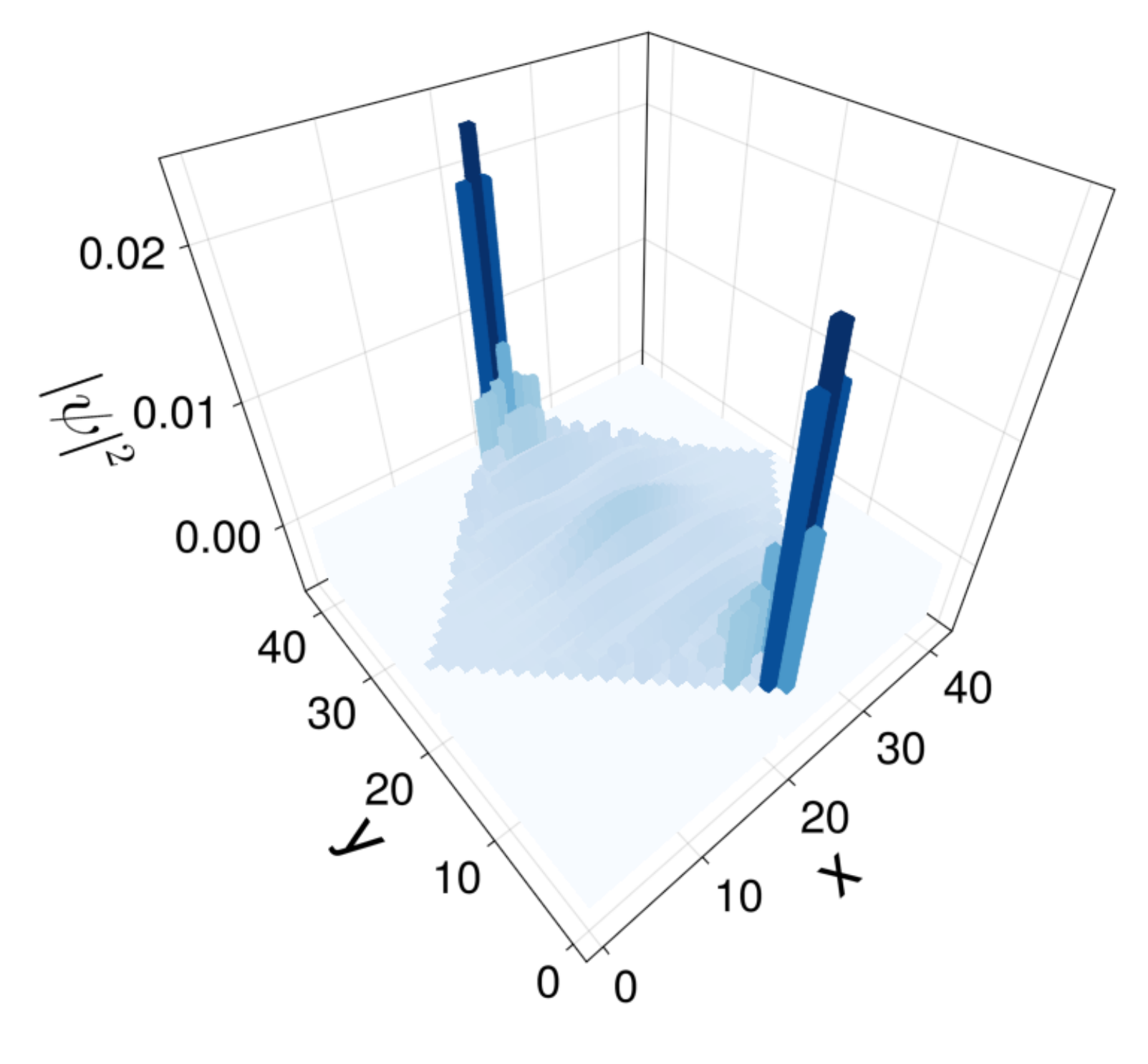}
	\caption{Real-space distribution of one of the %right 
    eigenstates at $E=0$ of the Hermitian Hamiltonian in Eq.~(\ref{eq:classDIII_Hermitized}) 
    under the full open boundary conditions for a square crystal with the edges perpendicular to the $(1,1)$ and $(1, -{1})$ directions ($m_1=0.1$ and $m_2=0.3$).
    The system size is $41$ in the $x$ and $y$ directions.
 }
	\label{fig:soti_DIII}
\end{figure}

We study the following Hermitized Hamiltonian ${\sf H}_{\rm DIII}(\boldsymbol{k})$ obtained from $H_{\tilde{\mathcal{M}}}(\boldsymbol{k})$ introduced in Eq.~(\ref{eq:classD_2D_pseud}) of Sec.~\ref{sec:2DETI}:
\begin{align}\label{eq:classDIII_Hermitized}
    {\sf H}_{\rm DIII}(\boldsymbol{k}) = &(1-\cos k_x- \cos k_y) \mu_x\tau_y + \sin k_x \mu_x \tau_x \sigma_z  \nonumber \\
    &- \sin k_y \mu_y \sigma_z-m_1 \mu_y \tau_x \sigma_y + m_2 \mu_x \tau_y \sigma_x.
\end{align}
This model respects reflection symmetry \begin{align}
    {\sf M}{\sf H}(k_x,k_y){\sf M}^{-1}={\sf H}(-k_x,k_y), \quad {\sf M}=\mu_x \sigma_x;
\end{align}
it also respects time-reversal symmetry, particle-hole symmetry, and chiral symmetry with
\begin{align} 
\quad {\sf T}=i\mu_y, \quad {\sf C}=\mu_x, \quad {\sf \Gamma} = \mu_z.
\end{align}
Thus, ${\sf H}_{\rm DIII}(\boldsymbol{k})$ belongs to class DIII $+$ ${\sf M}_{-+}$.
Figure~\ref{fig:soti_DIII} shows that this model supports corner states protected by reflection symmetry at $E=0$, consistent with the edge theory developed above.

%\clearpage

\section{2D exceptional topological insulator}\label{appendix: 2DETI}

We show that the Hamiltonian $H_{\rm 2D}^{(\pm)}(\boldsymbol{k})$ in Eq.~(\ref{eq: 2D ETI}) of Sec.~\ref{sec:3DETI} exhibits boundary states with a single exceptional point. 
The Hamiltonian $H_{\rm 2D}^{(\pm)}(\boldsymbol{k})$ respects chiral symmetry 
\begin{align}
        &\Gamma (H_{\rm 2D}^{(\pm)}(\boldsymbol{k}))^{\dagger}\Gamma^{-1} = - H_{\rm 2D}^{(\pm)}(\boldsymbol{k}), \\
        &\Gamma \coloneqq \tau_x \gamma, \quad \gamma \coloneqq \sigma_z,
\end{align}
and sublattice symmetry
\begin{align}
    \mathcal{S} H_{\rm 2D}^{(\pm)}(\boldsymbol{k})\mathcal{S}^{-1} = - H_{\rm 2D}^{(\pm)}(\boldsymbol{k}), \quad \mathcal{S} \coloneqq \tau_z.
\end{align}
The symmetry operators 
$\Gamma$ and $\mathcal{S}$ satisfy the anticommutation relation 
\begin{align}
\Gamma \mathcal{S}=-\mathcal{S} \Gamma.
\end{align}
Therefore, this model belongs to class AIII $+$ $\mathcal{S_{-}}$ in the 38-fold symmetry classification \cite{PhysRevX.9.041015}. 
In this symmetry class, the non-Hermitian Hamiltonian $H_{\rm 2D}^{(\pm)}(\boldsymbol{k})$ satisfies 
\begin{align}
    \mathcal{S}[i \Gamma H_{\rm 2D}^{(\pm)}(\boldsymbol{k})] \mathcal{S}^{-1} =  i \Gamma H_{\rm 2D}^{(\pm)}(\boldsymbol{k}),\quad \mathcal{S}^2=1,
\end{align}
which implies that the Hermitian matrix $i \Gamma H_{\rm 2D}^{(\pm)}(\boldsymbol{k})$ can be block-diagonalized into the two Hermitian matrices labeled by the eigenvalues $\pm1$ of $\mathcal{S}$: $ih^{(\pm)}_{1}\gamma$ and $ih^{(\pm)}_{2}\gamma$.
Topological invariants for these point-gap topological phases are given by a pair of the Chern numbers of the two Hermitian matrices $ih^{(\pm)}_{1}\gamma$ and $ih^{(\pm)}_{2}\gamma$  \cite{PhysRevX.9.041015}: 
\begin{align}
    ({\rm Ch}_1, {\rm Ch}_2)\coloneqq({\rm Ch}[ih^{(\pm)}_{1}\gamma], {\rm Ch}[ih^{(\pm)}_{2}\gamma])\in \mathbb{Z} \oplus \mathbb{Z}.
\end{align}
When $({\rm Ch}_1, {\rm Ch}_2)=(\pm1,0)$ or $({\rm Ch}_1, {\rm Ch}_2)=(0,\pm1)$ is satisfied, the system is a 2D exceptional topological insulator supporting edge states with a single  exceptional point \cite{PhysRevLett.132.136401, Denner_2023_infernal}. 

\begin{figure}
\includegraphics[width=1.\columnwidth]{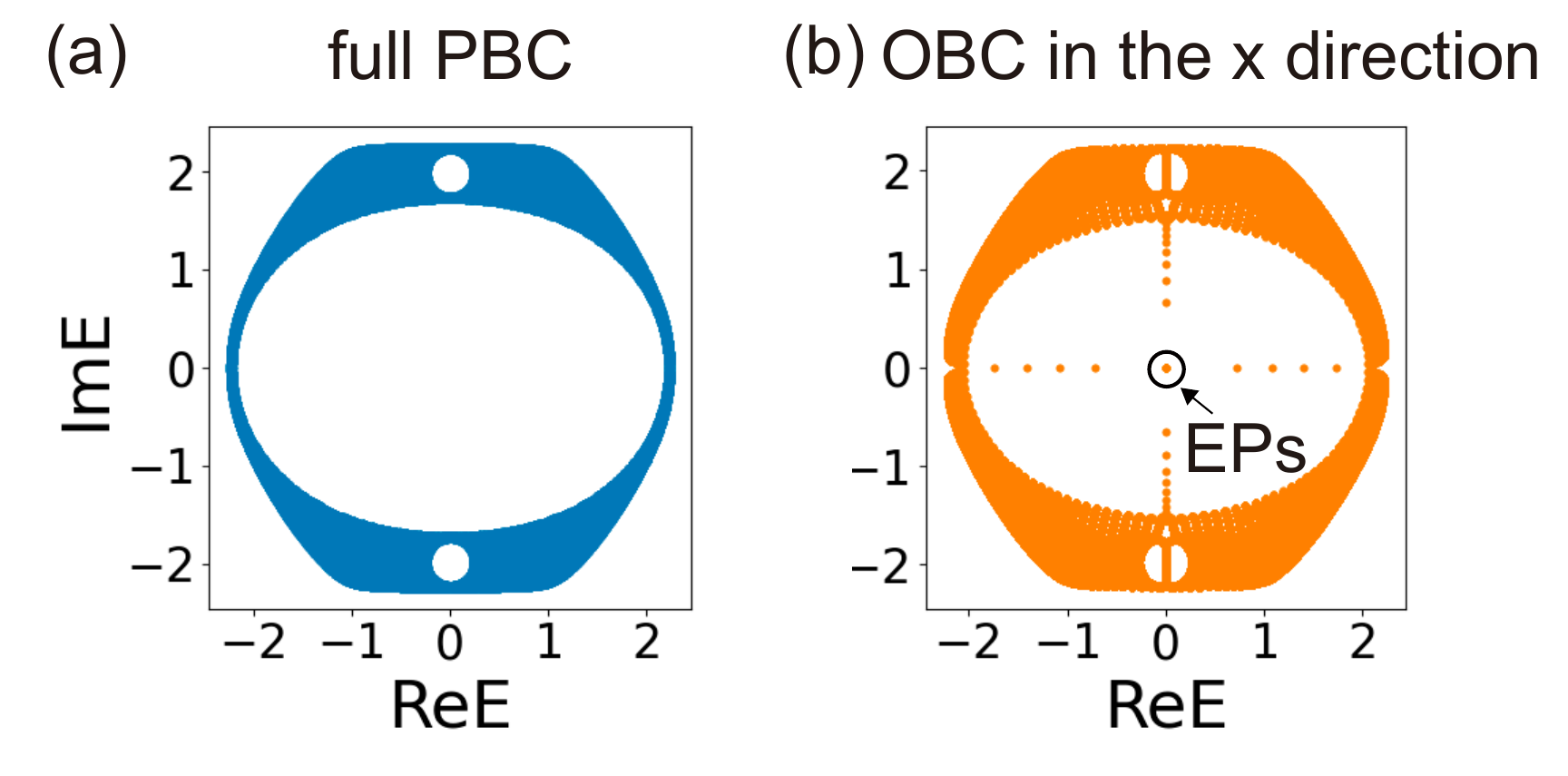}
	\caption{Complex energy spectra of the 2D exceptional topological insulator $H_{\rm 2D}^{(+)}(\boldsymbol{k})$ under the (a)~full periodic boundary conditions (PBC) and (b)~open boundary conditions (OBC) in the $x$ direction and PBC in the $y$ direction. 
    ``EPs" indicates exceptional points. 
    The system size in the $x$ direction is 30, and the momentum resolution is $\Delta k_y =2\pi/20000$ in (b). The parameters are the same as those in Fig.~\ref{fig:3d_intrinsic}.
 }
	\label{fig:2DETI}
\end{figure}

The Hamiltonian $H_{\rm 2D}^{(+)}(\boldsymbol{k})$ has a point gap at $E=0$ in the complex energy plane under the full periodic boundary conditions (PBC) [Fig.~\ref{fig:2DETI}(a)]. 
The pair of the Chern numbers is given by $({\rm Ch}_1, {\rm Ch}_2)=(1,0)$, and the model has edge states under the OBC in the $x$ direction [Fig.~\ref{fig:2DETI}(b)]. 
These edge states host a single exceptional point, as discussed in the following.

\subsection{Exceptional points }\label{subsec:proof_EP}
\begin{figure}
\includegraphics[width=1.\columnwidth]{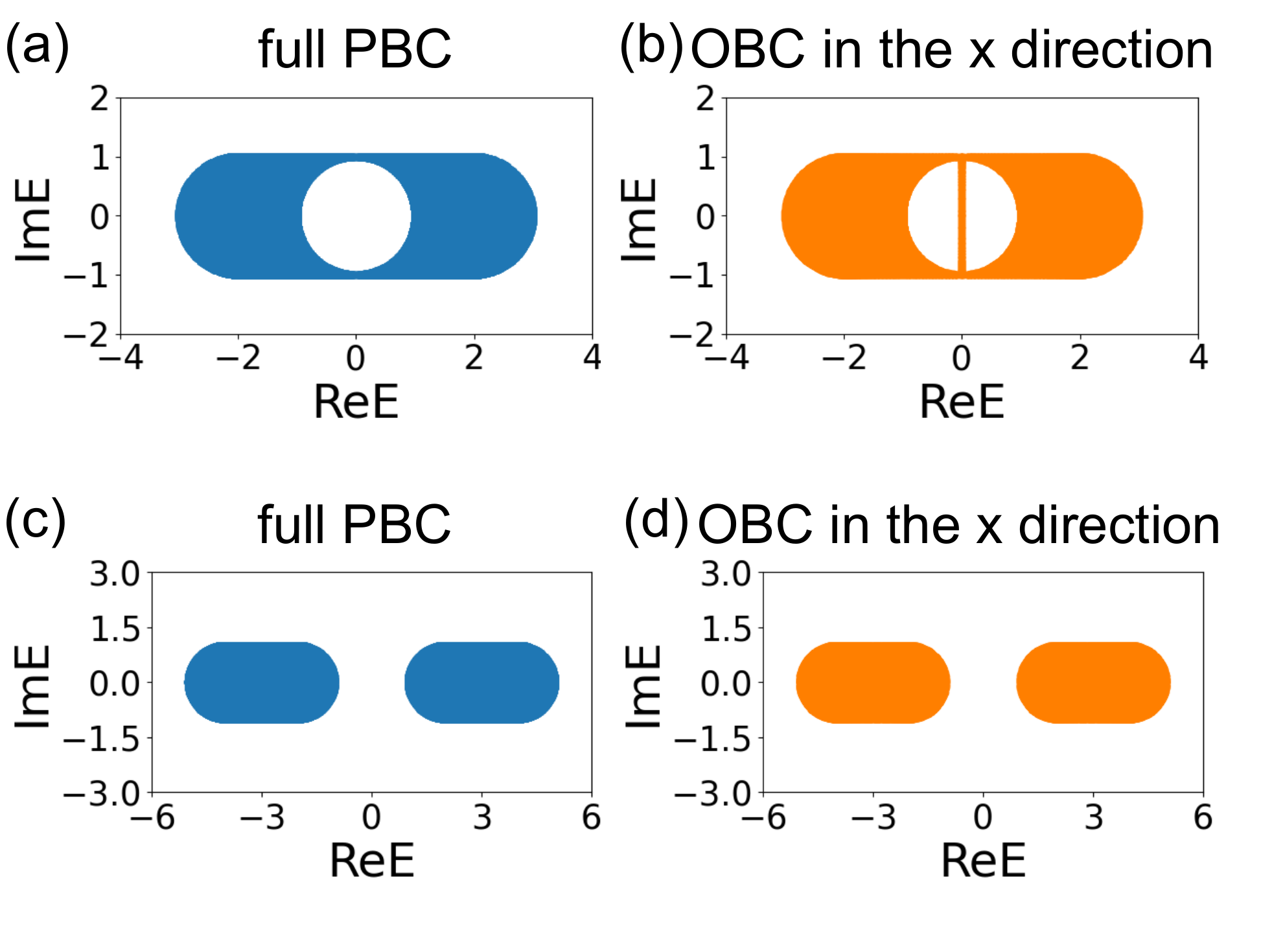}
	\caption{Complex energy spectra of (a,b) $h^{(\pm)}_{1}(\boldsymbol{k})$ and (c,d) $h^{(\pm)}_{2}(\boldsymbol{k})$. %in the complex plane.
    (a,c)~The boundary conditions are the full periodic boundary conditions (PBC). (b,d)~The boundary conditions are the open boundary conditions (OBC) in the $x$ direction and the PBC in the $y$ direction. The parameters are the same as those in Fig.~\ref{fig:3d_intrinsic}.
    The system size is 50 in the $x$ direction. 
 }
	\label{fig:calc_h1h2}
\end{figure}

Here, we show the emergence of edge states with exceptional points in $H^{(\pm)}_{\rm 2D}$ in Sec.~\ref{sec:3DETI}, based on the discussion in  Ref.~\cite{PhysRevLett.132.136401}.
Since our model $H^{(\pm)}_{\rm 2D}$ possesses sublattice symmetry with $\mathcal{S}=\tau_z$, the Hamiltonian is expressed as 
\begin{align}
    H^{(\pm)}_{\rm 2D}=\begin{pmatrix}
        0 & h^{(\pm)}_{1} \\
        h^{(\pm)}_{2} & 0
    \end{pmatrix}.
\end{align}
%%%%%%%%%%%%%%%%%%%%%%%%%%%%
The matrix $h^{(\pm)}_1$ has a point gap at $E=0$ in the full PBC [Fig.~\ref{fig:calc_h1h2}(a)] and hosts boundary states in the OBC in the $x$ direction [Fig.~\ref{fig:calc_h1h2}(b)].
Consequently, $h^{(\pm)}_1$ has an eigenvector $\ket{\psi_0}$ with zero eigenvalue under the OBC in the $x$ direction, 
\begin{align}
    h^{(\pm)}_1\ket{\psi_0}=0.
\end{align}
On the other hand, the matrix $h^{(\pm)}_2$ does not have a point gap to yield nontrivial topology [Fig.~\ref{fig:calc_h1h2}(c)] and therefore the system does not host boundary states [Fig.~\ref{fig:calc_h1h2}(d)].
Because $h^{(\pm)}_2$ does not have a zero eigenvalue, an inverse matrix $(h^{(\pm)}_2)^{-1}$ exists.

Here, we introduce the following basis,  
\begin{align}
        \ket{\Psi_0^1}\coloneqq\begin{pmatrix}
        0 \\
        \ket{\psi_0}
    \end{pmatrix}, \quad 
    \ket{\Psi_0^2}\coloneqq\begin{pmatrix}
        (h^{(\pm)}_2)^{-1}\ket{\psi_0} \\
        0
    \end{pmatrix},
\end{align}
and then we get
\begin{gather}
    H^{(\pm)}_{\rm 2D} \ket{\Psi_0^1}=0, %\nonumber \\
    \quad 
    H^{(\pm)}_{\rm 2D} \ket{\Psi_0^2}=\ket{\Psi_0^1}.
\end{gather}
We cannot introduce $\ket{\Psi_0^1}$ and $\ket{\Psi_0^2}$ under the full PBC because $h^{(\pm)}_1$ does not have $\ket{\psi_0}$ with a zero eigenvalue. 
Therefore, we find that the boundary states $\ket{\psi_0}$ of $h^{(\pm)}_1$ lead to the emergence of boundary states $\ket{\Psi_0^1}$ in $H^{(\pm)}_{\rm 2D}$.
Furthermore, $H^{(\pm)}_{\rm 2D}$ has a Jordan block with zero diagonal components:
\begin{align}
    H^{(\pm)}_{\rm 2D} \ket{\Psi_0^i} = \sum_{j=1,2}\ket{\Psi_0^j}J_{ji}, \quad J\coloneqq\begin{pmatrix}
        0 & 1\\
        0 & 0
    \end{pmatrix}.
\end{align}
It follows that $H^{(\pm)}_{\rm 2D}$ hosts boundary states with exceptional points.

%\begin{figure}
%\includegraphics[width=0.6\columnwidth]{breaking_inversion_l.pdf}
%	\caption{Band structures of (a) $H^{(+)}_{\rm 2D}(\boldsymbol{k})$ and (b) $H^{(+)}_{\rm 2D}(\boldsymbol{k})+\Delta H_{\rm 2D}$ under the open boundary conditions (OBC) in the $x$ direction and the periodic boundary conditions (PBC) in the $y$ direction. The system size is 200 in the $x$ direction. The momentum resolution is $\Delta k_y =2\pi/500$. The parameters of $H_{\rm 2D}^{(+)}(\boldsymbol{k})$ are the same as those in the main text. 
% }\label{fig:inversion_breaking}
%\end{figure}

\subsection{Energy dispersion and inversion-symmetry breaking}\label{subsec:wo_inversion}
\begin{figure}
\includegraphics[width=1.\columnwidth]{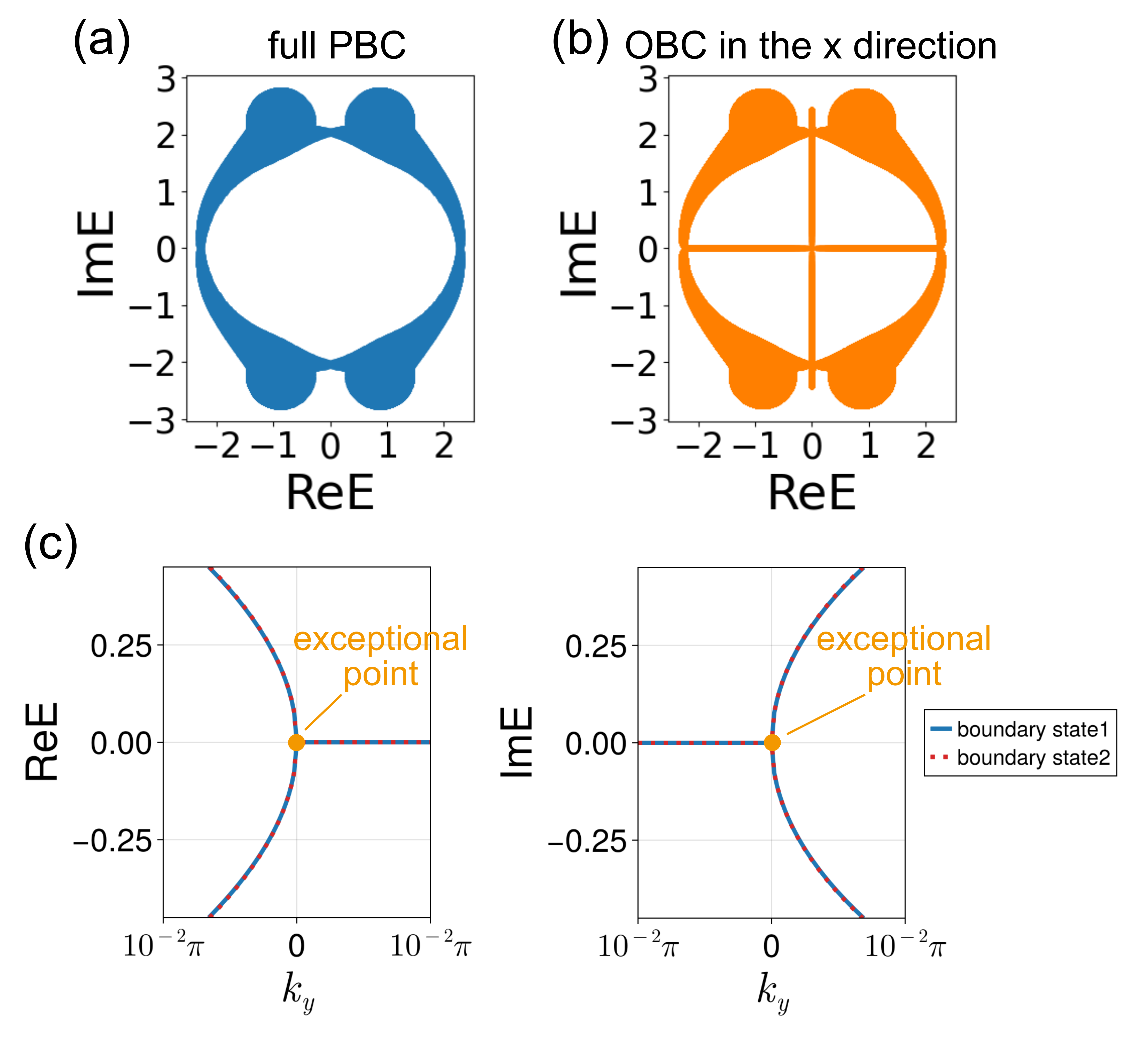}
	\caption{(a,b)~Complex energy spectra of $H_{\rm 2DETI}(\boldsymbol{k})$ under (a)~the full periodic boundary conditions (PBC) and (b)~the open boundary conditions (OBC) in the $x$ direction and the PBC in the $y$ direction. (c)~Energy dispersion of the boundary states in the point gap. The parameters are the same as those in Fig.~\ref{fig:3d_intrinsic}.
    The system size is 50 in the $x$ direction. 
 }
	\label{fig:2dETI_band}
\end{figure}

Here, we discuss the band dispersion of the 2D exceptional topological insulator $H^{(+)}_{\rm 2D}(\boldsymbol{k})$ and show that the boundary states possess an exceptional point. 
The Hamiltonian $H^{(+)}_{\rm 2D}(\boldsymbol{k})$ respects inversion symmetry
\begin{align}
    \mathcal{I} H_{\rm 2D}^{(\pm)}(\boldsymbol{k}) \mathcal{I}^{-1}=H_{\rm 2D}^{(\pm)}(-\boldsymbol{k}),
\end{align}
with $\mathcal{I}=\tau_z\sigma_y$.
Since inversion symmetry is not required for the protection of the 2D exceptional topological insulator, we introduce an inversion-symmetry-breaking term 
\begin{align}
    \Delta H_{\rm 2D} = 
    %0.5i
    %\begin{pmatrix}
    %    0 & 0 & 0 & 0 \\
    %    0 & 0 & 0 & 0 \\
    %    1 & 0 & 0 & 0 \\
    %    0 & 1 & 0 & 0
    %\end{pmatrix},
    \frac{i\lambda}{2}(\tau_x-i\tau_y),
\end{align}
with $\lambda=0.5$.
The overall Hamiltonian reads  
\begin{align}
    H_{\rm 2DETI}(\boldsymbol{k})=H^{(+)}_{\rm 2D}(\boldsymbol{k})+\Delta H_{\rm 2D}.
\end{align}
The Hamiltonian $H_{\rm 2DETI}(\boldsymbol{k})$ respects both sublattice and chiral symmetries.
Therefore, the topological invariants are still given by $({\rm Ch}_1, {\rm Ch}_2 ) = (1,0)$, resulting in the emergence of boundary states [Figs.~\ref{fig:2dETI_band}(a) and~\ref{fig:2dETI_band}(b)]. 
Figure~\ref{fig:2dETI_band}(c) shows that the energy dispersion of the boundary states of $H_{\rm 2DETI}(\boldsymbol{k})$
exhibits the square-root dependence on the wave vector. 
 The effective Hamiltonian for the boundary states of 
 %$H^{(+)}_{\rm 2D}(\boldsymbol{k})+\Delta H_{\rm 2D}$ 
 $H_{\rm 2DETI}(\boldsymbol{k})$ is given by
 \begin{align}
     H = \frac{i}{2}(\sigma_x+i\sigma_y)k_y + \frac{i}{2}(\sigma_x-i\sigma_y)\lambda. 
 \end{align}
 The energy dispersion is given by $\pm i\sqrt{\lambda k_y}$, which is consistent with the result of Fig.~\ref{fig:2dETI_band}(c). 

%\clearpage
\section{3D exceptional second-order topological insulators protected by inversion symmetry}\label{appendix: 3DESOTI_inversion}
We propose another 3D exceptional second-order topological insulator protected by inversion (parity) symmetry, which is different from the reflection-symmetric exceptional second-order topological insulator in Sec.~\ref{sec:3DETI}.

\subsection{Model}
We construct an inversion-symmetric 3D exceptional second-order topological insulator, using the layer construction in a similar manner to reflection-symmetric one in Sec.~\ref{sec:3DETI}.
We introduce the 2D exceptional topological insulators $H^{(+)}_{\rm 2D}(\boldsymbol{k})$ and $H^{(-)}_{\rm 2D}(\boldsymbol{k})$ given by Eq.~(\ref{eq: 2D ETI}) in Sec.~\ref{sec:3DETI}, and alternately stack the $H^{(+)}_{\rm 2D}(\boldsymbol{k})$ layer and the $H^{(-)}_{\rm 2D}(\boldsymbol{k})$ layer in the $z$ direction.
The Bloch Hamiltonian of the stacked layers is given by 
\begin{align}\label{eq:inversion-symmetric_}
    H_{\mathcal{I}}(\boldsymbol{k})=\begin{pmatrix}
        H_{\rm 2D}^{(+)}(\boldsymbol{k}) & H_{z, \mathcal{I}} (\boldsymbol{k}) \\
        H_{z, \mathcal{I}}^{\dagger} (\boldsymbol{k}) & H_{\rm 2D}^{(-)}(\boldsymbol{k})
    \end{pmatrix},
\end{align}
where $H_{z, \mathcal{I}} (\boldsymbol{k})$ is the interlayer hopping expressed as 
\begin{align}
    H_{z, \mathcal{I}} (\boldsymbol{k})= 
   \tau_x \otimes \begin{pmatrix}
      0 & t_1+t_1'e^{-ik_z} \\
      t_1e^{-ik_z}+t_1' & 0
   \end{pmatrix} \nonumber \\ 
    +i \tau_y \otimes \begin{pmatrix}
       t_2+t_2'e^{-ik_z} & 0 \\
      0 & -t_2e^{-ik_z}-t_2'  
   \end{pmatrix}.
\end{align}

\begin{figure}
\includegraphics[width=1.\columnwidth]{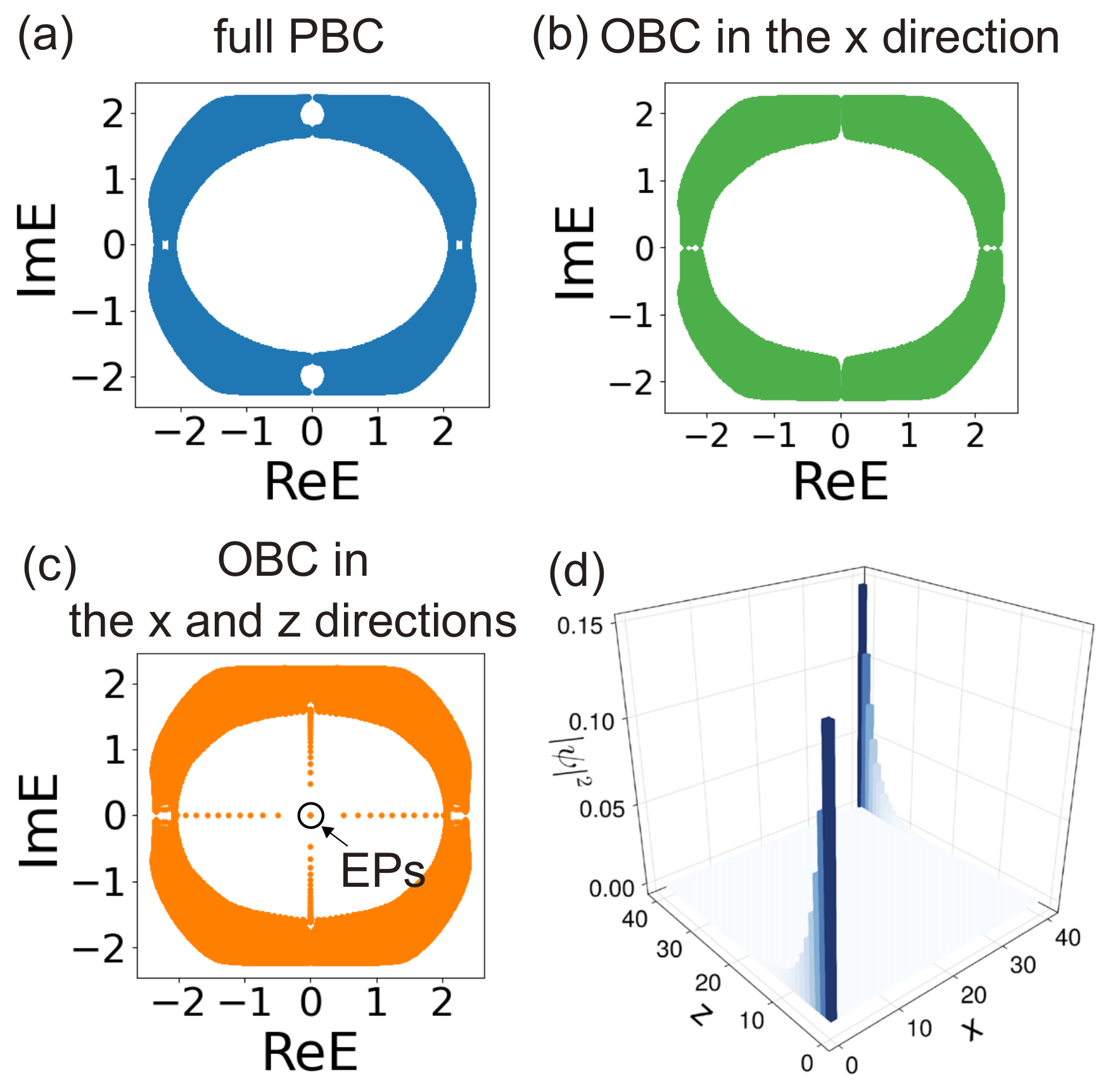}
	\caption{Complex energy spectra of the exceptional second-order topological insulator protected by inversion symmetry under the (a)~full periodic boundary conditions (PBC), (b)~open boundary conditions (OBC) in the $x$ direction and PBC in the $y$ and $z$ directions, and (c)~OBC in the $x$ and $z$ directions and PBC in the $y$ direction. ``EPs" in (c) indicates exceptional points. (d)~Real-space distribution of one of the right eigenstates encircled by the circle in (c). The parameters are $t_1=0.1$, $t_1'=0.2$, $t_2=0.3$, and $t_2'=0.25$. The system size in the $x$ direction is 50 in (b). The system size is $40$ in the $x$ direction, and the number of layers is $79$ in (c) and (d). The momentum resolutions in the $k_i$ ($i=y,z$) direction are $\Delta k_i=2\pi/80$ in the whole Brillouin zone in (b) and (c). For $-\pi/1000 \leq k_y \leq \pi/1000$, the momentum resolution is $\Delta k_y=2\pi/40000$ to capture the energy spectra of the boundary states.
 }
	\label{fig:model_calc}
\end{figure} 

In this model, each layer respects inversion symmetry
$
    \mathcal{I} H_{\rm 2D}^{(\pm)}(\boldsymbol{k}) \mathcal{I}^{-1}=H_{\rm 2D}^{(\pm)}(-\boldsymbol{k})
$
with $\mathcal{I}=\tau_z\sigma_y$.
By choosing the inversion center at $z=0$ in the unit cell, inversion symmetry for the overall Hamiltonian $H_{\mathcal{I}}(\boldsymbol{k})$ is written as 
\begin{align}
    \mathcal{I}_{\rm 3D} H_{\mathcal{I}}(\boldsymbol{k}) \mathcal{I}_{\rm 3D}^{-1}=H_{\mathcal{I}}(-\boldsymbol{k}), \quad \mathcal{I}_{\rm 3D} \coloneqq \begin{pmatrix}
          1 & 0 \\
          0 & e^{-ik_z}
      \end{pmatrix}\otimes \mathcal{I},
\end{align}
where the matrix ${\rm diag}(1, e^{-ik_z})$ acts on the sublattice degrees of freedom. 
Henceforth, we consider an odd number of layers to preserve inversion symmetry. 

Here, we focus on the point-gap topology at $E=0$ in the complex-energy spectrum [Fig.~\ref{fig:model_calc}(a)].
Figure~\ref{fig:model_calc}(b) shows that boundary states do not appear
in the point gap around $E=0$ under the OBC in the $x$ direction and the PBC in the other directions (slab geometry).
By contrast, Fig.~\ref{fig:model_calc}(c) demonstrates that the boundary states with two exceptional points appear in the point gap around $E=0$ under the OBC in the $x$ and $z$ directions and the PBC in the $y$ direction.  
Due to inversion symmetry, these boundary states are localized at opposite hinges [Fig.~\ref{fig:model_calc}(d)], each of which supports a single EP.

\subsection{Chern number in slab geometry}
To characterize the exceptional second-order topological insulator protected by inversion symmetry, we introduce a Chern number in the slab geometry with the OBC in the $z$ direction. 
Under weak interlayer coupling, the 3D model can be adiabatically connected to the 2D layers without couplings as long as the point gap is open.  
Thus, the Chern number in the slab geometry with such weak interlayer coupling is given by the sum of the Chern numbers $({\rm Ch}^{(l)}_{1}, {\rm Ch}^{(l)}_2)$ of each layer labeled by $l$, 
\begin{align}
    {\rm Ch}^{\rm slab}_{i}=\sum_l {\rm Ch}^{(l)}_{i},
\end{align}
with $i=1,2$, where $l$ runs over all the layers. 
Since the number of layers is odd, the Chern numbers under the slab geometry  are given by 
\begin{equation}
    ({\rm Ch}^{\rm slab}_{\rm 1}, {\rm Ch}^{\rm slab}_{\rm 2})=(1, 0).
\end{equation}
Therefore, our model can be regarded as a pseudo-2D exceptional topological insulator with a finite thickness in the $z$ direction.
Consequently, the hinge states with a single exceptional point appear under the OBC in the $x$ direction. 

\begin{figure}
\includegraphics[width=1.\columnwidth]{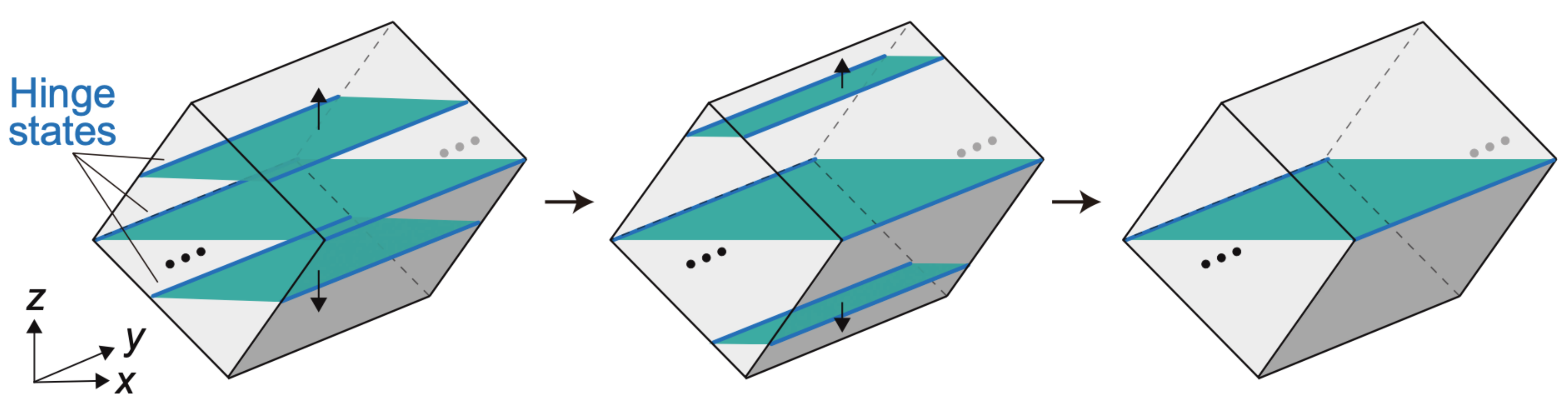}
	\caption{Removing of an even number of hinge channels in an inversion-symmetric exceptional second-order topological insulator. The hinge channels except for the middle state can be removed by moving them to the top hinge and bottom hinge without breaking inversion symmetry.}
	\label{fig:removing_hinge}
\end{figure}

The hinge state of the exceptional second-order topological insulator is protected against surface perturbations that preserve inversion symmetry in addition to chiral and sublattice symmetries. 
To see the importance of inversion symmetry, we consider an exceptional second-order topological insulator with an odd number of exceptional hinge channels.
The even number of channels can be removed by moving them to the top and bottom hinges while preserving inversion symmetry (Fig.~\ref{fig:removing_hinge}). On the other hand, the middle channel cannot be removed, and therefore this state is topologically protected by inversion symmetry. 
This scenario can also be applied to hinge states of the reflection-symmetric exceptional second-order topological insulators.

\subsection{Numbers of layers}

\begin{figure}
\includegraphics[width=1.\columnwidth]{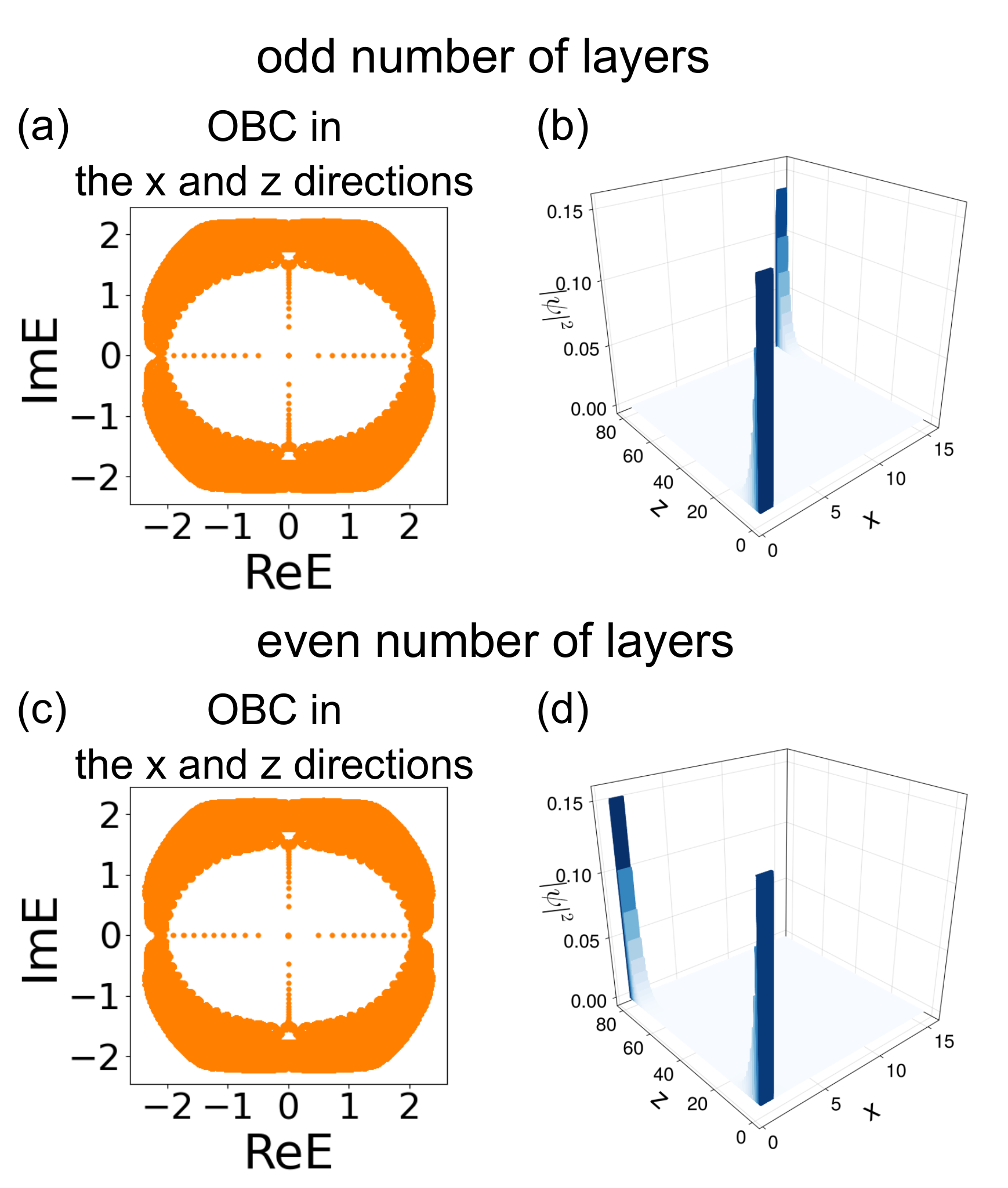}
	\caption{(a,c)~Complex energy spectra of the exceptional second-order topological insulator protected by inversion symmetry under the open boundary conditions (OBC) in the $x$ and $z$ directions and the periodic boundary conditions in the $y$ direction.
    (b)~Real-space distribution of one of the right eigenstates at $E=0$ in (a). (d)~Real-space distribution of one of the right eigenstates at $E=0$ in (c).
    The system size is $15$ in the $x$ direction, and the numbers of layers are $159$ in (a) and (b) and $160$ in (c) and (d). The parameters and the momentum resolutions are the same as Fig.~\ref{fig:model_calc}.
 }
	\label{fig:odd_even_layers}
\end{figure} 

We discuss the dependence of the positions of hinge states on the number of layers in the exceptional second-order topological insulator protected by inversion symmetry. 
When the number of layers is odd, the Hamiltonian with the finite size in the $z$ direction respects inversion symmetry.
Therefore, the hinge states appear at the inversion-symmetric positions [Figs.~\ref{fig:odd_even_layers}(a) and~\ref{fig:odd_even_layers}(b)]. 
On the other hand, the Hamiltonian with an even number of layers does not respect inversion symmetry. Therefore, the hinge states appear in the inversion-asymmetric positions [Figs.~\ref{fig:odd_even_layers}(c) and~\ref{fig:odd_even_layers}(d)].
The system with an even number of layers can be regarded as a 2D exceptional topological insulator attached to a 3D exceptional second-order topological insulator with an odd number of layers. When the 2D exceptional topological insulator is added, its edge states hybridize with the hinge states of the system with an odd number of layers, resulting in the hinge states at the inversion-asymmetric positions.
Consequently, the positions of the hinge states are switched by changing the parity of the number of layers. 

%\clearpage

\section{3D exceptional topological insulators protected by reflection symmetry}\label{appendix: 3DETCI}

\begin{figure}
\includegraphics[width=1.\columnwidth]{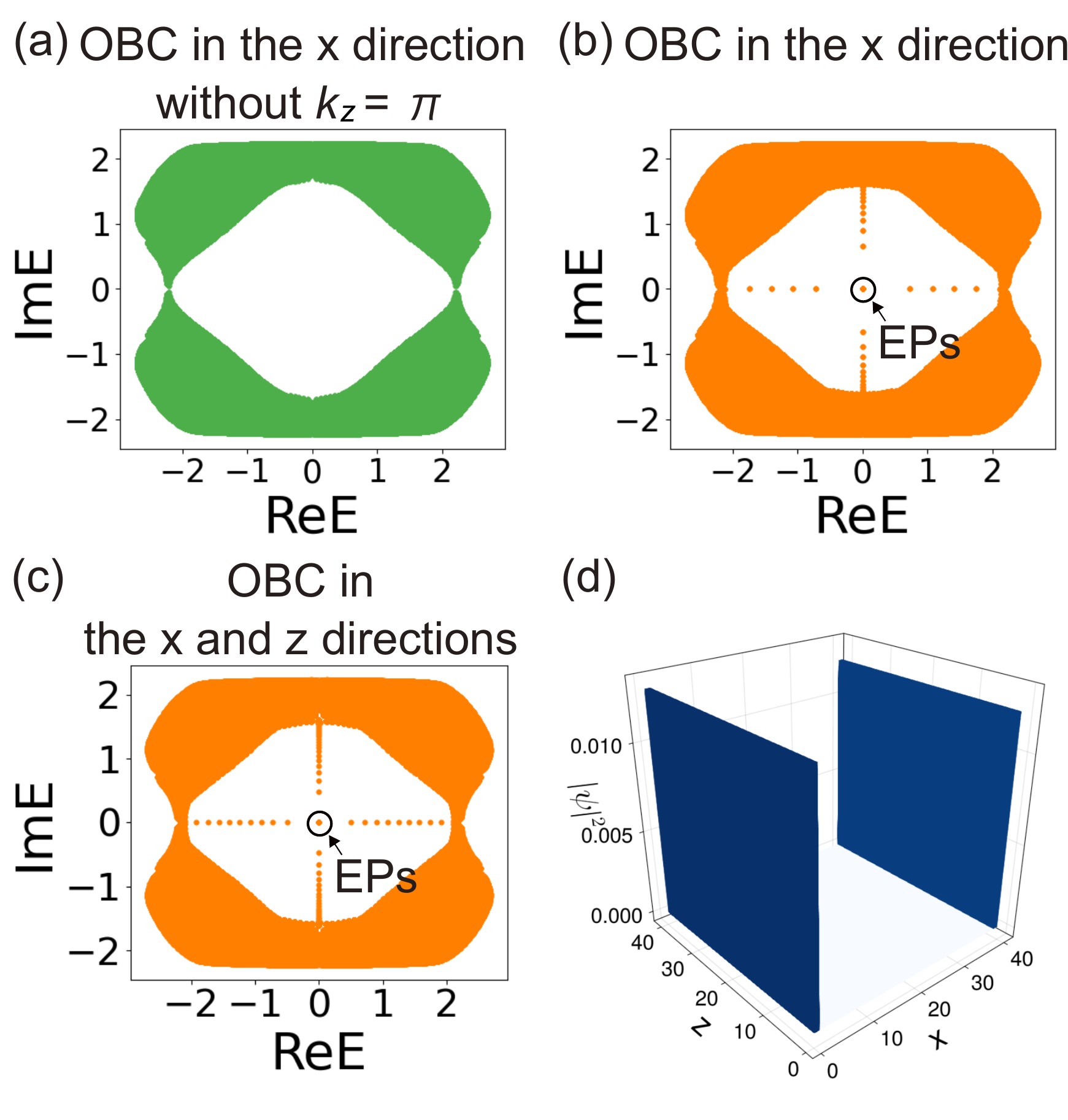}
	\caption{Complex energy spectra of the Hamiltonian $H_{\mathcal{M}}(\boldsymbol{k})$ under the open boundary conditions (OBC) in the $x$ direction and the periodic boundary conditions (PBC) in the $y$ and $z$ directions. (a)~Energy spectrum at various $k_z$ without $k_z=\pi$. (b)~Energy spectrum in the whole $k_x$-$k_y$ Brillouin zone including $k_z=\pi$, (c)~OBC in the $x$ and $z$ directions and PBC in the $y$ direction.
    (d)~Real-space distribution of one of the right eigenstates at $E=0$ in (c).
    ``EPs" in (b) and (c) indicates exceptional points. The parameters are the same as those in Fig.~\ref{fig:3d_intrinsic}.
    The system size is 50 in the $x$ direction.  The momentum resolutions in the $k_i$ ($i=y,z$) direction are $\Delta k_i=2\pi/80$ in (a,b). For $-\pi/100 < k_y< \pi/100$ and $k_z=\pi$, the momentum resolution is $\Delta k_y = 2\pi/20000$ in (a,b). The system size and the momentum resolutions in (c,d) are the same as Fig.~\ref{fig:model_calc}.}
	\label{fig:mirror_symm_model}
\end{figure} 

We discuss the surface states of the 3D reflection-symmetric exceptional second-order topological insulator $H_{\mathcal{M}}(\boldsymbol{k})$ given by Eq.~(\ref{eq:intrinsc3d}). In Sec.~\ref{sec:3DETI}, we have discussed the hinge states of $H_{\mathcal{M}}(\boldsymbol{k})$ under the OBC in the $x$ and $z$ directions with the surfaces perpendicular to the ($1,0,1$) and $(1,0,-1)$ directions.
Here, we show that boundary states appear on the surfaces perpendicular to the $x$ direction. 
A pair of the mirror Chern numbers for our model takes $({\rm Ch}_{\mathcal{M}, 1}, {\rm Ch}_{\mathcal{M}, 2})=(1,0)$ at $k_z=\pi$. 

We calculate the energy spectra of our model $H_{\mathcal{M}}(\boldsymbol{k})$.
Figure~\ref{fig:mirror_symm_model}(a) shows the energy spectrum without $k_z=\pi$ in the slab geometry with the OBC in the $x$ direction with the $y$-$z$ surface.
This result without $k_z=\pi$ shows that surface states do not appear at $E=0$ in the point gap.
We also calculate the energy spectrum with the whole Brillouin zone including  $k_z=\pi$ [Fig.~\ref{fig:mirror_symm_model}(b)] and find that the surface states with four exceptional points appear in the point gap at $E=0$. 
Since our model has inversion symmetry,  
two of the four exceptional points are related to each other via inversion symmetry. Therefore, each $y$-$z$ surface hosts two exceptional points.
In addition, one of the two exceptional points belongs to the block $H_{+}(k_x, k_y, \pi)$, and the other belongs to $H_{-}(k_x, k_y, \pi)$. 
Therefore, these exceptional points in the different mirror blocks cannot be annihilated without breaking $\mathcal{M}$ symmetry.
The pair of the two exceptional points 
is characterized by the mirror Chern number $({\rm Ch}_{\mathcal{M}, 1}, {\rm Ch}_{\mathcal{M}, 2})=(1,0)$ and hence topologically protected.
Figure~\ref{fig:mirror_symm_model}(c) shows the energy spectra under the OBC in the $x$ and $z$ directions. The point-gapless boundary states are localized at the $x$-$y$ surface [Fig.~\ref{fig:mirror_symm_model}(d)] in contrast to the boundary states localized at the hinges in Fig.~\ref{fig:model_calc}.

%\clearpage

%\bibliography{merge}

\begin{thebibliography}{111}%
\makeatletter
\providecommand \@ifxundefined [1]{%
 \@ifx{#1\undefined}
}%
\providecommand \@ifnum [1]{%
 \ifnum #1\expandafter \@firstoftwo
 \else \expandafter \@secondoftwo
 \fi
}%
\providecommand \@ifx [1]{%
 \ifx #1\expandafter \@firstoftwo
 \else \expandafter \@secondoftwo
 \fi
}%
\providecommand \natexlab [1]{#1}%
\providecommand \enquote  [1]{``#1''}%
\providecommand \bibnamefont  [1]{#1}%
\providecommand \bibfnamefont [1]{#1}%
\providecommand \citenamefont [1]{#1}%
\providecommand \href@noop [0]{\@secondoftwo}%
\providecommand \href [0]{\begingroup \@sanitize@url \@href}%
\providecommand \@href[1]{\@@startlink{#1}\@@href}%
\providecommand \@@href[1]{\endgroup#1\@@endlink}%
\providecommand \@sanitize@url [0]{\catcode `\\12\catcode `\$12\catcode
  `\&12\catcode `\#12\catcode `\^12\catcode `\_12\catcode `\%12\relax}%
\providecommand \@@startlink[1]{}%
\providecommand \@@endlink[0]{}%
\providecommand \url  [0]{\begingroup\@sanitize@url \@url }%
\providecommand \@url [1]{\endgroup\@href {#1}{\urlprefix }}%
\providecommand \urlprefix  [0]{URL }%
\providecommand \Eprint [0]{\href }%
\providecommand \doibase [0]{https://doi.org/}%
\providecommand \selectlanguage [0]{\@gobble}%
\providecommand \bibinfo  [0]{\@secondoftwo}%
\providecommand \bibfield  [0]{\@secondoftwo}%
\providecommand \translation [1]{[#1]}%
\providecommand \BibitemOpen [0]{}%
\providecommand \bibitemStop [0]{}%
\providecommand \bibitemNoStop [0]{.\EOS\space}%
\providecommand \EOS [0]{\spacefactor3000\relax}%
\providecommand \BibitemShut  [1]{\csname bibitem#1\endcsname}%
\let\auto@bib@innerbib\@empty
%</preamble>
\bibitem [{\citenamefont {Hasan}\ and\ \citenamefont
  {Kane}(2010)}]{RevModPhys.82.3045}%
  \BibitemOpen
  \bibfield  {author} {\bibinfo {author} {\bibfnamefont {M.~Z.}\ \bibnamefont
  {Hasan}}\ and\ \bibinfo {author} {\bibfnamefont {C.~L.}\ \bibnamefont
  {Kane}},\ }\bibfield  {title} {\bibinfo {title} {{Colloquium: Topological
  insulators}},\ }\href {https://doi.org/10.1103/RevModPhys.82.3045} {\bibfield
   {journal} {\bibinfo  {journal} {Rev. Mod. Phys.}\ }\textbf {\bibinfo
  {volume} {82}},\ \bibinfo {pages} {3045} (\bibinfo {year}
  {2010})}\BibitemShut {NoStop}%
\bibitem [{\citenamefont {Qi}\ and\ \citenamefont
  {Zhang}(2011)}]{RevModPhys.83.1057}%
  \BibitemOpen
  \bibfield  {author} {\bibinfo {author} {\bibfnamefont {X.-L.}\ \bibnamefont
  {Qi}}\ and\ \bibinfo {author} {\bibfnamefont {S.-C.}\ \bibnamefont {Zhang}},\
  }\bibfield  {title} {\bibinfo {title} {{Topological insulators and
  superconductors}},\ }\href {https://doi.org/10.1103/RevModPhys.83.1057}
  {\bibfield  {journal} {\bibinfo  {journal} {Rev. Mod. Phys.}\ }\textbf
  {\bibinfo {volume} {83}},\ \bibinfo {pages} {1057} (\bibinfo {year}
  {2011})}\BibitemShut {NoStop}%
\bibitem [{\citenamefont {Sitte}\ \emph {et~al.}(2012)\citenamefont {Sitte},
  \citenamefont {Rosch}, \citenamefont {Altman},\ and\ \citenamefont
  {Fritz}}]{PhysRevLett.108.126807}%
  \BibitemOpen
  \bibfield  {author} {\bibinfo {author} {\bibfnamefont {M.}~\bibnamefont
  {Sitte}}, \bibinfo {author} {\bibfnamefont {A.}~\bibnamefont {Rosch}},
  \bibinfo {author} {\bibfnamefont {E.}~\bibnamefont {Altman}},\ and\ \bibinfo
  {author} {\bibfnamefont {L.}~\bibnamefont {Fritz}},\ }\bibfield  {title}
  {\bibinfo {title} {{Topological Insulators in Magnetic Fields: Quantum Hall
  Effect and Edge Channels with a Nonquantized $\ensuremath{\theta}$ Term}},\
  }\href {https://doi.org/10.1103/PhysRevLett.108.126807} {\bibfield  {journal}
  {\bibinfo  {journal} {Phys. Rev. Lett.}\ }\textbf {\bibinfo {volume} {108}},\
  \bibinfo {pages} {126807} (\bibinfo {year} {2012})}\BibitemShut {NoStop}%
\bibitem [{\citenamefont {Zhang}\ \emph {et~al.}(2013)\citenamefont {Zhang},
  \citenamefont {Kane},\ and\ \citenamefont {Mele}}]{PhysRevLett.110.046404}%
  \BibitemOpen
  \bibfield  {author} {\bibinfo {author} {\bibfnamefont {F.}~\bibnamefont
  {Zhang}}, \bibinfo {author} {\bibfnamefont {C.~L.}\ \bibnamefont {Kane}},\
  and\ \bibinfo {author} {\bibfnamefont {E.~J.}\ \bibnamefont {Mele}},\
  }\bibfield  {title} {\bibinfo {title} {{Surface State Magnetization and
  Chiral Edge States on Topological Insulators}},\ }\href
  {https://doi.org/10.1103/PhysRevLett.110.046404} {\bibfield  {journal}
  {\bibinfo  {journal} {Phys. Rev. Lett.}\ }\textbf {\bibinfo {volume} {110}},\
  \bibinfo {pages} {046404} (\bibinfo {year} {2013})}\BibitemShut {NoStop}%
\bibitem [{\citenamefont {Benalcazar}\ \emph
  {et~al.}(2017{\natexlab{a}})\citenamefont {Benalcazar}, \citenamefont
  {Bernevig},\ and\ \citenamefont {Hughes}}]{benalcazar2017quantized}%
  \BibitemOpen
  \bibfield  {author} {\bibinfo {author} {\bibfnamefont {W.~A.}\ \bibnamefont
  {Benalcazar}}, \bibinfo {author} {\bibfnamefont {B.~A.}\ \bibnamefont
  {Bernevig}},\ and\ \bibinfo {author} {\bibfnamefont {T.~L.}\ \bibnamefont
  {Hughes}},\ }\bibfield  {title} {\bibinfo {title} {{Quantized electric
  multipole insulators}},\ }\href {https://doi.org/10.1126/science.aah6442}
  {\bibfield  {journal} {\bibinfo  {journal} {Science}\ }\textbf {\bibinfo
  {volume} {357}},\ \bibinfo {pages} {61} (\bibinfo {year}
  {2017}{\natexlab{a}})}\BibitemShut {NoStop}%
\bibitem [{\citenamefont {Benalcazar}\ \emph
  {et~al.}(2017{\natexlab{b}})\citenamefont {Benalcazar}, \citenamefont
  {Bernevig},\ and\ \citenamefont {Hughes}}]{PhysRevB.96.245115}%
  \BibitemOpen
  \bibfield  {author} {\bibinfo {author} {\bibfnamefont {W.~A.}\ \bibnamefont
  {Benalcazar}}, \bibinfo {author} {\bibfnamefont {B.~A.}\ \bibnamefont
  {Bernevig}},\ and\ \bibinfo {author} {\bibfnamefont {T.~L.}\ \bibnamefont
  {Hughes}},\ }\bibfield  {title} {\bibinfo {title} {{Electric multipole
  moments, topological multipole moment pumping, and chiral hinge states in
  crystalline insulators}},\ }\href
  {https://doi.org/10.1103/PhysRevB.96.245115} {\bibfield  {journal} {\bibinfo
  {journal} {Phys. Rev. B}\ }\textbf {\bibinfo {volume} {96}},\ \bibinfo
  {pages} {245115} (\bibinfo {year} {2017}{\natexlab{b}})}\BibitemShut
  {NoStop}%
\bibitem [{\citenamefont {Langbehn}\ \emph {et~al.}(2017)\citenamefont
  {Langbehn}, \citenamefont {Peng}, \citenamefont {Trifunovic}, \citenamefont
  {von Oppen},\ and\ \citenamefont {Brouwer}}]{PhysRevLett.119.246401}%
  \BibitemOpen
  \bibfield  {author} {\bibinfo {author} {\bibfnamefont {J.}~\bibnamefont
  {Langbehn}}, \bibinfo {author} {\bibfnamefont {Y.}~\bibnamefont {Peng}},
  \bibinfo {author} {\bibfnamefont {L.}~\bibnamefont {Trifunovic}}, \bibinfo
  {author} {\bibfnamefont {F.}~\bibnamefont {von Oppen}},\ and\ \bibinfo
  {author} {\bibfnamefont {P.~W.}\ \bibnamefont {Brouwer}},\ }\bibfield
  {title} {\bibinfo {title} {{Reflection-Symmetric Second-Order Topological
  Insulators and Superconductors}},\ }\href
  {https://doi.org/10.1103/PhysRevLett.119.246401} {\bibfield  {journal}
  {\bibinfo  {journal} {Phys. Rev. Lett.}\ }\textbf {\bibinfo {volume} {119}},\
  \bibinfo {pages} {246401} (\bibinfo {year} {2017})}\BibitemShut {NoStop}%
\bibitem [{\citenamefont {Song}\ \emph
  {et~al.}(2017{\natexlab{a}})\citenamefont {Song}, \citenamefont {Fang},\ and\
  \citenamefont {Fang}}]{PhysRevLett.119.246402}%
  \BibitemOpen
  \bibfield  {author} {\bibinfo {author} {\bibfnamefont {Z.}~\bibnamefont
  {Song}}, \bibinfo {author} {\bibfnamefont {Z.}~\bibnamefont {Fang}},\ and\
  \bibinfo {author} {\bibfnamefont {C.}~\bibnamefont {Fang}},\ }\bibfield
  {title} {\bibinfo {title} {{$(d\ensuremath{-}2)$-Dimensional Edge States of
  Rotation Symmetry Protected Topological States}},\ }\href
  {https://doi.org/10.1103/PhysRevLett.119.246402} {\bibfield  {journal}
  {\bibinfo  {journal} {Phys. Rev. Lett.}\ }\textbf {\bibinfo {volume} {119}},\
  \bibinfo {pages} {246402} (\bibinfo {year} {2017}{\natexlab{a}})}\BibitemShut
  {NoStop}%
\bibitem [{\citenamefont {Schindler}\ \emph {et~al.}(2018)\citenamefont
  {Schindler}, \citenamefont {Cook}, \citenamefont {Vergniory}, \citenamefont
  {Wang}, \citenamefont {Parkin}, \citenamefont {Bernevig},\ and\ \citenamefont
  {Neupert}}]{schindler2018higher}%
  \BibitemOpen
  \bibfield  {author} {\bibinfo {author} {\bibfnamefont {F.}~\bibnamefont
  {Schindler}}, \bibinfo {author} {\bibfnamefont {A.~M.}\ \bibnamefont {Cook}},
  \bibinfo {author} {\bibfnamefont {M.~G.}\ \bibnamefont {Vergniory}}, \bibinfo
  {author} {\bibfnamefont {Z.}~\bibnamefont {Wang}}, \bibinfo {author}
  {\bibfnamefont {S.~S.}\ \bibnamefont {Parkin}}, \bibinfo {author}
  {\bibfnamefont {B.~A.}\ \bibnamefont {Bernevig}},\ and\ \bibinfo {author}
  {\bibfnamefont {T.}~\bibnamefont {Neupert}},\ }\bibfield  {title} {\bibinfo
  {title} {{Higher-order topological insulators}},\ }\href
  {https://doi.org/10.1126/sciadv.aat0346} {\bibfield  {journal} {\bibinfo
  {journal} {Sci. Adv.}\ }\textbf {\bibinfo {volume} {4}},\ \bibinfo {pages}
  {eaat0346} (\bibinfo {year} {2018})}\BibitemShut {NoStop}%
\bibitem [{\citenamefont {Fang}\ and\ \citenamefont
  {Fu}(2019)}]{fang2017rotation}%
  \BibitemOpen
  \bibfield  {author} {\bibinfo {author} {\bibfnamefont {C.}~\bibnamefont
  {Fang}}\ and\ \bibinfo {author} {\bibfnamefont {L.}~\bibnamefont {Fu}},\
  }\bibfield  {title} {\bibinfo {title} {{New classes of topological
  crystalline insulators having surface rotation anomaly}},\ }\href
  {https://advances.sciencemag.org/content/5/12/eaat2374.abstract} {\bibfield
  {journal} {\bibinfo  {journal} {Sci. Adv.}\ }\textbf {\bibinfo {volume}
  {5}},\ \bibinfo {pages} {eaat2374} (\bibinfo {year} {2019})}\BibitemShut
  {NoStop}%
\bibitem [{\citenamefont {Gamow}(1928)}]{Gamow1928}%
  \BibitemOpen
  \bibfield  {author} {\bibinfo {author} {\bibfnamefont {G.}~\bibnamefont
  {Gamow}},\ }\bibfield  {title} {\bibinfo {title} {{Zur Quantentheorie des
  Atomkernes}},\ }\href {https://doi.org/10.1007/BF01343196} {\bibfield
  {journal} {\bibinfo  {journal} {{Z. Physik}}\ }\textbf {\bibinfo {volume}
  {51}},\ \bibinfo {pages} {204} (\bibinfo {year} {1928})}\BibitemShut
  {NoStop}%
\bibitem [{\citenamefont {Siegert}(1939)}]{PhysRev.56.750}%
  \BibitemOpen
  \bibfield  {author} {\bibinfo {author} {\bibfnamefont {A.~J.~F.}\
  \bibnamefont {Siegert}},\ }\bibfield  {title} {\bibinfo {title} {{On the
  Derivation of the Dispersion Formula for Nuclear Reactions}},\ }\href
  {https://doi.org/10.1103/PhysRev.56.750} {\bibfield  {journal} {\bibinfo
  {journal} {Phys. Rev.}\ }\textbf {\bibinfo {volume} {56}},\ \bibinfo {pages}
  {750} (\bibinfo {year} {1939})}\BibitemShut {NoStop}%
\bibitem [{\citenamefont {Feshbach}(1958)}]{FESHBACH1958357}%
  \BibitemOpen
  \bibfield  {author} {\bibinfo {author} {\bibfnamefont {H.}~\bibnamefont
  {Feshbach}},\ }\bibfield  {title} {\bibinfo {title} {{Unified theory of
  nuclear reactions}},\ }\href
  {https://doi.org/https://doi.org/10.1016/0003-4916(58)90007-1} {\bibfield
  {journal} {\bibinfo  {journal} {Ann. Phys.}\ }\textbf {\bibinfo {volume}
  {5}},\ \bibinfo {pages} {357} (\bibinfo {year} {1958})}\BibitemShut {NoStop}%
\bibitem [{\citenamefont {Feshbach}(1962)}]{FESHBACH1962287}%
  \BibitemOpen
  \bibfield  {author} {\bibinfo {author} {\bibfnamefont {H.}~\bibnamefont
  {Feshbach}},\ }\bibfield  {title} {\bibinfo {title} {{A unified theory of
  nuclear reactions. II}},\ }\href
  {https://doi.org/https://doi.org/10.1016/0003-4916(62)90221-X} {\bibfield
  {journal} {\bibinfo  {journal} {Ann. Phys.}\ }\textbf {\bibinfo {volume}
  {19}},\ \bibinfo {pages} {287} (\bibinfo {year} {1962})}\BibitemShut
  {NoStop}%
\bibitem [{\citenamefont {Kane}\ and\ \citenamefont
  {Lubensky}(2014)}]{Kane:2014uu}%
  \BibitemOpen
  \bibfield  {author} {\bibinfo {author} {\bibfnamefont {C.~L.}\ \bibnamefont
  {Kane}}\ and\ \bibinfo {author} {\bibfnamefont {T.~C.}\ \bibnamefont
  {Lubensky}},\ }\bibfield  {title} {\bibinfo {title} {Topological boundary
  modes in isostatic lattices},\ }\href {https://doi.org/10.1038/nphys2835}
  {\bibfield  {journal} {\bibinfo  {journal} {Nat. Phys.}\ }\textbf {\bibinfo
  {volume} {10}},\ \bibinfo {pages} {39} (\bibinfo {year} {2014})}\BibitemShut
  {NoStop}%
\bibitem [{\citenamefont {Huber}(2016)}]{Huber:2016tg}%
  \BibitemOpen
  \bibfield  {author} {\bibinfo {author} {\bibfnamefont {S.~D.}\ \bibnamefont
  {Huber}},\ }\bibfield  {title} {\bibinfo {title} {Topological mechanics},\
  }\href {https://doi.org/10.1038/nphys3801} {\bibfield  {journal} {\bibinfo
  {journal} {Nat. Phys.}\ }\textbf {\bibinfo {volume} {12}},\ \bibinfo {pages}
  {621} (\bibinfo {year} {2016})}\BibitemShut {NoStop}%
\bibitem [{\citenamefont {Konotop}\ \emph {et~al.}(2016)\citenamefont
  {Konotop}, \citenamefont {Yang},\ and\ \citenamefont
  {Zezyulin}}]{RevModPhys.88.035002}%
  \BibitemOpen
  \bibfield  {author} {\bibinfo {author} {\bibfnamefont {V.~V.}\ \bibnamefont
  {Konotop}}, \bibinfo {author} {\bibfnamefont {J.}~\bibnamefont {Yang}},\ and\
  \bibinfo {author} {\bibfnamefont {D.~A.}\ \bibnamefont {Zezyulin}},\
  }\bibfield  {title} {\bibinfo {title} {{Nonlinear waves in
  $\mathcal{PT}$-symmetric systems}},\ }\href
  {https://doi.org/10.1103/RevModPhys.88.035002} {\bibfield  {journal}
  {\bibinfo  {journal} {Rev. Mod. Phys.}\ }\textbf {\bibinfo {volume} {88}},\
  \bibinfo {pages} {035002} (\bibinfo {year} {2016})}\BibitemShut {NoStop}%
\bibitem [{\citenamefont {Feng}\ \emph {et~al.}(2017)\citenamefont {Feng},
  \citenamefont {El-Ganainy},\ and\ \citenamefont {Ge}}]{feng2017non}%
  \BibitemOpen
  \bibfield  {author} {\bibinfo {author} {\bibfnamefont {L.}~\bibnamefont
  {Feng}}, \bibinfo {author} {\bibfnamefont {R.}~\bibnamefont {El-Ganainy}},\
  and\ \bibinfo {author} {\bibfnamefont {L.}~\bibnamefont {Ge}},\ }\bibfield
  {title} {\bibinfo {title} {{Non-Hermitian photonics based on parity--time
  symmetry}},\ }\href {https://www.nature.com/articles/s41566-017-0031-1}
  {\bibfield  {journal} {\bibinfo  {journal} {Nat. Photon.}\ }\textbf {\bibinfo
  {volume} {11}},\ \bibinfo {pages} {752} (\bibinfo {year} {2017})}\BibitemShut
  {NoStop}%
\bibitem [{\citenamefont {El-Ganainy}\ \emph {et~al.}(2018)\citenamefont
  {El-Ganainy}, \citenamefont {Makris}, \citenamefont {Khajavikhan},
  \citenamefont {Musslimani}, \citenamefont {Rotter},\ and\ \citenamefont
  {Christodoulides}}]{el2018non}%
  \BibitemOpen
  \bibfield  {author} {\bibinfo {author} {\bibfnamefont {R.}~\bibnamefont
  {El-Ganainy}}, \bibinfo {author} {\bibfnamefont {K.~G.}\ \bibnamefont
  {Makris}}, \bibinfo {author} {\bibfnamefont {M.}~\bibnamefont {Khajavikhan}},
  \bibinfo {author} {\bibfnamefont {Z.~H.}\ \bibnamefont {Musslimani}},
  \bibinfo {author} {\bibfnamefont {S.}~\bibnamefont {Rotter}},\ and\ \bibinfo
  {author} {\bibfnamefont {D.~N.}\ \bibnamefont {Christodoulides}},\ }\bibfield
   {title} {\bibinfo {title} {{Non-Hermitian physics and PT symmetry}},\ }\href
  {https://www.nature.com/articles/nphys4323} {\bibfield  {journal} {\bibinfo
  {journal} {Nat. Phys.}\ }\textbf {\bibinfo {volume} {14}},\ \bibinfo {pages}
  {11} (\bibinfo {year} {2018})}\BibitemShut {NoStop}%
\bibitem [{\citenamefont {Kozii}\ and\ \citenamefont
  {Fu}(2024)}]{kozii2017non}%
  \BibitemOpen
  \bibfield  {author} {\bibinfo {author} {\bibfnamefont {V.}~\bibnamefont
  {Kozii}}\ and\ \bibinfo {author} {\bibfnamefont {L.}~\bibnamefont {Fu}},\
  }\bibfield  {title} {\bibinfo {title} {{Non-Hermitian topological theory of
  finite-lifetime quasiparticles: Prediction of bulk Fermi arc due to
  exceptional point}},\ }\href {https://doi.org/10.1103/PhysRevB.109.235139}
  {\bibfield  {journal} {\bibinfo  {journal} {Phys. Rev. B}\ }\textbf {\bibinfo
  {volume} {109}},\ \bibinfo {pages} {235139} (\bibinfo {year}
  {2024})}\BibitemShut {NoStop}%
\bibitem [{\citenamefont {Shen}\ and\ \citenamefont
  {Fu}(2018)}]{PhysRevLett.121.026403}%
  \BibitemOpen
  \bibfield  {author} {\bibinfo {author} {\bibfnamefont {H.}~\bibnamefont
  {Shen}}\ and\ \bibinfo {author} {\bibfnamefont {L.}~\bibnamefont {Fu}},\
  }\bibfield  {title} {\bibinfo {title} {{Quantum Oscillation from In-Gap
  States and a Non-Hermitian Landau Level Problem}},\ }\href
  {https://doi.org/10.1103/PhysRevLett.121.026403} {\bibfield  {journal}
  {\bibinfo  {journal} {Phys. Rev. Lett.}\ }\textbf {\bibinfo {volume} {121}},\
  \bibinfo {pages} {026403} (\bibinfo {year} {2018})}\BibitemShut {NoStop}%
\bibitem [{\citenamefont {Papaj}\ \emph {et~al.}(2019)\citenamefont {Papaj},
  \citenamefont {Isobe},\ and\ \citenamefont {Fu}}]{PhysRevB.99.201107}%
  \BibitemOpen
  \bibfield  {author} {\bibinfo {author} {\bibfnamefont {M.}~\bibnamefont
  {Papaj}}, \bibinfo {author} {\bibfnamefont {H.}~\bibnamefont {Isobe}},\ and\
  \bibinfo {author} {\bibfnamefont {L.}~\bibnamefont {Fu}},\ }\bibfield
  {title} {\bibinfo {title} {{Nodal arc of disordered Dirac fermions and
  non-Hermitian band theory}},\ }\href
  {https://doi.org/10.1103/PhysRevB.99.201107} {\bibfield  {journal} {\bibinfo
  {journal} {Phys. Rev. B}\ }\textbf {\bibinfo {volume} {99}},\ \bibinfo
  {pages} {201107(R)} (\bibinfo {year} {2019})}\BibitemShut {NoStop}%
\bibitem [{\citenamefont {Ashida}\ \emph {et~al.}(2020)\citenamefont {Ashida},
  \citenamefont {Gong},\ and\ \citenamefont
  {Ueda}}]{doi:10.1080/00018732.2021.1876991}%
  \BibitemOpen
  \bibfield  {author} {\bibinfo {author} {\bibfnamefont {Y.}~\bibnamefont
  {Ashida}}, \bibinfo {author} {\bibfnamefont {Z.}~\bibnamefont {Gong}},\ and\
  \bibinfo {author} {\bibfnamefont {M.}~\bibnamefont {Ueda}},\ }\bibfield
  {title} {\bibinfo {title} {{Non-Hermitian physics}},\ }\href
  {https://doi.org/10.1080/00018732.2021.1876991} {\bibfield  {journal}
  {\bibinfo  {journal} {Adv. Phys.}\ }\textbf {\bibinfo {volume} {69}},\
  \bibinfo {pages} {249} (\bibinfo {year} {2020})}\BibitemShut {NoStop}%
\bibitem [{\citenamefont {Bergholtz}\ \emph {et~al.}(2021)\citenamefont
  {Bergholtz}, \citenamefont {Budich},\ and\ \citenamefont
  {Kunst}}]{RevModPhys.93.015005}%
  \BibitemOpen
  \bibfield  {author} {\bibinfo {author} {\bibfnamefont {E.~J.}\ \bibnamefont
  {Bergholtz}}, \bibinfo {author} {\bibfnamefont {J.~C.}\ \bibnamefont
  {Budich}},\ and\ \bibinfo {author} {\bibfnamefont {F.~K.}\ \bibnamefont
  {Kunst}},\ }\bibfield  {title} {\bibinfo {title} {{Exceptional topology of
  non-Hermitian systems}},\ }\href
  {https://doi.org/10.1103/RevModPhys.93.015005} {\bibfield  {journal}
  {\bibinfo  {journal} {Rev. Mod. Phys.}\ }\textbf {\bibinfo {volume} {93}},\
  \bibinfo {pages} {015005} (\bibinfo {year} {2021})}\BibitemShut {NoStop}%
\bibitem [{\citenamefont {Shen}\ \emph {et~al.}(2018)\citenamefont {Shen},
  \citenamefont {Zhen},\ and\ \citenamefont {Fu}}]{PhysRevLett.120.146402}%
  \BibitemOpen
  \bibfield  {author} {\bibinfo {author} {\bibfnamefont {H.}~\bibnamefont
  {Shen}}, \bibinfo {author} {\bibfnamefont {B.}~\bibnamefont {Zhen}},\ and\
  \bibinfo {author} {\bibfnamefont {L.}~\bibnamefont {Fu}},\ }\bibfield
  {title} {\bibinfo {title} {{Topological Band Theory for Non-Hermitian
  Hamiltonians}},\ }\href {https://doi.org/10.1103/PhysRevLett.120.146402}
  {\bibfield  {journal} {\bibinfo  {journal} {Phys. Rev. Lett.}\ }\textbf
  {\bibinfo {volume} {120}},\ \bibinfo {pages} {146402} (\bibinfo {year}
  {2018})}\BibitemShut {NoStop}%
\bibitem [{\citenamefont {Gong}\ \emph {et~al.}(2018)\citenamefont {Gong},
  \citenamefont {Ashida}, \citenamefont {Kawabata}, \citenamefont {Takasan},
  \citenamefont {Higashikawa},\ and\ \citenamefont {Ueda}}]{PhysRevX.8.031079}%
  \BibitemOpen
  \bibfield  {author} {\bibinfo {author} {\bibfnamefont {Z.}~\bibnamefont
  {Gong}}, \bibinfo {author} {\bibfnamefont {Y.}~\bibnamefont {Ashida}},
  \bibinfo {author} {\bibfnamefont {K.}~\bibnamefont {Kawabata}}, \bibinfo
  {author} {\bibfnamefont {K.}~\bibnamefont {Takasan}}, \bibinfo {author}
  {\bibfnamefont {S.}~\bibnamefont {Higashikawa}},\ and\ \bibinfo {author}
  {\bibfnamefont {M.}~\bibnamefont {Ueda}},\ }\bibfield  {title} {\bibinfo
  {title} {{Topological Phases of Non-Hermitian Systems}},\ }\href
  {https://doi.org/10.1103/PhysRevX.8.031079} {\bibfield  {journal} {\bibinfo
  {journal} {Phys. Rev. X}\ }\textbf {\bibinfo {volume} {8}},\ \bibinfo {pages}
  {031079} (\bibinfo {year} {2018})}\BibitemShut {NoStop}%
\bibitem [{\citenamefont {Kawabata}\ \emph {et~al.}(2019)\citenamefont
  {Kawabata}, \citenamefont {Shiozaki}, \citenamefont {Ueda},\ and\
  \citenamefont {Sato}}]{PhysRevX.9.041015}%
  \BibitemOpen
  \bibfield  {author} {\bibinfo {author} {\bibfnamefont {K.}~\bibnamefont
  {Kawabata}}, \bibinfo {author} {\bibfnamefont {K.}~\bibnamefont {Shiozaki}},
  \bibinfo {author} {\bibfnamefont {M.}~\bibnamefont {Ueda}},\ and\ \bibinfo
  {author} {\bibfnamefont {M.}~\bibnamefont {Sato}},\ }\bibfield  {title}
  {\bibinfo {title} {{Symmetry and Topology in Non-Hermitian Physics}},\ }\href
  {https://doi.org/10.1103/PhysRevX.9.041015} {\bibfield  {journal} {\bibinfo
  {journal} {Phys. Rev. X}\ }\textbf {\bibinfo {volume} {9}},\ \bibinfo {pages}
  {041015} (\bibinfo {year} {2019})}\BibitemShut {NoStop}%
\bibitem [{\citenamefont {Okuma}\ \emph {et~al.}(2020)\citenamefont {Okuma},
  \citenamefont {Kawabata}, \citenamefont {Shiozaki},\ and\ \citenamefont
  {Sato}}]{PhysRevLett.124.086801}%
  \BibitemOpen
  \bibfield  {author} {\bibinfo {author} {\bibfnamefont {N.}~\bibnamefont
  {Okuma}}, \bibinfo {author} {\bibfnamefont {K.}~\bibnamefont {Kawabata}},
  \bibinfo {author} {\bibfnamefont {K.}~\bibnamefont {Shiozaki}},\ and\
  \bibinfo {author} {\bibfnamefont {M.}~\bibnamefont {Sato}},\ }\bibfield
  {title} {\bibinfo {title} {{Topological Origin of Non-Hermitian Skin
  Effects}},\ }\href {https://doi.org/10.1103/PhysRevLett.124.086801}
  {\bibfield  {journal} {\bibinfo  {journal} {Phys. Rev. Lett.}\ }\textbf
  {\bibinfo {volume} {124}},\ \bibinfo {pages} {086801} (\bibinfo {year}
  {2020})}\BibitemShut {NoStop}%
\bibitem [{\citenamefont {Nakamura}\ \emph {et~al.}(2024)\citenamefont
  {Nakamura}, \citenamefont {Bessho},\ and\ \citenamefont
  {Sato}}]{PhysRevLett.132.136401}%
  \BibitemOpen
  \bibfield  {author} {\bibinfo {author} {\bibfnamefont {D.}~\bibnamefont
  {Nakamura}}, \bibinfo {author} {\bibfnamefont {T.}~\bibnamefont {Bessho}},\
  and\ \bibinfo {author} {\bibfnamefont {M.}~\bibnamefont {Sato}},\ }\bibfield
  {title} {\bibinfo {title} {{Bulk-Boundary Correspondence in Point-Gap
  Topological Phases}},\ }\href
  {https://doi.org/10.1103/PhysRevLett.132.136401} {\bibfield  {journal}
  {\bibinfo  {journal} {Phys. Rev. Lett.}\ }\textbf {\bibinfo {volume} {132}},\
  \bibinfo {pages} {136401} (\bibinfo {year} {2024})}\BibitemShut {NoStop}%
\bibitem [{\citenamefont {Lee}(2016)}]{PhysRevLett.116.133903}%
  \BibitemOpen
  \bibfield  {author} {\bibinfo {author} {\bibfnamefont {T.~E.}\ \bibnamefont
  {Lee}},\ }\bibfield  {title} {\bibinfo {title} {{Anomalous Edge State in a
  Non-Hermitian Lattice}},\ }\href
  {https://doi.org/10.1103/PhysRevLett.116.133903} {\bibfield  {journal}
  {\bibinfo  {journal} {Phys. Rev. Lett.}\ }\textbf {\bibinfo {volume} {116}},\
  \bibinfo {pages} {133903} (\bibinfo {year} {2016})}\BibitemShut {NoStop}%
\bibitem [{\citenamefont {Martinez~Alvarez}\ \emph {et~al.}(2018)\citenamefont
  {Martinez~Alvarez}, \citenamefont {Barrios~Vargas},\ and\ \citenamefont
  {Foa~Torres}}]{PhysRevB.97.121401}%
  \BibitemOpen
  \bibfield  {author} {\bibinfo {author} {\bibfnamefont {V.~M.}\ \bibnamefont
  {Martinez~Alvarez}}, \bibinfo {author} {\bibfnamefont {J.~E.}\ \bibnamefont
  {Barrios~Vargas}},\ and\ \bibinfo {author} {\bibfnamefont {L.~E.~F.}\
  \bibnamefont {Foa~Torres}},\ }\bibfield  {title} {\bibinfo {title}
  {{Non-Hermitian robust edge states in one dimension: Anomalous localization
  and eigenspace condensation at exceptional points}},\ }\href
  {https://doi.org/10.1103/PhysRevB.97.121401} {\bibfield  {journal} {\bibinfo
  {journal} {Phys. Rev. B}\ }\textbf {\bibinfo {volume} {97}},\ \bibinfo
  {pages} {121401(R)} (\bibinfo {year} {2018})}\BibitemShut {NoStop}%
\bibitem [{\citenamefont {Yao}\ and\ \citenamefont
  {Wang}(2018)}]{PhysRevLett.121.086803}%
  \BibitemOpen
  \bibfield  {author} {\bibinfo {author} {\bibfnamefont {S.}~\bibnamefont
  {Yao}}\ and\ \bibinfo {author} {\bibfnamefont {Z.}~\bibnamefont {Wang}},\
  }\bibfield  {title} {\bibinfo {title} {{Edge States and Topological
  Invariants of Non-Hermitian Systems}},\ }\href
  {https://doi.org/10.1103/PhysRevLett.121.086803} {\bibfield  {journal}
  {\bibinfo  {journal} {Phys. Rev. Lett.}\ }\textbf {\bibinfo {volume} {121}},\
  \bibinfo {pages} {086803} (\bibinfo {year} {2018})}\BibitemShut {NoStop}%
\bibitem [{\citenamefont {Yao}\ \emph {et~al.}(2018)\citenamefont {Yao},
  \citenamefont {Song},\ and\ \citenamefont {Wang}}]{PhysRevLett.121.136802}%
  \BibitemOpen
  \bibfield  {author} {\bibinfo {author} {\bibfnamefont {S.}~\bibnamefont
  {Yao}}, \bibinfo {author} {\bibfnamefont {F.}~\bibnamefont {Song}},\ and\
  \bibinfo {author} {\bibfnamefont {Z.}~\bibnamefont {Wang}},\ }\bibfield
  {title} {\bibinfo {title} {{Non-Hermitian Chern Bands}},\ }\href
  {https://doi.org/10.1103/PhysRevLett.121.136802} {\bibfield  {journal}
  {\bibinfo  {journal} {Phys. Rev. Lett.}\ }\textbf {\bibinfo {volume} {121}},\
  \bibinfo {pages} {136802} (\bibinfo {year} {2018})}\BibitemShut {NoStop}%
\bibitem [{\citenamefont {Kunst}\ \emph {et~al.}(2018)\citenamefont {Kunst},
  \citenamefont {Edvardsson}, \citenamefont {Budich},\ and\ \citenamefont
  {Bergholtz}}]{PhysRevLett.121.026808}%
  \BibitemOpen
  \bibfield  {author} {\bibinfo {author} {\bibfnamefont {F.~K.}\ \bibnamefont
  {Kunst}}, \bibinfo {author} {\bibfnamefont {E.}~\bibnamefont {Edvardsson}},
  \bibinfo {author} {\bibfnamefont {J.~C.}\ \bibnamefont {Budich}},\ and\
  \bibinfo {author} {\bibfnamefont {E.~J.}\ \bibnamefont {Bergholtz}},\
  }\bibfield  {title} {\bibinfo {title} {{Biorthogonal Bulk-Boundary
  Correspondence in Non-Hermitian Systems}},\ }\href
  {https://doi.org/10.1103/PhysRevLett.121.026808} {\bibfield  {journal}
  {\bibinfo  {journal} {Phys. Rev. Lett.}\ }\textbf {\bibinfo {volume} {121}},\
  \bibinfo {pages} {026808} (\bibinfo {year} {2018})}\BibitemShut {NoStop}%
\bibitem [{\citenamefont {Lee}\ and\ \citenamefont
  {Thomale}(2019)}]{PhysRevB.99.201103}%
  \BibitemOpen
  \bibfield  {author} {\bibinfo {author} {\bibfnamefont {C.~H.}\ \bibnamefont
  {Lee}}\ and\ \bibinfo {author} {\bibfnamefont {R.}~\bibnamefont {Thomale}},\
  }\bibfield  {title} {\bibinfo {title} {{Anatomy of skin modes and topology in
  non-Hermitian systems}},\ }\href {https://doi.org/10.1103/PhysRevB.99.201103}
  {\bibfield  {journal} {\bibinfo  {journal} {Phys. Rev. B}\ }\textbf {\bibinfo
  {volume} {99}},\ \bibinfo {pages} {201103(R)} (\bibinfo {year}
  {2019})}\BibitemShut {NoStop}%
\bibitem [{\citenamefont {Zhou}\ and\ \citenamefont
  {Lee}(2019)}]{PhysRevB.99.235112}%
  \BibitemOpen
  \bibfield  {author} {\bibinfo {author} {\bibfnamefont {H.}~\bibnamefont
  {Zhou}}\ and\ \bibinfo {author} {\bibfnamefont {J.~Y.}\ \bibnamefont {Lee}},\
  }\bibfield  {title} {\bibinfo {title} {{Periodic table for topological bands
  with non-Hermitian symmetries}},\ }\href
  {https://doi.org/10.1103/PhysRevB.99.235112} {\bibfield  {journal} {\bibinfo
  {journal} {Phys. Rev. B}\ }\textbf {\bibinfo {volume} {99}},\ \bibinfo
  {pages} {235112} (\bibinfo {year} {2019})}\BibitemShut {NoStop}%
\bibitem [{\citenamefont {Lee}\ \emph {et~al.}(2019)\citenamefont {Lee},
  \citenamefont {Li},\ and\ \citenamefont {Gong}}]{PhysRevLett.123.016805}%
  \BibitemOpen
  \bibfield  {author} {\bibinfo {author} {\bibfnamefont {C.~H.}\ \bibnamefont
  {Lee}}, \bibinfo {author} {\bibfnamefont {L.}~\bibnamefont {Li}},\ and\
  \bibinfo {author} {\bibfnamefont {J.}~\bibnamefont {Gong}},\ }\bibfield
  {title} {\bibinfo {title} {{Hybrid Higher-Order Skin-Topological Modes in
  Nonreciprocal Systems}},\ }\href
  {https://doi.org/10.1103/PhysRevLett.123.016805} {\bibfield  {journal}
  {\bibinfo  {journal} {Phys. Rev. Lett.}\ }\textbf {\bibinfo {volume} {123}},\
  \bibinfo {pages} {016805} (\bibinfo {year} {2019})}\BibitemShut {NoStop}%
\bibitem [{\citenamefont {Yokomizo}\ and\ \citenamefont
  {Murakami}(2019)}]{PhysRevLett.123.066404}%
  \BibitemOpen
  \bibfield  {author} {\bibinfo {author} {\bibfnamefont {K.}~\bibnamefont
  {Yokomizo}}\ and\ \bibinfo {author} {\bibfnamefont {S.}~\bibnamefont
  {Murakami}},\ }\bibfield  {title} {\bibinfo {title} {{Non-Bloch Band Theory
  of Non-Hermitian Systems}},\ }\href
  {https://doi.org/10.1103/PhysRevLett.123.066404} {\bibfield  {journal}
  {\bibinfo  {journal} {Phys. Rev. Lett.}\ }\textbf {\bibinfo {volume} {123}},\
  \bibinfo {pages} {066404} (\bibinfo {year} {2019})}\BibitemShut {NoStop}%
\bibitem [{\citenamefont {Song}\ \emph
  {et~al.}(2019{\natexlab{a}})\citenamefont {Song}, \citenamefont {Yao},\ and\
  \citenamefont {Wang}}]{PhysRevLett.123.170401}%
  \BibitemOpen
  \bibfield  {author} {\bibinfo {author} {\bibfnamefont {F.}~\bibnamefont
  {Song}}, \bibinfo {author} {\bibfnamefont {S.}~\bibnamefont {Yao}},\ and\
  \bibinfo {author} {\bibfnamefont {Z.}~\bibnamefont {Wang}},\ }\bibfield
  {title} {\bibinfo {title} {{Non-Hermitian Skin Effect and Chiral Damping in
  Open Quantum Systems}},\ }\href
  {https://doi.org/10.1103/PhysRevLett.123.170401} {\bibfield  {journal}
  {\bibinfo  {journal} {Phys. Rev. Lett.}\ }\textbf {\bibinfo {volume} {123}},\
  \bibinfo {pages} {170401} (\bibinfo {year} {2019}{\natexlab{a}})}\BibitemShut
  {NoStop}%
\bibitem [{\citenamefont {Song}\ \emph
  {et~al.}(2019{\natexlab{b}})\citenamefont {Song}, \citenamefont {Yao},\ and\
  \citenamefont {Wang}}]{PhysRevLett.123.246801}%
  \BibitemOpen
  \bibfield  {author} {\bibinfo {author} {\bibfnamefont {F.}~\bibnamefont
  {Song}}, \bibinfo {author} {\bibfnamefont {S.}~\bibnamefont {Yao}},\ and\
  \bibinfo {author} {\bibfnamefont {Z.}~\bibnamefont {Wang}},\ }\bibfield
  {title} {\bibinfo {title} {{Non-Hermitian Topological Invariants in Real
  Space}},\ }\href {https://doi.org/10.1103/PhysRevLett.123.246801} {\bibfield
  {journal} {\bibinfo  {journal} {Phys. Rev. Lett.}\ }\textbf {\bibinfo
  {volume} {123}},\ \bibinfo {pages} {246801} (\bibinfo {year}
  {2019}{\natexlab{b}})}\BibitemShut {NoStop}%
\bibitem [{\citenamefont {Longhi}(2019)}]{PhysRevResearch.1.023013}%
  \BibitemOpen
  \bibfield  {author} {\bibinfo {author} {\bibfnamefont {S.}~\bibnamefont
  {Longhi}},\ }\bibfield  {title} {\bibinfo {title} {{Probing non-Hermitian
  skin effect and non-Bloch phase transitions}},\ }\href
  {https://doi.org/10.1103/PhysRevResearch.1.023013} {\bibfield  {journal}
  {\bibinfo  {journal} {Phys. Rev. Res.}\ }\textbf {\bibinfo {volume} {1}},\
  \bibinfo {pages} {023013} (\bibinfo {year} {2019})}\BibitemShut {NoStop}%
\bibitem [{\citenamefont {Borgnia}\ \emph {et~al.}(2020)\citenamefont
  {Borgnia}, \citenamefont {Kruchkov},\ and\ \citenamefont
  {Slager}}]{PhysRevLett.124.056802}%
  \BibitemOpen
  \bibfield  {author} {\bibinfo {author} {\bibfnamefont {D.~S.}\ \bibnamefont
  {Borgnia}}, \bibinfo {author} {\bibfnamefont {A.~J.}\ \bibnamefont
  {Kruchkov}},\ and\ \bibinfo {author} {\bibfnamefont {R.-J.}\ \bibnamefont
  {Slager}},\ }\bibfield  {title} {\bibinfo {title} {{Non-Hermitian Boundary
  Modes and Topology}},\ }\href
  {https://doi.org/10.1103/PhysRevLett.124.056802} {\bibfield  {journal}
  {\bibinfo  {journal} {Phys. Rev. Lett.}\ }\textbf {\bibinfo {volume} {124}},\
  \bibinfo {pages} {056802} (\bibinfo {year} {2020})}\BibitemShut {NoStop}%
\bibitem [{\citenamefont {Kawabata}\ \emph
  {et~al.}(2020{\natexlab{a}})\citenamefont {Kawabata}, \citenamefont {Okuma},\
  and\ \citenamefont {Sato}}]{PhysRevB.101.195147}%
  \BibitemOpen
  \bibfield  {author} {\bibinfo {author} {\bibfnamefont {K.}~\bibnamefont
  {Kawabata}}, \bibinfo {author} {\bibfnamefont {N.}~\bibnamefont {Okuma}},\
  and\ \bibinfo {author} {\bibfnamefont {M.}~\bibnamefont {Sato}},\ }\bibfield
  {title} {\bibinfo {title} {{Non-Bloch band theory of non-Hermitian
  Hamiltonians in the symplectic class}},\ }\href
  {https://doi.org/10.1103/PhysRevB.101.195147} {\bibfield  {journal} {\bibinfo
   {journal} {Phys. Rev. B}\ }\textbf {\bibinfo {volume} {101}},\ \bibinfo
  {pages} {195147} (\bibinfo {year} {2020}{\natexlab{a}})}\BibitemShut
  {NoStop}%
\bibitem [{\citenamefont {Zhang}\ \emph {et~al.}(2020)\citenamefont {Zhang},
  \citenamefont {Yang},\ and\ \citenamefont {Fang}}]{PhysRevLett.125.126402}%
  \BibitemOpen
  \bibfield  {author} {\bibinfo {author} {\bibfnamefont {K.}~\bibnamefont
  {Zhang}}, \bibinfo {author} {\bibfnamefont {Z.}~\bibnamefont {Yang}},\ and\
  \bibinfo {author} {\bibfnamefont {C.}~\bibnamefont {Fang}},\ }\bibfield
  {title} {\bibinfo {title} {{Correspondence between Winding Numbers and Skin
  Modes in Non-Hermitian Systems}},\ }\href
  {https://doi.org/10.1103/PhysRevLett.125.126402} {\bibfield  {journal}
  {\bibinfo  {journal} {Phys. Rev. Lett.}\ }\textbf {\bibinfo {volume} {125}},\
  \bibinfo {pages} {126402} (\bibinfo {year} {2020})}\BibitemShut {NoStop}%
\bibitem [{\citenamefont {Yoshida}\ \emph {et~al.}(2020)\citenamefont
  {Yoshida}, \citenamefont {Mizoguchi},\ and\ \citenamefont
  {Hatsugai}}]{PhysRevResearch.2.022062}%
  \BibitemOpen
  \bibfield  {author} {\bibinfo {author} {\bibfnamefont {T.}~\bibnamefont
  {Yoshida}}, \bibinfo {author} {\bibfnamefont {T.}~\bibnamefont {Mizoguchi}},\
  and\ \bibinfo {author} {\bibfnamefont {Y.}~\bibnamefont {Hatsugai}},\
  }\bibfield  {title} {\bibinfo {title} {{Mirror skin effect and its electric
  circuit simulation}},\ }\href
  {https://doi.org/10.1103/PhysRevResearch.2.022062} {\bibfield  {journal}
  {\bibinfo  {journal} {Phys. Rev. Res.}\ }\textbf {\bibinfo {volume} {2}},\
  \bibinfo {pages} {022062(R)} (\bibinfo {year} {2020})}\BibitemShut {NoStop}%
\bibitem [{\citenamefont {Yokomizo}\ and\ \citenamefont
  {Murakami}(2021{\natexlab{a}})}]{PhysRevB.103.165123}%
  \BibitemOpen
  \bibfield  {author} {\bibinfo {author} {\bibfnamefont {K.}~\bibnamefont
  {Yokomizo}}\ and\ \bibinfo {author} {\bibfnamefont {S.}~\bibnamefont
  {Murakami}},\ }\bibfield  {title} {\bibinfo {title} {{Non-Bloch band theory
  in bosonic Bogoliubov--de Gennes systems}},\ }\href
  {https://doi.org/10.1103/PhysRevB.103.165123} {\bibfield  {journal} {\bibinfo
   {journal} {Phys. Rev. B}\ }\textbf {\bibinfo {volume} {103}},\ \bibinfo
  {pages} {165123} (\bibinfo {year} {2021}{\natexlab{a}})}\BibitemShut
  {NoStop}%
\bibitem [{\citenamefont {Yokomizo}\ and\ \citenamefont
  {Murakami}(2021{\natexlab{b}})}]{PhysRevB.104.165117}%
  \BibitemOpen
  \bibfield  {author} {\bibinfo {author} {\bibfnamefont {K.}~\bibnamefont
  {Yokomizo}}\ and\ \bibinfo {author} {\bibfnamefont {S.}~\bibnamefont
  {Murakami}},\ }\bibfield  {title} {\bibinfo {title} {{Scaling rule for the
  critical non-Hermitian skin effect}},\ }\href
  {https://doi.org/10.1103/PhysRevB.104.165117} {\bibfield  {journal} {\bibinfo
   {journal} {Phys. Rev. B}\ }\textbf {\bibinfo {volume} {104}},\ \bibinfo
  {pages} {165117} (\bibinfo {year} {2021}{\natexlab{b}})}\BibitemShut
  {NoStop}%
\bibitem [{\citenamefont {Xiao}\ and\ \citenamefont
  {Chan}(2022)}]{PhysRevB.105.075128}%
  \BibitemOpen
  \bibfield  {author} {\bibinfo {author} {\bibfnamefont {Y.-X.}\ \bibnamefont
  {Xiao}}\ and\ \bibinfo {author} {\bibfnamefont {C.~T.}\ \bibnamefont
  {Chan}},\ }\bibfield  {title} {\bibinfo {title} {{Topology in non-Hermitian
  Chern insulators with skin effect}},\ }\href
  {https://doi.org/10.1103/PhysRevB.105.075128} {\bibfield  {journal} {\bibinfo
   {journal} {Phys. Rev. B}\ }\textbf {\bibinfo {volume} {105}},\ \bibinfo
  {pages} {075128} (\bibinfo {year} {2022})}\BibitemShut {NoStop}%
\bibitem [{\citenamefont {Yokomizo}\ \emph {et~al.}(2022)\citenamefont
  {Yokomizo}, \citenamefont {Yoda},\ and\ \citenamefont
  {Murakami}}]{yokomizo2021non}%
  \BibitemOpen
  \bibfield  {author} {\bibinfo {author} {\bibfnamefont {K.}~\bibnamefont
  {Yokomizo}}, \bibinfo {author} {\bibfnamefont {T.}~\bibnamefont {Yoda}},\
  and\ \bibinfo {author} {\bibfnamefont {S.}~\bibnamefont {Murakami}},\
  }\bibfield  {title} {\bibinfo {title} {{Non-Hermitian waves in a continuous
  periodic model and application to photonic crystals}},\ }\href
  {https://doi.org/10.1103/PhysRevResearch.4.023089} {\bibfield  {journal}
  {\bibinfo  {journal} {Phys. Rev. Res.}\ }\textbf {\bibinfo {volume} {4}},\
  \bibinfo {pages} {023089} (\bibinfo {year} {2022})}\BibitemShut {NoStop}%
\bibitem [{\citenamefont {Deng}\ and\ \citenamefont
  {Flebus}(2022)}]{PhysRevB.105.L180406}%
  \BibitemOpen
  \bibfield  {author} {\bibinfo {author} {\bibfnamefont {K.}~\bibnamefont
  {Deng}}\ and\ \bibinfo {author} {\bibfnamefont {B.}~\bibnamefont {Flebus}},\
  }\bibfield  {title} {\bibinfo {title} {{Non-Hermitian skin effect in magnetic
  systems}},\ }\href {https://doi.org/10.1103/PhysRevB.105.L180406} {\bibfield
  {journal} {\bibinfo  {journal} {Phys. Rev. B}\ }\textbf {\bibinfo {volume}
  {105}},\ \bibinfo {pages} {L180406} (\bibinfo {year} {2022})}\BibitemShut
  {NoStop}%
\bibitem [{\citenamefont {Longhi}(2022)}]{PhysRevB.105.245143}%
  \BibitemOpen
  \bibfield  {author} {\bibinfo {author} {\bibfnamefont {S.}~\bibnamefont
  {Longhi}},\ }\bibfield  {title} {\bibinfo {title} {{Non-Hermitian skin effect
  and self-acceleration}},\ }\href
  {https://doi.org/10.1103/PhysRevB.105.245143} {\bibfield  {journal} {\bibinfo
   {journal} {Phys. Rev. B}\ }\textbf {\bibinfo {volume} {105}},\ \bibinfo
  {pages} {245143} (\bibinfo {year} {2022})}\BibitemShut {NoStop}%
\bibitem [{\citenamefont {Liang}\ \emph {et~al.}(2022)\citenamefont {Liang},
  \citenamefont {Xie}, \citenamefont {Dong}, \citenamefont {Li}, \citenamefont
  {Li}, \citenamefont {Gadway}, \citenamefont {Yi},\ and\ \citenamefont
  {Yan}}]{PhysRevLett.129.070401}%
  \BibitemOpen
  \bibfield  {author} {\bibinfo {author} {\bibfnamefont {Q.}~\bibnamefont
  {Liang}}, \bibinfo {author} {\bibfnamefont {D.}~\bibnamefont {Xie}}, \bibinfo
  {author} {\bibfnamefont {Z.}~\bibnamefont {Dong}}, \bibinfo {author}
  {\bibfnamefont {H.}~\bibnamefont {Li}}, \bibinfo {author} {\bibfnamefont
  {H.}~\bibnamefont {Li}}, \bibinfo {author} {\bibfnamefont {B.}~\bibnamefont
  {Gadway}}, \bibinfo {author} {\bibfnamefont {W.}~\bibnamefont {Yi}},\ and\
  \bibinfo {author} {\bibfnamefont {B.}~\bibnamefont {Yan}},\ }\bibfield
  {title} {\bibinfo {title} {{Dynamic Signatures of Non-Hermitian Skin Effect
  and Topology in Ultracold Atoms}},\ }\href
  {https://doi.org/10.1103/PhysRevLett.129.070401} {\bibfield  {journal}
  {\bibinfo  {journal} {Phys. Rev. Lett.}\ }\textbf {\bibinfo {volume} {129}},\
  \bibinfo {pages} {070401} (\bibinfo {year} {2022})}\BibitemShut {NoStop}%
\bibitem [{\citenamefont {Franca}\ \emph {et~al.}(2022)\citenamefont {Franca},
  \citenamefont {K\"onye}, \citenamefont {Hassler}, \citenamefont {van~den
  Brink},\ and\ \citenamefont {Fulga}}]{PhysRevLett.129.086601}%
  \BibitemOpen
  \bibfield  {author} {\bibinfo {author} {\bibfnamefont {S.}~\bibnamefont
  {Franca}}, \bibinfo {author} {\bibfnamefont {V.}~\bibnamefont {K\"onye}},
  \bibinfo {author} {\bibfnamefont {F.}~\bibnamefont {Hassler}}, \bibinfo
  {author} {\bibfnamefont {J.}~\bibnamefont {van~den Brink}},\ and\ \bibinfo
  {author} {\bibfnamefont {C.}~\bibnamefont {Fulga}},\ }\bibfield  {title}
  {\bibinfo {title} {{Non-Hermitian Physics without Gain or Loss: The Skin
  Effect of Reflected Waves}},\ }\href
  {https://doi.org/10.1103/PhysRevLett.129.086601} {\bibfield  {journal}
  {\bibinfo  {journal} {Phys. Rev. Lett.}\ }\textbf {\bibinfo {volume} {129}},\
  \bibinfo {pages} {086601} (\bibinfo {year} {2022})}\BibitemShut {NoStop}%
\bibitem [{\citenamefont {Zhang}\ \emph {et~al.}(2022)\citenamefont {Zhang},
  \citenamefont {Yang},\ and\ \citenamefont {Fang}}]{zhang2022universal}%
  \BibitemOpen
  \bibfield  {author} {\bibinfo {author} {\bibfnamefont {K.}~\bibnamefont
  {Zhang}}, \bibinfo {author} {\bibfnamefont {Z.}~\bibnamefont {Yang}},\ and\
  \bibinfo {author} {\bibfnamefont {C.}~\bibnamefont {Fang}},\ }\bibfield
  {title} {\bibinfo {title} {{Universal non-Hermitian skin effect in two and
  higher dimensions}},\ }\href
  {https://www.nature.com/articles/s41467-022-30161-6} {\bibfield  {journal}
  {\bibinfo  {journal} {Nat. Commun.}\ }\textbf {\bibinfo {volume} {13}},\
  \bibinfo {pages} {2496} (\bibinfo {year} {2022})}\BibitemShut {NoStop}%
\bibitem [{\citenamefont {Jin}\ \emph {et~al.}(2022)\citenamefont {Jin},
  \citenamefont {Zhong}, \citenamefont {Cai}, \citenamefont {Zhuang},
  \citenamefont {Pennec},\ and\ \citenamefont
  {Djafari-Rouhani}}]{doi:10.1063/5.0097530}%
  \BibitemOpen
  \bibfield  {author} {\bibinfo {author} {\bibfnamefont {Y.}~\bibnamefont
  {Jin}}, \bibinfo {author} {\bibfnamefont {W.}~\bibnamefont {Zhong}}, \bibinfo
  {author} {\bibfnamefont {R.}~\bibnamefont {Cai}}, \bibinfo {author}
  {\bibfnamefont {X.}~\bibnamefont {Zhuang}}, \bibinfo {author} {\bibfnamefont
  {Y.}~\bibnamefont {Pennec}},\ and\ \bibinfo {author} {\bibfnamefont
  {B.}~\bibnamefont {Djafari-Rouhani}},\ }\bibfield  {title} {\bibinfo {title}
  {{Non-Hermitian skin effect in a phononic beam based on piezoelectric
  feedback control}},\ }\href {https://doi.org/10.1063/5.0097530} {\bibfield
  {journal} {\bibinfo  {journal} {Appl. Phys. Lett.}\ }\textbf {\bibinfo
  {volume} {121}},\ \bibinfo {pages} {022202} (\bibinfo {year}
  {2022})}\BibitemShut {NoStop}%
\bibitem [{\citenamefont {Alsallom}\ \emph {et~al.}(2022)\citenamefont
  {Alsallom}, \citenamefont {Herviou}, \citenamefont {Yazyev},\ and\
  \citenamefont {Brzezi\ifmmode~\acute{n}\else
  \'{n}\fi{}ska}}]{PhysRevResearch.4.033122}%
  \BibitemOpen
  \bibfield  {author} {\bibinfo {author} {\bibfnamefont {F.}~\bibnamefont
  {Alsallom}}, \bibinfo {author} {\bibfnamefont {L.}~\bibnamefont {Herviou}},
  \bibinfo {author} {\bibfnamefont {O.~V.}\ \bibnamefont {Yazyev}},\ and\
  \bibinfo {author} {\bibfnamefont {M.}~\bibnamefont
  {Brzezi\ifmmode~\acute{n}\else \'{n}\fi{}ska}},\ }\bibfield  {title}
  {\bibinfo {title} {{Fate of the non-Hermitian skin effect in many-body
  fermionic systems}},\ }\href
  {https://doi.org/10.1103/PhysRevResearch.4.033122} {\bibfield  {journal}
  {\bibinfo  {journal} {Phys. Rev. Res.}\ }\textbf {\bibinfo {volume} {4}},\
  \bibinfo {pages} {033122} (\bibinfo {year} {2022})}\BibitemShut {NoStop}%
\bibitem [{\citenamefont {Kawabata}\ \emph {et~al.}(2023)\citenamefont
  {Kawabata}, \citenamefont {Numasawa},\ and\ \citenamefont
  {Ryu}}]{PhysRevX.13.021007}%
  \BibitemOpen
  \bibfield  {author} {\bibinfo {author} {\bibfnamefont {K.}~\bibnamefont
  {Kawabata}}, \bibinfo {author} {\bibfnamefont {T.}~\bibnamefont {Numasawa}},\
  and\ \bibinfo {author} {\bibfnamefont {S.}~\bibnamefont {Ryu}},\ }\bibfield
  {title} {\bibinfo {title} {{Entanglement Phase Transition Induced by the
  Non-Hermitian Skin Effect}},\ }\href
  {https://doi.org/10.1103/PhysRevX.13.021007} {\bibfield  {journal} {\bibinfo
  {journal} {Phys. Rev. X}\ }\textbf {\bibinfo {volume} {13}},\ \bibinfo
  {pages} {021007} (\bibinfo {year} {2023})}\BibitemShut {NoStop}%
\bibitem [{\citenamefont {Tanaka}\ \emph {et~al.}(2024)\citenamefont {Tanaka},
  \citenamefont {Takahashi},\ and\ \citenamefont
  {Okugawa}}]{PhysRevB.109.035131}%
  \BibitemOpen
  \bibfield  {author} {\bibinfo {author} {\bibfnamefont {Y.}~\bibnamefont
  {Tanaka}}, \bibinfo {author} {\bibfnamefont {R.}~\bibnamefont {Takahashi}},\
  and\ \bibinfo {author} {\bibfnamefont {R.}~\bibnamefont {Okugawa}},\
  }\bibfield  {title} {\bibinfo {title} {{Non-Hermitian skin effect enforced by
  nonsymmorphic symmetries}},\ }\href
  {https://doi.org/10.1103/PhysRevB.109.035131} {\bibfield  {journal} {\bibinfo
   {journal} {Phys. Rev. B}\ }\textbf {\bibinfo {volume} {109}},\ \bibinfo
  {pages} {035131} (\bibinfo {year} {2024})}\BibitemShut {NoStop}%
\bibitem [{\citenamefont {Nakai}\ \emph {et~al.}(2024)\citenamefont {Nakai},
  \citenamefont {Okuma}, \citenamefont {Nakamura}, \citenamefont {Shimomura},\
  and\ \citenamefont {Sato}}]{PhysRevB.109.144203}%
  \BibitemOpen
  \bibfield  {author} {\bibinfo {author} {\bibfnamefont {Y.~O.}\ \bibnamefont
  {Nakai}}, \bibinfo {author} {\bibfnamefont {N.}~\bibnamefont {Okuma}},
  \bibinfo {author} {\bibfnamefont {D.}~\bibnamefont {Nakamura}}, \bibinfo
  {author} {\bibfnamefont {K.}~\bibnamefont {Shimomura}},\ and\ \bibinfo
  {author} {\bibfnamefont {M.}~\bibnamefont {Sato}},\ }\bibfield  {title}
  {\bibinfo {title} {{Topological enhancement of nonnormality in non-Hermitian
  skin effects}},\ }\href {https://doi.org/10.1103/PhysRevB.109.144203}
  {\bibfield  {journal} {\bibinfo  {journal} {Phys. Rev. B}\ }\textbf {\bibinfo
  {volume} {109}},\ \bibinfo {pages} {144203} (\bibinfo {year}
  {2024})}\BibitemShut {NoStop}%
\bibitem [{\citenamefont {Shimomura}\ and\ \citenamefont
  {Sato}(2024)}]{PhysRevLett.133.136502}%
  \BibitemOpen
  \bibfield  {author} {\bibinfo {author} {\bibfnamefont {K.}~\bibnamefont
  {Shimomura}}\ and\ \bibinfo {author} {\bibfnamefont {M.}~\bibnamefont
  {Sato}},\ }\bibfield  {title} {\bibinfo {title} {{General Criterion for
  Non-Hermitian Skin Effects and Application: Fock Space Skin Effects in
  Many-Body Systems}},\ }\href {https://doi.org/10.1103/PhysRevLett.133.136502}
  {\bibfield  {journal} {\bibinfo  {journal} {Phys. Rev. Lett.}\ }\textbf
  {\bibinfo {volume} {133}},\ \bibinfo {pages} {136502} (\bibinfo {year}
  {2024})}\BibitemShut {NoStop}%
\bibitem [{\citenamefont {Peters}\ and\ \citenamefont
  {Yoshida}(2024)}]{peters2024hinge}%
  \BibitemOpen
  \bibfield  {author} {\bibinfo {author} {\bibfnamefont {R.}~\bibnamefont
  {Peters}}\ and\ \bibinfo {author} {\bibfnamefont {T.}~\bibnamefont
  {Yoshida}},\ }\bibfield  {title} {\bibinfo {title} {{Hinge non-Hermitian skin
  effect in the single-particle properties of a strongly correlated
  $f$-electron system}},\ }\href {https://doi.org/10.1103/PhysRevB.110.125114}
  {\bibfield  {journal} {\bibinfo  {journal} {Phys. Rev. B}\ }\textbf {\bibinfo
  {volume} {110}},\ \bibinfo {pages} {125114} (\bibinfo {year}
  {2024})}\BibitemShut {NoStop}%
\bibitem [{\citenamefont {Ishikawa}\ and\ \citenamefont
  {Yoshida}(2024)}]{ishikawa2024z4skin}%
  \BibitemOpen
  \bibfield  {author} {\bibinfo {author} {\bibfnamefont {S.}~\bibnamefont
  {Ishikawa}}\ and\ \bibinfo {author} {\bibfnamefont {T.}~\bibnamefont
  {Yoshida}},\ }\bibfield  {title} {\bibinfo {title} {{Non-Hermitian
  ${\mathbb{Z}}_{4}$ skin effect protected by glide symmetry}},\ }\href
  {https://doi.org/10.1103/PhysRevB.110.115301} {\bibfield  {journal} {\bibinfo
   {journal} {Phys. Rev. B}\ }\textbf {\bibinfo {volume} {110}},\ \bibinfo
  {pages} {115301} (\bibinfo {year} {2024})}\BibitemShut {NoStop}%
\bibitem [{\citenamefont {Kawabata}\ \emph
  {et~al.}(2020{\natexlab{b}})\citenamefont {Kawabata}, \citenamefont {Sato},\
  and\ \citenamefont {Shiozaki}}]{PhysRevB.102.205118}%
  \BibitemOpen
  \bibfield  {author} {\bibinfo {author} {\bibfnamefont {K.}~\bibnamefont
  {Kawabata}}, \bibinfo {author} {\bibfnamefont {M.}~\bibnamefont {Sato}},\
  and\ \bibinfo {author} {\bibfnamefont {K.}~\bibnamefont {Shiozaki}},\
  }\bibfield  {title} {\bibinfo {title} {{Higher-order non-Hermitian skin
  effect}},\ }\href {https://doi.org/10.1103/PhysRevB.102.205118} {\bibfield
  {journal} {\bibinfo  {journal} {Phys. Rev. B}\ }\textbf {\bibinfo {volume}
  {102}},\ \bibinfo {pages} {205118} (\bibinfo {year}
  {2020}{\natexlab{b}})}\BibitemShut {NoStop}%
\bibitem [{\citenamefont {Okugawa}\ \emph {et~al.}(2020)\citenamefont
  {Okugawa}, \citenamefont {Takahashi},\ and\ \citenamefont
  {Yokomizo}}]{PhysRevB.102.241202}%
  \BibitemOpen
  \bibfield  {author} {\bibinfo {author} {\bibfnamefont {R.}~\bibnamefont
  {Okugawa}}, \bibinfo {author} {\bibfnamefont {R.}~\bibnamefont {Takahashi}},\
  and\ \bibinfo {author} {\bibfnamefont {K.}~\bibnamefont {Yokomizo}},\
  }\bibfield  {title} {\bibinfo {title} {{Second-order topological
  non-Hermitian skin effects}},\ }\href
  {https://doi.org/10.1103/PhysRevB.102.241202} {\bibfield  {journal} {\bibinfo
   {journal} {Phys. Rev. B}\ }\textbf {\bibinfo {volume} {102}},\ \bibinfo
  {pages} {241202(R)} (\bibinfo {year} {2020})}\BibitemShut {NoStop}%
\bibitem [{\citenamefont {Fu}\ \emph {et~al.}(2021)\citenamefont {Fu},
  \citenamefont {Hu},\ and\ \citenamefont {Wan}}]{PhysRevB.103.045420}%
  \BibitemOpen
  \bibfield  {author} {\bibinfo {author} {\bibfnamefont {Y.}~\bibnamefont
  {Fu}}, \bibinfo {author} {\bibfnamefont {J.}~\bibnamefont {Hu}},\ and\
  \bibinfo {author} {\bibfnamefont {S.}~\bibnamefont {Wan}},\ }\bibfield
  {title} {\bibinfo {title} {{Non-Hermitian second-order skin and topological
  modes}},\ }\href {https://doi.org/10.1103/PhysRevB.103.045420} {\bibfield
  {journal} {\bibinfo  {journal} {Phys. Rev. B}\ }\textbf {\bibinfo {volume}
  {103}},\ \bibinfo {pages} {045420} (\bibinfo {year} {2021})}\BibitemShut
  {NoStop}%
\bibitem [{\citenamefont {Okugawa}\ \emph {et~al.}(2021)\citenamefont
  {Okugawa}, \citenamefont {Takahashi},\ and\ \citenamefont
  {Yokomizo}}]{PhysRevB.103.205205}%
  \BibitemOpen
  \bibfield  {author} {\bibinfo {author} {\bibfnamefont {R.}~\bibnamefont
  {Okugawa}}, \bibinfo {author} {\bibfnamefont {R.}~\bibnamefont {Takahashi}},\
  and\ \bibinfo {author} {\bibfnamefont {K.}~\bibnamefont {Yokomizo}},\
  }\bibfield  {title} {\bibinfo {title} {{Non-Hermitian band topology with
  generalized inversion symmetry}},\ }\href
  {https://doi.org/10.1103/PhysRevB.103.205205} {\bibfield  {journal} {\bibinfo
   {journal} {Phys. Rev. B}\ }\textbf {\bibinfo {volume} {103}},\ \bibinfo
  {pages} {205205} (\bibinfo {year} {2021})}\BibitemShut {NoStop}%
\bibitem [{\citenamefont {Shiozaki}\ and\ \citenamefont
  {Ono}(2021)}]{PhysRevB.104.035424}%
  \BibitemOpen
  \bibfield  {author} {\bibinfo {author} {\bibfnamefont {K.}~\bibnamefont
  {Shiozaki}}\ and\ \bibinfo {author} {\bibfnamefont {S.}~\bibnamefont {Ono}},\
  }\bibfield  {title} {\bibinfo {title} {{Symmetry indicator in non-Hermitian
  systems}},\ }\href {https://doi.org/10.1103/PhysRevB.104.035424} {\bibfield
  {journal} {\bibinfo  {journal} {Phys. Rev. B}\ }\textbf {\bibinfo {volume}
  {104}},\ \bibinfo {pages} {035424} (\bibinfo {year} {2021})}\BibitemShut
  {NoStop}%
\bibitem [{\citenamefont {Li}\ \emph {et~al.}(2023)\citenamefont {Li},
  \citenamefont {Trauzettel}, \citenamefont {Neupert},\ and\ \citenamefont
  {Zhang}}]{PhysRevLett.131.116601}%
  \BibitemOpen
  \bibfield  {author} {\bibinfo {author} {\bibfnamefont {C.-A.}\ \bibnamefont
  {Li}}, \bibinfo {author} {\bibfnamefont {B.}~\bibnamefont {Trauzettel}},
  \bibinfo {author} {\bibfnamefont {T.}~\bibnamefont {Neupert}},\ and\ \bibinfo
  {author} {\bibfnamefont {S.-B.}\ \bibnamefont {Zhang}},\ }\bibfield  {title}
  {\bibinfo {title} {{Enhancement of Second-Order Non-Hermitian Skin Effect by
  Magnetic Fields}},\ }\href {https://doi.org/10.1103/PhysRevLett.131.116601}
  {\bibfield  {journal} {\bibinfo  {journal} {Phys. Rev. Lett.}\ }\textbf
  {\bibinfo {volume} {131}},\ \bibinfo {pages} {116601} (\bibinfo {year}
  {2023})}\BibitemShut {NoStop}%
\bibitem [{\citenamefont {Liu}\ \emph {et~al.}(2023)\citenamefont {Liu},
  \citenamefont {Hu}, \citenamefont {Chen},\ and\ \citenamefont
  {Liu}}]{PhysRevB.108.174307}%
  \BibitemOpen
  \bibfield  {author} {\bibinfo {author} {\bibfnamefont {C.-H.}\ \bibnamefont
  {Liu}}, \bibinfo {author} {\bibfnamefont {H.}~\bibnamefont {Hu}}, \bibinfo
  {author} {\bibfnamefont {S.}~\bibnamefont {Chen}},\ and\ \bibinfo {author}
  {\bibfnamefont {X.-J.}\ \bibnamefont {Liu}},\ }\bibfield  {title} {\bibinfo
  {title} {{Anomalous second-order skin modes in Floquet non-Hermitian
  systems}},\ }\href {https://doi.org/10.1103/PhysRevB.108.174307} {\bibfield
  {journal} {\bibinfo  {journal} {Phys. Rev. B}\ }\textbf {\bibinfo {volume}
  {108}},\ \bibinfo {pages} {174307} (\bibinfo {year} {2023})}\BibitemShut
  {NoStop}%
\bibitem [{\citenamefont {Sun}\ \emph {et~al.}(2021)\citenamefont {Sun},
  \citenamefont {Zhu},\ and\ \citenamefont {Hughes}}]{PhysRevLett.127.066401}%
  \BibitemOpen
  \bibfield  {author} {\bibinfo {author} {\bibfnamefont {X.-Q.}\ \bibnamefont
  {Sun}}, \bibinfo {author} {\bibfnamefont {P.}~\bibnamefont {Zhu}},\ and\
  \bibinfo {author} {\bibfnamefont {T.~L.}\ \bibnamefont {Hughes}},\ }\bibfield
   {title} {\bibinfo {title} {{Geometric Response and Disclination-Induced Skin
  Effects in Non-Hermitian Systems}},\ }\href
  {https://doi.org/10.1103/PhysRevLett.127.066401} {\bibfield  {journal}
  {\bibinfo  {journal} {Phys. Rev. Lett.}\ }\textbf {\bibinfo {volume} {127}},\
  \bibinfo {pages} {066401} (\bibinfo {year} {2021})}\BibitemShut {NoStop}%
\bibitem [{\citenamefont {Schindler}\ and\ \citenamefont
  {Prem}(2021)}]{PhysRevB.104.L161106}%
  \BibitemOpen
  \bibfield  {author} {\bibinfo {author} {\bibfnamefont {F.}~\bibnamefont
  {Schindler}}\ and\ \bibinfo {author} {\bibfnamefont {A.}~\bibnamefont
  {Prem}},\ }\bibfield  {title} {\bibinfo {title} {{Dislocation non-Hermitian
  skin effect}},\ }\href {https://doi.org/10.1103/PhysRevB.104.L161106}
  {\bibfield  {journal} {\bibinfo  {journal} {Phys. Rev. B}\ }\textbf {\bibinfo
  {volume} {104}},\ \bibinfo {pages} {L161106} (\bibinfo {year}
  {2021})}\BibitemShut {NoStop}%
\bibitem [{\citenamefont {Bhargava}\ \emph {et~al.}(2021)\citenamefont
  {Bhargava}, \citenamefont {Fulga}, \citenamefont {van~den Brink},\ and\
  \citenamefont {Moghaddam}}]{PhysRevB.104.L241402}%
  \BibitemOpen
  \bibfield  {author} {\bibinfo {author} {\bibfnamefont {B.~A.}\ \bibnamefont
  {Bhargava}}, \bibinfo {author} {\bibfnamefont {I.~C.}\ \bibnamefont {Fulga}},
  \bibinfo {author} {\bibfnamefont {J.}~\bibnamefont {van~den Brink}},\ and\
  \bibinfo {author} {\bibfnamefont {A.~G.}\ \bibnamefont {Moghaddam}},\
  }\bibfield  {title} {\bibinfo {title} {{Non-Hermitian skin effect of
  dislocations and its topological origin}},\ }\href
  {https://doi.org/10.1103/PhysRevB.104.L241402} {\bibfield  {journal}
  {\bibinfo  {journal} {Phys. Rev. B}\ }\textbf {\bibinfo {volume} {104}},\
  \bibinfo {pages} {L241402} (\bibinfo {year} {2021})}\BibitemShut {NoStop}%
\bibitem [{\citenamefont {Xiao}\ \emph {et~al.}(2020)\citenamefont {Xiao},
  \citenamefont {Deng}, \citenamefont {Wang}, \citenamefont {Zhu},
  \citenamefont {Wang}, \citenamefont {Yi},\ and\ \citenamefont
  {Xue}}]{xiao2020non}%
  \BibitemOpen
  \bibfield  {author} {\bibinfo {author} {\bibfnamefont {L.}~\bibnamefont
  {Xiao}}, \bibinfo {author} {\bibfnamefont {T.}~\bibnamefont {Deng}}, \bibinfo
  {author} {\bibfnamefont {K.}~\bibnamefont {Wang}}, \bibinfo {author}
  {\bibfnamefont {G.}~\bibnamefont {Zhu}}, \bibinfo {author} {\bibfnamefont
  {Z.}~\bibnamefont {Wang}}, \bibinfo {author} {\bibfnamefont {W.}~\bibnamefont
  {Yi}},\ and\ \bibinfo {author} {\bibfnamefont {P.}~\bibnamefont {Xue}},\
  }\bibfield  {title} {\bibinfo {title} {{Non-Hermitian bulk--boundary
  correspondence in quantum dynamics}},\ }\href
  {https://www.nature.com/articles/s41567-020-0836-6} {\bibfield  {journal}
  {\bibinfo  {journal} {Nat. Phys.}\ }\textbf {\bibinfo {volume} {16}},\
  \bibinfo {pages} {761} (\bibinfo {year} {2020})}\BibitemShut {NoStop}%
\bibitem [{\citenamefont {Weidemann}\ \emph {et~al.}(2020)\citenamefont
  {Weidemann}, \citenamefont {Kremer}, \citenamefont {Helbig}, \citenamefont
  {Hofmann}, \citenamefont {Stegmaier}, \citenamefont {Greiter}, \citenamefont
  {Thomale},\ and\ \citenamefont {Szameit}}]{weidemann2020topological}%
  \BibitemOpen
  \bibfield  {author} {\bibinfo {author} {\bibfnamefont {S.}~\bibnamefont
  {Weidemann}}, \bibinfo {author} {\bibfnamefont {M.}~\bibnamefont {Kremer}},
  \bibinfo {author} {\bibfnamefont {T.}~\bibnamefont {Helbig}}, \bibinfo
  {author} {\bibfnamefont {T.}~\bibnamefont {Hofmann}}, \bibinfo {author}
  {\bibfnamefont {A.}~\bibnamefont {Stegmaier}}, \bibinfo {author}
  {\bibfnamefont {M.}~\bibnamefont {Greiter}}, \bibinfo {author} {\bibfnamefont
  {R.}~\bibnamefont {Thomale}},\ and\ \bibinfo {author} {\bibfnamefont
  {A.}~\bibnamefont {Szameit}},\ }\bibfield  {title} {\bibinfo {title}
  {{Topological funneling of light}},\ }\href
  {https://www.science.org/doi/full/10.1126/science.aaz8727} {\bibfield
  {journal} {\bibinfo  {journal} {Science}\ }\textbf {\bibinfo {volume}
  {368}},\ \bibinfo {pages} {311} (\bibinfo {year} {2020})}\BibitemShut
  {NoStop}%
\bibitem [{\citenamefont {Helbig}\ \emph {et~al.}(2020)\citenamefont {Helbig},
  \citenamefont {Hofmann}, \citenamefont {Imhof}, \citenamefont {Abdelghany},
  \citenamefont {Kiessling}, \citenamefont {Molenkamp}, \citenamefont {Lee},
  \citenamefont {Szameit}, \citenamefont {Greiter},\ and\ \citenamefont
  {Thomale}}]{helbig2020generalized}%
  \BibitemOpen
  \bibfield  {author} {\bibinfo {author} {\bibfnamefont {T.}~\bibnamefont
  {Helbig}}, \bibinfo {author} {\bibfnamefont {T.}~\bibnamefont {Hofmann}},
  \bibinfo {author} {\bibfnamefont {S.}~\bibnamefont {Imhof}}, \bibinfo
  {author} {\bibfnamefont {M.}~\bibnamefont {Abdelghany}}, \bibinfo {author}
  {\bibfnamefont {T.}~\bibnamefont {Kiessling}}, \bibinfo {author}
  {\bibfnamefont {L.}~\bibnamefont {Molenkamp}}, \bibinfo {author}
  {\bibfnamefont {C.}~\bibnamefont {Lee}}, \bibinfo {author} {\bibfnamefont
  {A.}~\bibnamefont {Szameit}}, \bibinfo {author} {\bibfnamefont
  {M.}~\bibnamefont {Greiter}},\ and\ \bibinfo {author} {\bibfnamefont
  {R.}~\bibnamefont {Thomale}},\ }\bibfield  {title} {\bibinfo {title}
  {{Generalized bulk--boundary correspondence in non-Hermitian topolectrical
  circuits}},\ }\href {https://www.nature.com/articles/s41567-020-0922-9}
  {\bibfield  {journal} {\bibinfo  {journal} {Nat. Phys.}\ }\textbf {\bibinfo
  {volume} {16}},\ \bibinfo {pages} {747} (\bibinfo {year} {2020})}\BibitemShut
  {NoStop}%
\bibitem [{\citenamefont {Brandenbourger}\ \emph {et~al.}(2019)\citenamefont
  {Brandenbourger}, \citenamefont {Locsin}, \citenamefont {Lerner},\ and\
  \citenamefont {Coulais}}]{brandenbourger2019non}%
  \BibitemOpen
  \bibfield  {author} {\bibinfo {author} {\bibfnamefont {M.}~\bibnamefont
  {Brandenbourger}}, \bibinfo {author} {\bibfnamefont {X.}~\bibnamefont
  {Locsin}}, \bibinfo {author} {\bibfnamefont {E.}~\bibnamefont {Lerner}},\
  and\ \bibinfo {author} {\bibfnamefont {C.}~\bibnamefont {Coulais}},\
  }\bibfield  {title} {\bibinfo {title} {{Non-reciprocal robotic
  metamaterials}},\ }\href {https://www.nature.com/articles/s41467-019-12599-3}
  {\bibfield  {journal} {\bibinfo  {journal} {Nat. Commun.}\ }\textbf {\bibinfo
  {volume} {10}},\ \bibinfo {pages} {4608} (\bibinfo {year}
  {2019})}\BibitemShut {NoStop}%
\bibitem [{\citenamefont {Chen}\ \emph {et~al.}(2021)\citenamefont {Chen},
  \citenamefont {Li}, \citenamefont {Scheibner}, \citenamefont {Vitelli},\ and\
  \citenamefont {Huang}}]{chen2021realization}%
  \BibitemOpen
  \bibfield  {author} {\bibinfo {author} {\bibfnamefont {Y.}~\bibnamefont
  {Chen}}, \bibinfo {author} {\bibfnamefont {X.}~\bibnamefont {Li}}, \bibinfo
  {author} {\bibfnamefont {C.}~\bibnamefont {Scheibner}}, \bibinfo {author}
  {\bibfnamefont {V.}~\bibnamefont {Vitelli}},\ and\ \bibinfo {author}
  {\bibfnamefont {G.}~\bibnamefont {Huang}},\ }\bibfield  {title} {\bibinfo
  {title} {{Realization of active metamaterials with odd micropolar
  elasticity}},\ }\href {https://www.nature.com/articles/s41467-021-26034-z}
  {\bibfield  {journal} {\bibinfo  {journal} {Nat. Commun.}\ }\textbf {\bibinfo
  {volume} {12}},\ \bibinfo {pages} {5935} (\bibinfo {year}
  {2021})}\BibitemShut {NoStop}%
\bibitem [{\citenamefont {Zhang}\ \emph {et~al.}(2021)\citenamefont {Zhang},
  \citenamefont {Yang}, \citenamefont {Ge}, \citenamefont {Guan}, \citenamefont
  {Chen}, \citenamefont {Yan}, \citenamefont {Chen}, \citenamefont {Xi},
  \citenamefont {Li}, \citenamefont {Jia}, \citenamefont {Yuan}, \citenamefont
  {Sun}, \citenamefont {Chen},\ and\ \citenamefont
  {Zhang}}]{zhang2021acoustic}%
  \BibitemOpen
  \bibfield  {author} {\bibinfo {author} {\bibfnamefont {L.}~\bibnamefont
  {Zhang}}, \bibinfo {author} {\bibfnamefont {Y.}~\bibnamefont {Yang}},
  \bibinfo {author} {\bibfnamefont {Y.}~\bibnamefont {Ge}}, \bibinfo {author}
  {\bibfnamefont {Y.-J.}\ \bibnamefont {Guan}}, \bibinfo {author}
  {\bibfnamefont {Q.}~\bibnamefont {Chen}}, \bibinfo {author} {\bibfnamefont
  {Q.}~\bibnamefont {Yan}}, \bibinfo {author} {\bibfnamefont {F.}~\bibnamefont
  {Chen}}, \bibinfo {author} {\bibfnamefont {R.}~\bibnamefont {Xi}}, \bibinfo
  {author} {\bibfnamefont {Y.}~\bibnamefont {Li}}, \bibinfo {author}
  {\bibfnamefont {D.}~\bibnamefont {Jia}}, \bibinfo {author} {\bibfnamefont
  {S.-Q.}\ \bibnamefont {Yuan}}, \bibinfo {author} {\bibfnamefont {H.-X.}\
  \bibnamefont {Sun}}, \bibinfo {author} {\bibfnamefont {H.}~\bibnamefont
  {Chen}},\ and\ \bibinfo {author} {\bibfnamefont {B.}~\bibnamefont {Zhang}},\
  }\bibfield  {title} {\bibinfo {title} {{Acoustic non-Hermitian skin effect
  from twisted winding topology}},\ }\href
  {https://www.nature.com/articles/s41467-021-26619-8} {\bibfield  {journal}
  {\bibinfo  {journal} {Nat. Commun.}\ }\textbf {\bibinfo {volume} {12}},\
  \bibinfo {pages} {6297} (\bibinfo {year} {2021})}\BibitemShut {NoStop}%
\bibitem [{\citenamefont {Hofmann}\ \emph {et~al.}(2020)\citenamefont
  {Hofmann}, \citenamefont {Helbig}, \citenamefont {Schindler}, \citenamefont
  {Salgo}, \citenamefont {Brzezi\ifmmode~\acute{n}\else \'{n}\fi{}ska},
  \citenamefont {Greiter}, \citenamefont {Kiessling}, \citenamefont {Wolf},
  \citenamefont {Vollhardt}, \citenamefont {Kaba\ifmmode~\check{s}\else
  \v{s}\fi{}i}, \citenamefont {Lee}, \citenamefont {Bilu\ifmmode \check{s}\else
  \v{s}\fi{}i\ifmmode~\acute{c}\else \'{c}\fi{}}, \citenamefont {Thomale},\
  and\ \citenamefont {Neupert}}]{PhysRevResearch.2.023265}%
  \BibitemOpen
  \bibfield  {author} {\bibinfo {author} {\bibfnamefont {T.}~\bibnamefont
  {Hofmann}}, \bibinfo {author} {\bibfnamefont {T.}~\bibnamefont {Helbig}},
  \bibinfo {author} {\bibfnamefont {F.}~\bibnamefont {Schindler}}, \bibinfo
  {author} {\bibfnamefont {N.}~\bibnamefont {Salgo}}, \bibinfo {author}
  {\bibfnamefont {M.}~\bibnamefont {Brzezi\ifmmode~\acute{n}\else
  \'{n}\fi{}ska}}, \bibinfo {author} {\bibfnamefont {M.}~\bibnamefont
  {Greiter}}, \bibinfo {author} {\bibfnamefont {T.}~\bibnamefont {Kiessling}},
  \bibinfo {author} {\bibfnamefont {D.}~\bibnamefont {Wolf}}, \bibinfo {author}
  {\bibfnamefont {A.}~\bibnamefont {Vollhardt}}, \bibinfo {author}
  {\bibfnamefont {A.}~\bibnamefont {Kaba\ifmmode~\check{s}\else \v{s}\fi{}i}},
  \bibinfo {author} {\bibfnamefont {C.~H.}\ \bibnamefont {Lee}}, \bibinfo
  {author} {\bibfnamefont {A.}~\bibnamefont {Bilu\ifmmode \check{s}\else
  \v{s}\fi{}i\ifmmode~\acute{c}\else \'{c}\fi{}}}, \bibinfo {author}
  {\bibfnamefont {R.}~\bibnamefont {Thomale}},\ and\ \bibinfo {author}
  {\bibfnamefont {T.}~\bibnamefont {Neupert}},\ }\bibfield  {title} {\bibinfo
  {title} {{Reciprocal skin effect and its realization in a topolectrical
  circuit}},\ }\href {https://doi.org/10.1103/PhysRevResearch.2.023265}
  {\bibfield  {journal} {\bibinfo  {journal} {Phys. Rev. Res.}\ }\textbf
  {\bibinfo {volume} {2}},\ \bibinfo {pages} {023265} (\bibinfo {year}
  {2020})}\BibitemShut {NoStop}%
\bibitem [{\citenamefont {Palacios}\ \emph {et~al.}(2021)\citenamefont
  {Palacios}, \citenamefont {Tchoumakov}, \citenamefont {Guix}, \citenamefont
  {Pagonabarraga}, \citenamefont {S{\'a}nchez},\ and\ \citenamefont
  {G~Grushin}}]{palacios2021guided}%
  \BibitemOpen
  \bibfield  {author} {\bibinfo {author} {\bibfnamefont {L.~S.}\ \bibnamefont
  {Palacios}}, \bibinfo {author} {\bibfnamefont {S.}~\bibnamefont
  {Tchoumakov}}, \bibinfo {author} {\bibfnamefont {M.}~\bibnamefont {Guix}},
  \bibinfo {author} {\bibfnamefont {I.}~\bibnamefont {Pagonabarraga}}, \bibinfo
  {author} {\bibfnamefont {S.}~\bibnamefont {S{\'a}nchez}},\ and\ \bibinfo
  {author} {\bibfnamefont {A.}~\bibnamefont {G~Grushin}},\ }\bibfield  {title}
  {\bibinfo {title} {{Guided accumulation of active particles by topological
  design of a second-order skin effect}},\ }\href
  {https://www.nature.com/articles/s41467-021-24948-2} {\bibfield  {journal}
  {\bibinfo  {journal} {Nat. Commun.}\ }\textbf {\bibinfo {volume} {12}},\
  \bibinfo {pages} {4691} (\bibinfo {year} {2021})}\BibitemShut {NoStop}%
\bibitem [{\citenamefont {Shang}\ \emph {et~al.}(2022)\citenamefont {Shang},
  \citenamefont {Liu}, \citenamefont {Shao}, \citenamefont {Han}, \citenamefont
  {Zang}, \citenamefont {Zhang}, \citenamefont {Salama}, \citenamefont {Gao},
  \citenamefont {Lee}, \citenamefont {Thomale}, \citenamefont {Manchon},
  \citenamefont {Zhang}, \citenamefont {Cui},\ and\ \citenamefont
  {Schwingenschlögl}}]{shang2022experimental}%
  \BibitemOpen
  \bibfield  {author} {\bibinfo {author} {\bibfnamefont {C.}~\bibnamefont
  {Shang}}, \bibinfo {author} {\bibfnamefont {S.}~\bibnamefont {Liu}}, \bibinfo
  {author} {\bibfnamefont {R.}~\bibnamefont {Shao}}, \bibinfo {author}
  {\bibfnamefont {P.}~\bibnamefont {Han}}, \bibinfo {author} {\bibfnamefont
  {X.}~\bibnamefont {Zang}}, \bibinfo {author} {\bibfnamefont {X.}~\bibnamefont
  {Zhang}}, \bibinfo {author} {\bibfnamefont {K.~N.}\ \bibnamefont {Salama}},
  \bibinfo {author} {\bibfnamefont {W.}~\bibnamefont {Gao}}, \bibinfo {author}
  {\bibfnamefont {C.~H.}\ \bibnamefont {Lee}}, \bibinfo {author} {\bibfnamefont
  {R.}~\bibnamefont {Thomale}}, \bibinfo {author} {\bibfnamefont
  {A.}~\bibnamefont {Manchon}}, \bibinfo {author} {\bibfnamefont
  {S.}~\bibnamefont {Zhang}}, \bibinfo {author} {\bibfnamefont {T.~J.}\
  \bibnamefont {Cui}},\ and\ \bibinfo {author} {\bibfnamefont {U.}~\bibnamefont
  {Schwingenschlögl}},\ }\bibfield  {title} {\bibinfo {title} {{Experimental
  Identification of the Second-Order Non-Hermitian Skin Effect with
  Physics-Graph-Informed Machine Learning}},\ }\href
  {https://doi.org/https://doi.org/10.1002/advs.202202922} {\bibfield
  {journal} {\bibinfo  {journal} {Adv. Sci.}\ }\textbf {\bibinfo {volume}
  {9}},\ \bibinfo {pages} {2202922} (\bibinfo {year} {2022})}\BibitemShut
  {NoStop}%
\bibitem [{\citenamefont {Wu}\ \emph {et~al.}(2024)\citenamefont {Wu},
  \citenamefont {Zheng}, \citenamefont {Liang}, \citenamefont {Ke},
  \citenamefont {Lu}, \citenamefont {Deng}, \citenamefont {Huang},\ and\
  \citenamefont {Liu}}]{wu2023spin}%
  \BibitemOpen
  \bibfield  {author} {\bibinfo {author} {\bibfnamefont {J.}~\bibnamefont
  {Wu}}, \bibinfo {author} {\bibfnamefont {R.}~\bibnamefont {Zheng}}, \bibinfo
  {author} {\bibfnamefont {J.}~\bibnamefont {Liang}}, \bibinfo {author}
  {\bibfnamefont {M.}~\bibnamefont {Ke}}, \bibinfo {author} {\bibfnamefont
  {J.}~\bibnamefont {Lu}}, \bibinfo {author} {\bibfnamefont {W.}~\bibnamefont
  {Deng}}, \bibinfo {author} {\bibfnamefont {X.}~\bibnamefont {Huang}},\ and\
  \bibinfo {author} {\bibfnamefont {Z.}~\bibnamefont {Liu}},\ }\bibfield
  {title} {\bibinfo {title} {{Spin-Dependent Localization of Helical Edge
  States in a Non-Hermitian Phononic Crystal}},\ }\href
  {https://doi.org/10.1103/PhysRevLett.133.126601} {\bibfield  {journal}
  {\bibinfo  {journal} {Phys. Rev. Lett.}\ }\textbf {\bibinfo {volume} {133}},\
  \bibinfo {pages} {126601} (\bibinfo {year} {2024})}\BibitemShut {NoStop}%
\bibitem [{\citenamefont {Sun}\ \emph {et~al.}(2024)\citenamefont {Sun},
  \citenamefont {Hou}, \citenamefont {Wan}, \citenamefont {Wang}, \citenamefont
  {Zhu}, \citenamefont {Ruan},\ and\ \citenamefont
  {Yang}}]{PhysRevLett.132.063804}%
  \BibitemOpen
  \bibfield  {author} {\bibinfo {author} {\bibfnamefont {Y.}~\bibnamefont
  {Sun}}, \bibinfo {author} {\bibfnamefont {X.}~\bibnamefont {Hou}}, \bibinfo
  {author} {\bibfnamefont {T.}~\bibnamefont {Wan}}, \bibinfo {author}
  {\bibfnamefont {F.}~\bibnamefont {Wang}}, \bibinfo {author} {\bibfnamefont
  {S.}~\bibnamefont {Zhu}}, \bibinfo {author} {\bibfnamefont {Z.}~\bibnamefont
  {Ruan}},\ and\ \bibinfo {author} {\bibfnamefont {Z.}~\bibnamefont {Yang}},\
  }\bibfield  {title} {\bibinfo {title} {{Photonic Floquet Skin-Topological
  Effect}},\ }\href {https://doi.org/10.1103/PhysRevLett.132.063804} {\bibfield
   {journal} {\bibinfo  {journal} {Phys. Rev. Lett.}\ }\textbf {\bibinfo
  {volume} {132}},\ \bibinfo {pages} {063804} (\bibinfo {year}
  {2024})}\BibitemShut {NoStop}%
\bibitem [{\citenamefont {Liu}\ \emph {et~al.}(2024)\citenamefont {Liu},
  \citenamefont {Mandal}, \citenamefont {Zhou}, \citenamefont {Xi},
  \citenamefont {Banerjee}, \citenamefont {Hu}, \citenamefont {Wei},
  \citenamefont {Wang}, \citenamefont {Wang}, \citenamefont {Gao},
  \citenamefont {Chen}, \citenamefont {Yang}, \citenamefont {Chong},\ and\
  \citenamefont {Zhang}}]{PhysRevLett.132.113802}%
  \BibitemOpen
  \bibfield  {author} {\bibinfo {author} {\bibfnamefont {G.-G.}\ \bibnamefont
  {Liu}}, \bibinfo {author} {\bibfnamefont {S.}~\bibnamefont {Mandal}},
  \bibinfo {author} {\bibfnamefont {P.}~\bibnamefont {Zhou}}, \bibinfo {author}
  {\bibfnamefont {X.}~\bibnamefont {Xi}}, \bibinfo {author} {\bibfnamefont
  {R.}~\bibnamefont {Banerjee}}, \bibinfo {author} {\bibfnamefont {Y.-H.}\
  \bibnamefont {Hu}}, \bibinfo {author} {\bibfnamefont {M.}~\bibnamefont
  {Wei}}, \bibinfo {author} {\bibfnamefont {M.}~\bibnamefont {Wang}}, \bibinfo
  {author} {\bibfnamefont {Q.}~\bibnamefont {Wang}}, \bibinfo {author}
  {\bibfnamefont {Z.}~\bibnamefont {Gao}}, \bibinfo {author} {\bibfnamefont
  {H.}~\bibnamefont {Chen}}, \bibinfo {author} {\bibfnamefont {Y.}~\bibnamefont
  {Yang}}, \bibinfo {author} {\bibfnamefont {Y.}~\bibnamefont {Chong}},\ and\
  \bibinfo {author} {\bibfnamefont {B.}~\bibnamefont {Zhang}},\ }\bibfield
  {title} {\bibinfo {title} {{Localization of Chiral Edge States by the
  Non-Hermitian Skin Effect}},\ }\href
  {https://doi.org/10.1103/PhysRevLett.132.113802} {\bibfield  {journal}
  {\bibinfo  {journal} {Phys. Rev. Lett.}\ }\textbf {\bibinfo {volume} {132}},\
  \bibinfo {pages} {113802} (\bibinfo {year} {2024})}\BibitemShut {NoStop}%
\bibitem [{\citenamefont {Zhong}\ \emph {et~al.}(2024)\citenamefont {Zhong},
  \citenamefont {de~Castro}, \citenamefont {Lu}, \citenamefont {Kim},
  \citenamefont {Oudich}, \citenamefont {Ji}, \citenamefont {Shi},
  \citenamefont {Chen}, \citenamefont {Lu}, \citenamefont {Jing},\ and\
  \citenamefont {Benalcazar}}]{zhong2024higher}%
  \BibitemOpen
  \bibfield  {author} {\bibinfo {author} {\bibfnamefont {J.-X.}\ \bibnamefont
  {Zhong}}, \bibinfo {author} {\bibfnamefont {P.~F.}\ \bibnamefont
  {de~Castro}}, \bibinfo {author} {\bibfnamefont {T.}~\bibnamefont {Lu}},
  \bibinfo {author} {\bibfnamefont {J.}~\bibnamefont {Kim}}, \bibinfo {author}
  {\bibfnamefont {M.}~\bibnamefont {Oudich}}, \bibinfo {author} {\bibfnamefont
  {J.}~\bibnamefont {Ji}}, \bibinfo {author} {\bibfnamefont {L.}~\bibnamefont
  {Shi}}, \bibinfo {author} {\bibfnamefont {K.}~\bibnamefont {Chen}}, \bibinfo
  {author} {\bibfnamefont {J.}~\bibnamefont {Lu}}, \bibinfo {author}
  {\bibfnamefont {Y.}~\bibnamefont {Jing}},\ and\ \bibinfo {author}
  {\bibfnamefont {W.~A.}\ \bibnamefont {Benalcazar}},\ }\bibfield  {title}
  {\bibinfo {title} {{Higher-order Skin Effect through a
  Hermitian-non-Hermitian Correspondence and Its Observation in an Acoustic
  Kagome Lattice}},\ }\href {https://arxiv.org/abs/2409.01516} {\bibfield
  {journal} {\bibinfo  {journal} {arXiv:2409.01516}\ } (\bibinfo {year}
  {2024})}\BibitemShut {NoStop}%
\bibitem [{\citenamefont {Terrier}\ and\ \citenamefont
  {Kunst}(2020)}]{PhysRevResearch.2.023364}%
  \BibitemOpen
  \bibfield  {author} {\bibinfo {author} {\bibfnamefont {F.}~\bibnamefont
  {Terrier}}\ and\ \bibinfo {author} {\bibfnamefont {F.~K.}\ \bibnamefont
  {Kunst}},\ }\bibfield  {title} {\bibinfo {title} {{Dissipative analog of
  four-dimensional quantum Hall physics}},\ }\href
  {https://doi.org/10.1103/PhysRevResearch.2.023364} {\bibfield  {journal}
  {\bibinfo  {journal} {Phys. Rev. Res.}\ }\textbf {\bibinfo {volume} {2}},\
  \bibinfo {pages} {023364} (\bibinfo {year} {2020})}\BibitemShut {NoStop}%
\bibitem [{\citenamefont {Kawabata}\ \emph {et~al.}(2021)\citenamefont
  {Kawabata}, \citenamefont {Shiozaki},\ and\ \citenamefont
  {Ryu}}]{PhysRevLett.126.216405}%
  \BibitemOpen
  \bibfield  {author} {\bibinfo {author} {\bibfnamefont {K.}~\bibnamefont
  {Kawabata}}, \bibinfo {author} {\bibfnamefont {K.}~\bibnamefont {Shiozaki}},\
  and\ \bibinfo {author} {\bibfnamefont {S.}~\bibnamefont {Ryu}},\ }\bibfield
  {title} {\bibinfo {title} {{Topological Field Theory of Non-Hermitian
  Systems}},\ }\href {https://doi.org/10.1103/PhysRevLett.126.216405}
  {\bibfield  {journal} {\bibinfo  {journal} {Phys. Rev. Lett.}\ }\textbf
  {\bibinfo {volume} {126}},\ \bibinfo {pages} {216405} (\bibinfo {year}
  {2021})}\BibitemShut {NoStop}%
\bibitem [{\citenamefont {Denner}\ \emph {et~al.}(2021)\citenamefont {Denner},
  \citenamefont {Skurativska}, \citenamefont {Schindler}, \citenamefont
  {Fischer}, \citenamefont {Thomale}, \citenamefont {Bzdu{\v{s}}ek},\ and\
  \citenamefont {Neupert}}]{denner2021exceptional}%
  \BibitemOpen
  \bibfield  {author} {\bibinfo {author} {\bibfnamefont {M.~M.}\ \bibnamefont
  {Denner}}, \bibinfo {author} {\bibfnamefont {A.}~\bibnamefont {Skurativska}},
  \bibinfo {author} {\bibfnamefont {F.}~\bibnamefont {Schindler}}, \bibinfo
  {author} {\bibfnamefont {M.~H.}\ \bibnamefont {Fischer}}, \bibinfo {author}
  {\bibfnamefont {R.}~\bibnamefont {Thomale}}, \bibinfo {author} {\bibfnamefont
  {T.}~\bibnamefont {Bzdu{\v{s}}ek}},\ and\ \bibinfo {author} {\bibfnamefont
  {T.}~\bibnamefont {Neupert}},\ }\bibfield  {title} {\bibinfo {title}
  {Exceptional topological insulators},\ }\href
  {https://www.nature.com/articles/s41467-021-25947-z} {\bibfield  {journal}
  {\bibinfo  {journal} {Nat. Commun.}\ }\textbf {\bibinfo {volume} {12}},\
  \bibinfo {pages} {5681} (\bibinfo {year} {2021})}\BibitemShut {NoStop}%
\bibitem [{\citenamefont {Denner}\ \emph {et~al.}(2023)\citenamefont {Denner},
  \citenamefont {Neupert},\ and\ \citenamefont
  {Schindler}}]{Denner_2023_infernal}%
  \BibitemOpen
  \bibfield  {author} {\bibinfo {author} {\bibfnamefont {M.~M.}\ \bibnamefont
  {Denner}}, \bibinfo {author} {\bibfnamefont {T.}~\bibnamefont {Neupert}},\
  and\ \bibinfo {author} {\bibfnamefont {F.}~\bibnamefont {Schindler}},\
  }\bibfield  {title} {\bibinfo {title} {{Infernal and exceptional edge modes:
  non-Hermitian topology beyond the skin effect}},\ }\href
  {https://doi.org/10.1088/2515-7639/acf2ca} {\bibfield  {journal} {\bibinfo
  {journal} {J. Phys. Mater.}\ }\textbf {\bibinfo {volume} {6}},\ \bibinfo
  {pages} {045006} (\bibinfo {year} {2023})}\BibitemShut {NoStop}%
\bibitem [{\citenamefont {Schindler}\ \emph {et~al.}(2023)\citenamefont
  {Schindler}, \citenamefont {Gu}, \citenamefont {Lian},\ and\ \citenamefont
  {Kawabata}}]{PRXQuantum.4.030315}%
  \BibitemOpen
  \bibfield  {author} {\bibinfo {author} {\bibfnamefont {F.}~\bibnamefont
  {Schindler}}, \bibinfo {author} {\bibfnamefont {K.}~\bibnamefont {Gu}},
  \bibinfo {author} {\bibfnamefont {B.}~\bibnamefont {Lian}},\ and\ \bibinfo
  {author} {\bibfnamefont {K.}~\bibnamefont {Kawabata}},\ }\bibfield  {title}
  {\bibinfo {title} {{Hermitian Bulk -- Non-Hermitian Boundary
  Correspondence}},\ }\href {https://doi.org/10.1103/PRXQuantum.4.030315}
  {\bibfield  {journal} {\bibinfo  {journal} {PRX Quantum}\ }\textbf {\bibinfo
  {volume} {4}},\ \bibinfo {pages} {030315} (\bibinfo {year}
  {2023})}\BibitemShut {NoStop}%
\bibitem [{\citenamefont {Nakamura}\ \emph {et~al.}(2023)\citenamefont
  {Nakamura}, \citenamefont {Inaka}, \citenamefont {Okuma},\ and\ \citenamefont
  {Sato}}]{PhysRevLett.131.256602}%
  \BibitemOpen
  \bibfield  {author} {\bibinfo {author} {\bibfnamefont {D.}~\bibnamefont
  {Nakamura}}, \bibinfo {author} {\bibfnamefont {K.}~\bibnamefont {Inaka}},
  \bibinfo {author} {\bibfnamefont {N.}~\bibnamefont {Okuma}},\ and\ \bibinfo
  {author} {\bibfnamefont {M.}~\bibnamefont {Sato}},\ }\bibfield  {title}
  {\bibinfo {title} {{Universal Platform of Point-Gap Topological Phases from
  Topological Materials}},\ }\href
  {https://doi.org/10.1103/PhysRevLett.131.256602} {\bibfield  {journal}
  {\bibinfo  {journal} {Phys. Rev. Lett.}\ }\textbf {\bibinfo {volume} {131}},\
  \bibinfo {pages} {256602} (\bibinfo {year} {2023})}\BibitemShut {NoStop}%
\bibitem [{\citenamefont {Dash}\ \emph {et~al.}(2024)\citenamefont {Dash},
  \citenamefont {Bid},\ and\ \citenamefont {Thakurathi}}]{PhysRevB.109.035418}%
  \BibitemOpen
  \bibfield  {author} {\bibinfo {author} {\bibfnamefont {G.~K.}\ \bibnamefont
  {Dash}}, \bibinfo {author} {\bibfnamefont {S.}~\bibnamefont {Bid}},\ and\
  \bibinfo {author} {\bibfnamefont {M.}~\bibnamefont {Thakurathi}},\ }\bibfield
   {title} {\bibinfo {title} {{Floquet exceptional topological insulator}},\
  }\href {https://doi.org/10.1103/PhysRevB.109.035418} {\bibfield  {journal}
  {\bibinfo  {journal} {Phys. Rev. B}\ }\textbf {\bibinfo {volume} {109}},\
  \bibinfo {pages} {035418} (\bibinfo {year} {2024})}\BibitemShut {NoStop}%
\bibitem [{\citenamefont {Hamanaka}\ \emph {et~al.}(2024)\citenamefont
  {Hamanaka}, \citenamefont {Yoshida},\ and\ \citenamefont
  {Kawabata}}]{Hamanaka2024PRL}%
  \BibitemOpen
  \bibfield  {author} {\bibinfo {author} {\bibfnamefont {S.}~\bibnamefont
  {Hamanaka}}, \bibinfo {author} {\bibfnamefont {T.}~\bibnamefont {Yoshida}},\
  and\ \bibinfo {author} {\bibfnamefont {K.}~\bibnamefont {Kawabata}},\
  }\bibfield  {title} {\bibinfo {title} {{Non-Hermitian Topology in Hermitian
  Topological Matter}},\ }\href
  {https://doi.org/10.1103/PhysRevLett.133.266604} {\bibfield  {journal}
  {\bibinfo  {journal} {Phys. Rev. Lett.}\ }\textbf {\bibinfo {volume} {133}},\
  \bibinfo {pages} {266604} (\bibinfo {year} {2024})}\BibitemShut {NoStop}%
\bibitem [{\citenamefont {Geier}\ \emph {et~al.}(2018)\citenamefont {Geier},
  \citenamefont {Trifunovic}, \citenamefont {Hoskam},\ and\ \citenamefont
  {Brouwer}}]{physrevb.97.205135}%
  \BibitemOpen
  \bibfield  {author} {\bibinfo {author} {\bibfnamefont {M.}~\bibnamefont
  {Geier}}, \bibinfo {author} {\bibfnamefont {L.}~\bibnamefont {Trifunovic}},
  \bibinfo {author} {\bibfnamefont {M.}~\bibnamefont {Hoskam}},\ and\ \bibinfo
  {author} {\bibfnamefont {P.~W.}\ \bibnamefont {Brouwer}},\ }\bibfield
  {title} {\bibinfo {title} {{Second-order topological insulators and
  superconductors with an order-two crystalline symmetry}},\ }\href
  {https://doi.org/10.1103/PhysRevB.97.205135} {\bibfield  {journal} {\bibinfo
  {journal} {Phys. Rev. B}\ }\textbf {\bibinfo {volume} {97}},\ \bibinfo
  {pages} {205135} (\bibinfo {year} {2018})}\BibitemShut {NoStop}%
\bibitem [{\citenamefont {Fulga}\ \emph {et~al.}(2016)\citenamefont {Fulga},
  \citenamefont {Avraham}, \citenamefont {Beidenkopf},\ and\ \citenamefont
  {Stern}}]{PhysRevB.94.125405}%
  \BibitemOpen
  \bibfield  {author} {\bibinfo {author} {\bibfnamefont {I.~C.}\ \bibnamefont
  {Fulga}}, \bibinfo {author} {\bibfnamefont {N.}~\bibnamefont {Avraham}},
  \bibinfo {author} {\bibfnamefont {H.}~\bibnamefont {Beidenkopf}},\ and\
  \bibinfo {author} {\bibfnamefont {A.}~\bibnamefont {Stern}},\ }\bibfield
  {title} {\bibinfo {title} {{Coupled-layer description of topological
  crystalline insulators}},\ }\href
  {https://doi.org/10.1103/PhysRevB.94.125405} {\bibfield  {journal} {\bibinfo
  {journal} {Phys. Rev. B}\ }\textbf {\bibinfo {volume} {94}},\ \bibinfo
  {pages} {125405} (\bibinfo {year} {2016})}\BibitemShut {NoStop}%
\bibitem [{\citenamefont {Ezawa}(2016)}]{PhysRevB.94.155148}%
  \BibitemOpen
  \bibfield  {author} {\bibinfo {author} {\bibfnamefont {M.}~\bibnamefont
  {Ezawa}},\ }\bibfield  {title} {\bibinfo {title} {{Hourglass fermion surface
  states in stacked topological insulators with nonsymmorphic symmetry}},\
  }\href {https://doi.org/10.1103/PhysRevB.94.155148} {\bibfield  {journal}
  {\bibinfo  {journal} {Phys. Rev. B}\ }\textbf {\bibinfo {volume} {94}},\
  \bibinfo {pages} {155148} (\bibinfo {year} {2016})}\BibitemShut {NoStop}%
\bibitem [{\citenamefont {Song}\ \emph
  {et~al.}(2017{\natexlab{b}})\citenamefont {Song}, \citenamefont {Huang},
  \citenamefont {Fu},\ and\ \citenamefont {Hermele}}]{PhysRevX.7.011020}%
  \BibitemOpen
  \bibfield  {author} {\bibinfo {author} {\bibfnamefont {H.}~\bibnamefont
  {Song}}, \bibinfo {author} {\bibfnamefont {S.-J.}\ \bibnamefont {Huang}},
  \bibinfo {author} {\bibfnamefont {L.}~\bibnamefont {Fu}},\ and\ \bibinfo
  {author} {\bibfnamefont {M.}~\bibnamefont {Hermele}},\ }\bibfield  {title}
  {\bibinfo {title} {{Topological Phases Protected by Point Group Symmetry}},\
  }\href {https://doi.org/10.1103/PhysRevX.7.011020} {\bibfield  {journal}
  {\bibinfo  {journal} {Phys. Rev. X}\ }\textbf {\bibinfo {volume} {7}},\
  \bibinfo {pages} {011020} (\bibinfo {year} {2017}{\natexlab{b}})}\BibitemShut
  {NoStop}%
\bibitem [{\citenamefont {Huang}\ \emph {et~al.}(2017)\citenamefont {Huang},
  \citenamefont {Song}, \citenamefont {Huang},\ and\ \citenamefont
  {Hermele}}]{PhysRevB.96.205106}%
  \BibitemOpen
  \bibfield  {author} {\bibinfo {author} {\bibfnamefont {S.-J.}\ \bibnamefont
  {Huang}}, \bibinfo {author} {\bibfnamefont {H.}~\bibnamefont {Song}},
  \bibinfo {author} {\bibfnamefont {Y.-P.}\ \bibnamefont {Huang}},\ and\
  \bibinfo {author} {\bibfnamefont {M.}~\bibnamefont {Hermele}},\ }\bibfield
  {title} {\bibinfo {title} {{Building crystalline topological phases from
  lower-dimensional states}},\ }\href
  {https://doi.org/10.1103/PhysRevB.96.205106} {\bibfield  {journal} {\bibinfo
  {journal} {Phys. Rev. B}\ }\textbf {\bibinfo {volume} {96}},\ \bibinfo
  {pages} {205106} (\bibinfo {year} {2017})}\BibitemShut {NoStop}%
\bibitem [{\citenamefont {Song}\ \emph {et~al.}(2018)\citenamefont {Song},
  \citenamefont {Zhang}, \citenamefont {Fang},\ and\ \citenamefont
  {Fang}}]{song2018quantitative}%
  \BibitemOpen
  \bibfield  {author} {\bibinfo {author} {\bibfnamefont {Z.}~\bibnamefont
  {Song}}, \bibinfo {author} {\bibfnamefont {T.}~\bibnamefont {Zhang}},
  \bibinfo {author} {\bibfnamefont {Z.}~\bibnamefont {Fang}},\ and\ \bibinfo
  {author} {\bibfnamefont {C.}~\bibnamefont {Fang}},\ }\bibfield  {title}
  {\bibinfo {title} {{Quantitative mappings between symmetry and topology in
  solids}},\ }\href {https://www.nature.com/articles/s41467-018-06010-w}
  {\bibfield  {journal} {\bibinfo  {journal} {Nat. Commun.}\ }\textbf {\bibinfo
  {volume} {9}},\ \bibinfo {pages} {3530} (\bibinfo {year} {2018})}\BibitemShut
  {NoStop}%
\bibitem [{\citenamefont {Trifunovic}\ and\ \citenamefont
  {Brouwer}(2019)}]{PhysRevX.9.011012}%
  \BibitemOpen
  \bibfield  {author} {\bibinfo {author} {\bibfnamefont {L.}~\bibnamefont
  {Trifunovic}}\ and\ \bibinfo {author} {\bibfnamefont {P.~W.}\ \bibnamefont
  {Brouwer}},\ }\bibfield  {title} {\bibinfo {title} {{Higher-Order
  Bulk-Boundary Correspondence for Topological Crystalline Phases}},\ }\href
  {https://doi.org/10.1103/PhysRevX.9.011012} {\bibfield  {journal} {\bibinfo
  {journal} {Phys. Rev. X}\ }\textbf {\bibinfo {volume} {9}},\ \bibinfo {pages}
  {011012} (\bibinfo {year} {2019})}\BibitemShut {NoStop}%
\bibitem [{\citenamefont {Khalaf}(2018)}]{PhysRevB.97.205136}%
  \BibitemOpen
  \bibfield  {author} {\bibinfo {author} {\bibfnamefont {E.}~\bibnamefont
  {Khalaf}},\ }\bibfield  {title} {\bibinfo {title} {{Higher-order topological
  insulators and superconductors protected by inversion symmetry}},\ }\href
  {https://doi.org/10.1103/PhysRevB.97.205136} {\bibfield  {journal} {\bibinfo
  {journal} {Phys. Rev. B}\ }\textbf {\bibinfo {volume} {97}},\ \bibinfo
  {pages} {205136} (\bibinfo {year} {2018})}\BibitemShut {NoStop}%
\bibitem [{\citenamefont {Ono}\ and\ \citenamefont
  {Watanabe}(2018)}]{PhysRevB.98.115150}%
  \BibitemOpen
  \bibfield  {author} {\bibinfo {author} {\bibfnamefont {S.}~\bibnamefont
  {Ono}}\ and\ \bibinfo {author} {\bibfnamefont {H.}~\bibnamefont {Watanabe}},\
  }\bibfield  {title} {\bibinfo {title} {{Unified understanding of symmetry
  indicators for all internal symmetry classes}},\ }\href
  {https://doi.org/10.1103/PhysRevB.98.115150} {\bibfield  {journal} {\bibinfo
  {journal} {Phys. Rev. B}\ }\textbf {\bibinfo {volume} {98}},\ \bibinfo
  {pages} {115150} (\bibinfo {year} {2018})}\BibitemShut {NoStop}%
\bibitem [{\citenamefont {Matsugatani}\ and\ \citenamefont
  {Watanabe}(2018)}]{PhysRevB.98.205129}%
  \BibitemOpen
  \bibfield  {author} {\bibinfo {author} {\bibfnamefont {A.}~\bibnamefont
  {Matsugatani}}\ and\ \bibinfo {author} {\bibfnamefont {H.}~\bibnamefont
  {Watanabe}},\ }\bibfield  {title} {\bibinfo {title} {{Connecting higher-order
  topological insulators to lower-dimensional topological insulators}},\ }\href
  {https://doi.org/10.1103/PhysRevB.98.205129} {\bibfield  {journal} {\bibinfo
  {journal} {Phys. Rev. B}\ }\textbf {\bibinfo {volume} {98}},\ \bibinfo
  {pages} {205129} (\bibinfo {year} {2018})}\BibitemShut {NoStop}%
\bibitem [{\citenamefont {Kooi}\ \emph {et~al.}(2018)\citenamefont {Kooi},
  \citenamefont {van Miert},\ and\ \citenamefont {Ortix}}]{PhysRevB.98.245102}%
  \BibitemOpen
  \bibfield  {author} {\bibinfo {author} {\bibfnamefont {S.~H.}\ \bibnamefont
  {Kooi}}, \bibinfo {author} {\bibfnamefont {G.}~\bibnamefont {van Miert}},\
  and\ \bibinfo {author} {\bibfnamefont {C.}~\bibnamefont {Ortix}},\ }\bibfield
   {title} {\bibinfo {title} {{Inversion-symmetry protected chiral hinge states
  in stacks of doped quantum Hall layers}},\ }\href
  {https://doi.org/10.1103/PhysRevB.98.245102} {\bibfield  {journal} {\bibinfo
  {journal} {Phys. Rev. B}\ }\textbf {\bibinfo {volume} {98}},\ \bibinfo
  {pages} {245102} (\bibinfo {year} {2018})}\BibitemShut {NoStop}%
\bibitem [{\citenamefont {Tanaka}\ \emph {et~al.}(2020)\citenamefont {Tanaka},
  \citenamefont {Takahashi}, \citenamefont {Zhang},\ and\ \citenamefont
  {Murakami}}]{PhysRevResearch.2.043274}%
  \BibitemOpen
  \bibfield  {author} {\bibinfo {author} {\bibfnamefont {Y.}~\bibnamefont
  {Tanaka}}, \bibinfo {author} {\bibfnamefont {R.}~\bibnamefont {Takahashi}},
  \bibinfo {author} {\bibfnamefont {T.}~\bibnamefont {Zhang}},\ and\ \bibinfo
  {author} {\bibfnamefont {S.}~\bibnamefont {Murakami}},\ }\bibfield  {title}
  {\bibinfo {title} {{Theory of inversion-${\mathbb{Z}}_{4}$ protected
  topological chiral hinge states and its applications to layered
  antiferromagnets}},\ }\href
  {https://doi.org/10.1103/PhysRevResearch.2.043274} {\bibfield  {journal}
  {\bibinfo  {journal} {Phys. Rev. Res.}\ }\textbf {\bibinfo {volume} {2}},\
  \bibinfo {pages} {043274} (\bibinfo {year} {2020})}\BibitemShut {NoStop}%
\bibitem [{\citenamefont {Feng}\ \emph {et~al.}(2014)\citenamefont {Feng},
  \citenamefont {Wong}, \citenamefont {Ma}, \citenamefont {Wang},\ and\
  \citenamefont {Zhang}}]{Feng2014Science}%
  \BibitemOpen
  \bibfield  {author} {\bibinfo {author} {\bibfnamefont {L.}~\bibnamefont
  {Feng}}, \bibinfo {author} {\bibfnamefont {Z.~J.}\ \bibnamefont {Wong}},
  \bibinfo {author} {\bibfnamefont {R.-M.}\ \bibnamefont {Ma}}, \bibinfo
  {author} {\bibfnamefont {Y.}~\bibnamefont {Wang}},\ and\ \bibinfo {author}
  {\bibfnamefont {X.}~\bibnamefont {Zhang}},\ }\bibfield  {title} {\bibinfo
  {title} {{Single-mode laser by parity-time symmetry breaking}},\ }\href
  {https://doi.org/10.1126/science.1258479} {\bibfield  {journal} {\bibinfo
  {journal} {Science}\ }\textbf {\bibinfo {volume} {346}},\ \bibinfo {pages}
  {972} (\bibinfo {year} {2014})}\BibitemShut {NoStop}%
\bibitem [{\citenamefont {Hodaei}\ \emph {et~al.}(2014)\citenamefont {Hodaei},
  \citenamefont {Miri}, \citenamefont {Heinrich}, \citenamefont
  {Christodoulides},\ and\ \citenamefont {Khajavikhan}}]{Hodaei2014Science}%
  \BibitemOpen
  \bibfield  {author} {\bibinfo {author} {\bibfnamefont {H.}~\bibnamefont
  {Hodaei}}, \bibinfo {author} {\bibfnamefont {M.-A.}\ \bibnamefont {Miri}},
  \bibinfo {author} {\bibfnamefont {M.}~\bibnamefont {Heinrich}}, \bibinfo
  {author} {\bibfnamefont {D.~N.}\ \bibnamefont {Christodoulides}},\ and\
  \bibinfo {author} {\bibfnamefont {M.}~\bibnamefont {Khajavikhan}},\
  }\bibfield  {title} {\bibinfo {title} {{Parity-time--symmetric microring
  lasers}},\ }\href {https://www.science.org/doi/full/10.1126/science.1258480}
  {\bibfield  {journal} {\bibinfo  {journal} {Science}\ }\textbf {\bibinfo
  {volume} {346}},\ \bibinfo {pages} {975} (\bibinfo {year}
  {2014})}\BibitemShut {NoStop}%
\bibitem [{\citenamefont {Doppler}\ \emph {et~al.}(2016)\citenamefont
  {Doppler}, \citenamefont {Mailybaev}, \citenamefont {B{\"o}hm}, \citenamefont
  {Kuhl}, \citenamefont {Girschik}, \citenamefont {Libisch}, \citenamefont
  {Milburn}, \citenamefont {Rabl}, \citenamefont {Moiseyev},\ and\
  \citenamefont {Rotter}}]{Doppler2016Nature}%
  \BibitemOpen
  \bibfield  {author} {\bibinfo {author} {\bibfnamefont {J.}~\bibnamefont
  {Doppler}}, \bibinfo {author} {\bibfnamefont {A.~A.}\ \bibnamefont
  {Mailybaev}}, \bibinfo {author} {\bibfnamefont {J.}~\bibnamefont {B{\"o}hm}},
  \bibinfo {author} {\bibfnamefont {U.}~\bibnamefont {Kuhl}}, \bibinfo {author}
  {\bibfnamefont {A.}~\bibnamefont {Girschik}}, \bibinfo {author}
  {\bibfnamefont {F.}~\bibnamefont {Libisch}}, \bibinfo {author} {\bibfnamefont
  {T.~J.}\ \bibnamefont {Milburn}}, \bibinfo {author} {\bibfnamefont
  {P.}~\bibnamefont {Rabl}}, \bibinfo {author} {\bibfnamefont {N.}~\bibnamefont
  {Moiseyev}},\ and\ \bibinfo {author} {\bibfnamefont {S.}~\bibnamefont
  {Rotter}},\ }\bibfield  {title} {\bibinfo {title} {{Dynamically encircling an
  exceptional point for asymmetric mode switching}},\ }\href
  {https://doi.org/10.1038/nature18605} {\bibfield  {journal} {\bibinfo
  {journal} {Nature}\ }\textbf {\bibinfo {volume} {537}},\ \bibinfo {pages}
  {76} (\bibinfo {year} {2016})}\BibitemShut {NoStop}%
\bibitem [{\citenamefont {Xu}\ \emph {et~al.}(2016)\citenamefont {Xu},
  \citenamefont {Mason}, \citenamefont {Jiang},\ and\ \citenamefont
  {Harris}}]{Xu2016Nature}%
  \BibitemOpen
  \bibfield  {author} {\bibinfo {author} {\bibfnamefont {H.}~\bibnamefont
  {Xu}}, \bibinfo {author} {\bibfnamefont {D.}~\bibnamefont {Mason}}, \bibinfo
  {author} {\bibfnamefont {L.}~\bibnamefont {Jiang}},\ and\ \bibinfo {author}
  {\bibfnamefont {J.~G.~E.}\ \bibnamefont {Harris}},\ }\bibfield  {title}
  {\bibinfo {title} {{Topological energy transfer in an optomechanical system
  with exceptional points}},\ }\href {https://doi.org/10.1038/nature18604}
  {\bibfield  {journal} {\bibinfo  {journal} {Nature}\ }\textbf {\bibinfo
  {volume} {537}},\ \bibinfo {pages} {80} (\bibinfo {year} {2016})}\BibitemShut
  {NoStop}%
\bibitem [{\citenamefont {Hodaei}\ \emph {et~al.}(2017)\citenamefont {Hodaei},
  \citenamefont {Hassan}, \citenamefont {Wittek}, \citenamefont
  {Garcia-Gracia}, \citenamefont {El-Ganainy}, \citenamefont
  {Christodoulides},\ and\ \citenamefont {Khajavikhan}}]{Hodaei2017Nature}%
  \BibitemOpen
  \bibfield  {author} {\bibinfo {author} {\bibfnamefont {H.}~\bibnamefont
  {Hodaei}}, \bibinfo {author} {\bibfnamefont {A.~U.}\ \bibnamefont {Hassan}},
  \bibinfo {author} {\bibfnamefont {S.}~\bibnamefont {Wittek}}, \bibinfo
  {author} {\bibfnamefont {H.}~\bibnamefont {Garcia-Gracia}}, \bibinfo {author}
  {\bibfnamefont {R.}~\bibnamefont {El-Ganainy}}, \bibinfo {author}
  {\bibfnamefont {D.~N.}\ \bibnamefont {Christodoulides}},\ and\ \bibinfo
  {author} {\bibfnamefont {M.}~\bibnamefont {Khajavikhan}},\ }\bibfield
  {title} {\bibinfo {title} {{Enhanced sensitivity at higher-order exceptional
  points}},\ }\href {https://doi.org/10.1038/nature23280} {\bibfield  {journal}
  {\bibinfo  {journal} {Nature}\ }\textbf {\bibinfo {volume} {548}},\ \bibinfo
  {pages} {187} (\bibinfo {year} {2017})}\BibitemShut {NoStop}%
\bibitem [{\citenamefont {Chen}\ \emph {et~al.}(2017)\citenamefont {Chen},
  \citenamefont {Kaya~{\"O}zdemir}, \citenamefont {Zhao}, \citenamefont
  {Wiersig},\ and\ \citenamefont {Yang}}]{Chen2017Nature}%
  \BibitemOpen
  \bibfield  {author} {\bibinfo {author} {\bibfnamefont {W.}~\bibnamefont
  {Chen}}, \bibinfo {author} {\bibfnamefont {{\c{S}}.}~\bibnamefont
  {Kaya~{\"O}zdemir}}, \bibinfo {author} {\bibfnamefont {G.}~\bibnamefont
  {Zhao}}, \bibinfo {author} {\bibfnamefont {J.}~\bibnamefont {Wiersig}},\ and\
  \bibinfo {author} {\bibfnamefont {L.}~\bibnamefont {Yang}},\ }\bibfield
  {title} {\bibinfo {title} {{Exceptional points enhance sensing in an optical
  microcavity}},\ }\href {https://doi.org/10.1038/nature23281} {\bibfield
  {journal} {\bibinfo  {journal} {Nature}\ }\textbf {\bibinfo {volume} {548}},\
  \bibinfo {pages} {192} (\bibinfo {year} {2017})}\BibitemShut {NoStop}%
\end{thebibliography}

%\begin{comment}
%
%\end{comment}

\end{document}